\documentstyle[12pt,axodraw]{article}

\newcommand{\mysection}{\setcounter{equation}{0}\section}

\begin{document}
%-----------------------------
%\format=latex
%\documentstyle[12pt]{article}
%\begin{document}
%--------------------------------------------
%\pagenumbering{}
\vskip 0.2cm
\hfill{YITP-SB-99-15}
\vskip 0.2cm
\hfill{INLO-PUB-12/99}\\[1cm]
\vskip 0.2cm
\centerline{\large\bf {Comparison between variable flavor number }}
\centerline{\large\bf {schemes for charm quark electroproduction}}
\vskip 0.2cm
%\centerline {\sc A. Chuvakin }
\centerline {A. Chuvakin, J. Smith 
%\footnote{partially supported
%by the National Science Foundation grant PHY-9722101}
}
\centerline{\it C.N. Yang Institute for Theoretical Physics,}
\centerline{\it State University of New York at Stony Brook,
New York 11794-3840, USA.}
\vskip 0.2cm
\centerline {W.L. van Neerven 
%\footnote{Work supported
%by the EC network `QCD and Particle Structure' under contract 
%\hspace*{5mm} No.~FMRX--CT98--0194.}
}
\centerline{\it Instituut-Lorentz,
University of Leiden, PO Box 9506, 2300 RA Leiden,}
%\centerline{\it University of Leiden,}
%\centerline{\it PO Box 9506, 2300 RA Leiden,}
\centerline{\it The Netherlands.}
\vskip 0.2cm
\centerline{October 1999}
\vskip 0.2cm
\centerline{\bf Abstract}
\vskip 0.3cm
A comparison is made between two variable flavor number schemes
which describe charm quark production in deep inelastic electron-proton
scattering. In these schemes the coefficient functions are derived 
from mass factorization of
the heavy quark coefficient functions presented in a fixed flavor number 
scheme. Since the coefficient functions in the variable
flavor number schemes have to be finite in the limit $m\rightarrow 0$
we have defined a prescription for those processes where the
virtual photon is attached to a light quark.
Furthermore one has to
construct a parton density set with four active flavors (u,d,s,c)
out of a set which only contains three light flavors (u,d,s). In order
$\alpha_s^2$ the two sets are discontinuous at $\mu=m_c$ which follows from 
mass factorization of the heavy quark coefficient functions.
The charm component of the structure function $F_{2,c}$ is insensitive
to the different variable flavor number schemes. 
In particular in the threshold region they both agree with the description
in fixed order perturbation theory presented in a three flavor scheme.
However one version does not lead to a correct description of the threshold 
behavior of the longitudinal structure function $F_{L,c}$. This
happens when one requires a non-vanishing zeroth order longitudinal
coefficient function.

\vfill
%\end{document}

%\format=latex
%\documentstyle[12pt]{article}
%\pagestyle{myheadings}  
%\begin{document}
%------------------This is Section 1---------------------------------
\mysection{Introduction}
%----------------------------------------------------------
%\topmargin=0in
%\headheight=0in
%\headsep=0in
%\oddsidemargin=7.2pt
%\evensidemargin=7.2pt
%\footheight=1in
%\marginparwidth=0in
%\marginparsep=0in
%\textheight=9in
%\textwidth=6in
\newcommand{\be}{\begin{eqnarray}}
\newcommand{\ee}{\end{eqnarray}}

Charm quark production is one of the important reactions used to extract the
gluon density $f_g(x,\mu^2)$ of the proton in deep inelastic lepton-hadron
scattering, especially when the Bjorken scaling variable $x$ is small.
However this is only true when the deep inelastic process
is of the neutral current type and the charm component of the proton wave
function is negligible. In this case the charm quark is produced in 
the so-called extrinsic way. For neutral current processes with 
only light partons in the initial state this means that the Born 
approximation in perturbative QCD is given by the virtual vector-boson 
gluon-fusion process \cite{Wit}. Notice that the light partons consist of 
the gluon and the three light flavors u, d, s together 
with their anti-particles. Furthermore if the virtuality of the exchanged
vector boson in deep inelastic lepton-hadron scattering satisfies
$Q^2 \ll M_Z^2$ then the vector boson is represented by the photon only 
and the contribution of the $Z$-boson is negligible. 
Extrinsic charm production also receives next-to-leading order
(NLO) contributions from boson-quark subprocesses, which could 
hamper the extraction of the gluon density. Fortunately this is not the case
at HERA, where the experiments \cite{ZEUS}, \cite {H1} are carried out at 
small $x$, because the gluon density overwhelms the light flavor densities 
completely. Moreover the NLO quark initiated processes are 
suppressed by at least one power of the strong coupling constant 
$\alpha_s(\mu^2)$ with respect to the Born contribution to the 
boson-gluon fusion reaction.
The quantity $\mu$ in the running coupling constant and
the parton densities represents both the renormalization and 
factorization scales respectively, because it is convenient to chose them
to be equal. 

In the literature one has adopted two different 
treatments of extrinsic charm production, which are known as the massive and 
massless charm descriptions. The former, advocated in \cite{grs}, treats the
charm quark as a heavy quark (with mass $m_c$) 
and the cross sections or coefficient functions
have to be described by fixed order perturbation theory. Notice that due to
the work in \cite{lrsn} the perturbation series is now known up to second
order and the NLO massive charm approach agrees with the recent
data in \cite{ZEUS} and \cite{H1}. The latter treatment, 
which has been rather popular among groups which
fitted parton densities to experimental data, 
treats the charm quark as a massless quark
so that it can be represented by a parton density $f_c(x,\mu^2)$, with
the boundary condition $f_c(x,\mu^2)=0$ for $\mu \le m_c$. Although at first
sight these approaches are completely different they are actually
intimately related. It was shown in \cite{bmsn1} that the
large logarithms of the type $\ln(Q^2/m_c^2)$, which appear in the 
perturbation series when $Q^2 \gg m_c^2$,
can be resummed in all orders. The upshot of this 
procedure is that the charm components of the deep inelastic structure 
functions $F_{i,c}(x,Q^2,m_c^2)$, where $i=2,L$, which in the 
first approach are written as convolutions of heavy quark coefficient 
functions with light parton densities, become,
after resummation, convolutions of light parton coefficient functions
with light parton densities which also include a charm quark density. 
This procedure leads to the so-called zero mass variable flavor number scheme
(ZM-VFNS) for $F_{i,c}(x,Q^2)$ where the mass 
of the charm quark is absorbed into the new four flavor densities. 
To implement this scheme one has to be careful to use quantities which
are collinearly finite in the limit $m_c \rightarrow 0$.
From the above considerations it is clear that the first 
approach is better when the charm quark pair is
produced near threshold because the mass of the quark is important 
in this region and it cannot be neglected. On the other hand far away from 
threshold, where also $Q^2\gg m_c^2$, 
the large logarithms above dominate the structure 
functions so that the second approach should be more appropriate. Both
approaches are characterized by the number of active flavors involved in the
description of the parton densities which are given by three and four 
respectively. Therefore one can also speak of three and four flavor number 
schemes (TFNS and FFNS respectively). Each scheme has a different gluon 
density so that the momentum sum rule is always satisfied. 

As most of the experimental data occur in the kinematical regime which
is between the threshold and the region of large $Q^2$  
a third approach has been introduced to describe the charm components of 
the structure functions.
This is called the variable flavor number scheme (VFNS). A first 
discussion was given by Aivasis, Collins, Olness and Tung \cite{acot},
where a VFNS prescription called ACOT was given in lowest order
only. The ACOT results were compared with the NLO results in
\cite{olri}. We will give our NLO version of a VFNS scheme in this paper and
we call it the CSN scheme to distinguish it.
A different approach, generalized to all orders, 
was given in the papers by Buza, Matiounine, Smith and van Neerven 
\cite{bmsn1},\cite{bmsn2}, which we denote by BMSN. 
Finally another version of a VFNS for the charm component of the
structure function was presented by Thorne and Roberts
in \cite{thro}, which will be called the TR scheme. 
Note that a proof of factorization to all orders for the total 
structure function, which includes charm and light parton production,
was recently given in \cite{col}.

The difference between the various versions can be attributed to
two ingredients entering the construction of a VFNS. 
The first one is the mass factorization procedure carried out before the 
large logarithms can be resummed. The second one is the matching condition 
imposed on the charm quark density, which has to vanish in the threshold region
of the production process. It will be one of our goals to elucidate these 
differences
in the next Section. Another problem, which was not clarified in the papers
above is that the mass factorization cannot be carried
out on the level of the charm components of the structure functions alone,
because one also needs contributions coming from the light parton components
of the structure functions. The latter can be attributed to all heavy charm 
quark loop contributions to gluon self energies, which appear in 
the virtual corrections
to the light parton coefficient functions. These corrections have to be 
combined with contributions from gluon splitting into heavy charm anti-charm 
quark pairs, which belong to the charm components 
(not the light quark components) of the
structure functions. In this paper we will give a much more careful analysis
than has been done previously in the literature. Another aspect of any VFNS
approach is that one needs two sets of parton densities. One set only
contains densities in a three flavor number scheme whereas the second one,
which also includes a charm quark density, is parametrized in a four flavor 
number scheme. Both parameterizations have to satisfy the relations quoted 
in \cite{bmsn1}. At this moment the latter set is not available
in the literature and we would like to fill in this gap. 
Starting from a three flavor number set of parton densities recently 
published in \cite{grv98} we will construct 
a four flavor number set of densities satisfying the relations in 
\cite{bmsn1}.

In Sec.II we give a general discussion of the CSN description
for heavy quark electroproduction, and explain the problems with mass 
factorization, collinear singularities and threshold dependence in the heavy 
flavor components of the structure functions. We then specialize to charm quark
electroproduction in Sec.III, working to second order in the running
coupling constant $\alpha_s(\mu^2)$. We first present details about the
charm quark density. Next numerical results are shown for
the structure functions in the various schemes. Analytic results for the 
contributions from the Compton scattering reaction with an invariant mass
cut are relegated to an Appendix. Finally we want to emphasize that we
only consider inclusive charm quark production in this paper. Exclusive
charm production which involves transverse momentum and rapidity distributions
will be dealt with in another paper.

%\end{document}

%\documentstyle[12pt,axodraw]{article}
\pagestyle{myheadings}  
%\begin{document}
%------------------This is Section 2---------------------------------
\mysection{Discussion of variable flavor number schemes}
%----------------------------------------------------------
%\topmargin=0in
%\headheight=0in
%\headsep=0in
%\oddsidemargin=7.2pt
%\evensidemargin=7.2pt
%\footheight=1in
%\marginparwidth=0in
%\marginparsep=0in
%\textheight=9in
%\textwidth=6in
%\newcommand{\be}{\begin{eqnarray}}
%\newcommand{\ee}{\end{eqnarray}}

In this section we discuss two different representations of the
deep inelastic structure functions in  
variable flavor number schemes. One is proposed here (CSN). 
The other (BMSN) was proposed in \cite{bmsn1} and \cite{bmsn2}.
The former starts from mass factorization of the exact heavy 
quark coefficient functions whereas the latter only applies this procedure
to the asymptotic expressions for these functions. In both schemes the special
role of the heavy quark loop contributions to the light quark coefficient
functions in combination with heavy quark production via gluon splitting 
was overlooked. This will be repaired in this paper.
Furthermore in both schemes there is a lot of freedom in the choice of matching
conditions, which are needed to connect the structure functions
presented for $n_f$ and $n_f+1$ light flavors. 
Different matching conditions 
lead to different threshold behaviors, which have consequences 
for the description of the structure functions at small $Q^2$ and large $x$.

Limiting ourselves to electroproduction, where deep inelastic lepton-hadron 
scattering is only mediated by a photon, the light parton components
of the structure functions are defined by
\begin{eqnarray}
\label{eqn2.1}
&& F_i^{\rm LIGHT}(n_f,Q^2,m^2) =  
\nonumber\\[2ex]
&& \sum\limits_{k=1}^{n_f} e_k^2 \left [
f_q^{\rm S}(n_f,\mu^2) \otimes \left (
\tilde {\cal C}_{i,q}^{\rm PS}\Big(n_f,\frac{Q^2}{\mu^2}\Big)
+\tilde {\cal C}_{i,q}^{\rm VIRT,PS}\Big(n_f,\frac{Q^2}{m^2},\frac{Q^2}{\mu^2}
\Big)\right ) \right.
\nonumber\\[2ex] 
&& \left. +  f_g^{\rm S}(n_f,\mu^2) \otimes \left (
\tilde {\cal C}_{i,g}^{\rm S}\Big(n_f,\frac{Q^2}{\mu^2}\Big)
+\tilde {\cal C}_{i,g}^{\rm VIRT,S}\Big(n_f,\frac{Q^2}{m^2},\frac{Q^2}{\mu^2} 
\Big) \right ) \right.
\nonumber\\[2ex] 
&& \left. + f_{k+\bar k}(n_f, \mu^2) \otimes \left (
{\cal C}_{i,q}^{\rm NS}\Big(n_f,\frac{Q^2}{\mu^2}\Big) 
+{\cal C}_{i,q}^{\rm VIRT,NS}\Big(n_f,\frac{Q^2}{m^2},\frac{Q^2}{\mu^2}\Big) 
\right ) \right ]
\,,
%\nonumber\\[2ex]
\end{eqnarray}
where $\otimes$ denotes the convolution symbol in the parton Bjorken 
scaling variable $z$. In this expression the  ${\cal C}_{i,k}$ ($i=2,L; k=q,g$) 
denote the light parton coefficient functions and the $e_k$ represent the 
charges of the light flavor quarks. The quantities
${\cal C}_{i,k}^{\rm VIRT}$  only contain the heavy quark loop contributions
to the light parton coefficient functions. Furthermore 
$f_g^{\rm S}(n_f,\mu^2)$ stands for the gluon density while the singlet (S) 
and non-singlet (NS) light quark densities, with respect to the SU$(n_f)$
flavor group, are defined by
\begin{eqnarray}
\label{eqn2.2}
&& f_{k+\bar k}(n_f,\mu^2) \equiv f_k(n_f,\mu^2) + f_{\bar k}(n_f,\mu^2)
\nonumber\\[2ex]
&& f_q^{\rm S}(n_f,\mu^2) = \sum_{k=1}^{n_f} f_{k+\bar k}(n_f,\mu^2) 
\nonumber\\[2ex]
&& f_q^{\rm NS}(n_f,\mu^2) = f_{k+\bar k}(n_f,\mu^2) 
-\frac{1}{n_f} f_q^{\rm S}(n_f,\mu^2) \,.
\end{eqnarray}
Finally we have set the factorization scale equal to the 
renormalization scale $\mu$. The light parton coefficient functions 
have been calculated up to order $\alpha_s^2$ in \cite{zn}.
The contributions to ${\cal C}_{i,k}^{\rm VIRT}$ appear for the first time in
second order perturbation theory and can be found in \cite{rijk}.
For our further discussion it will be convenient to distinguish between
the numbers of external and internal flavors. The former refers
to the number of light flavor densities whereas the latter denotes
the number of light flavors in the quark loop contributions to the virtual
corrections. They are not necessarily equal. Some of the coefficient
functions have the external flavor number as an overall factor.
To explicitly cancel this factor we have defined the quark and gluon 
coefficient functions in Eq. (\ref{eqn2.1})
as follows
\begin{eqnarray}
\label{eqn2.3}
 {\cal C}_{i,q}^{\rm S}(n_f,\frac{Q^2}{\mu^2})&=&
{\cal C}_{i,q}^{\rm NS}(n_f,\frac{Q^2}{\mu^2})
+{\cal C}_{i,q}^{\rm PS}(n_f,\frac{Q^2}{\mu^2})\,,
\nonumber\\[2ex]
{\cal C}_{i,q}^{\rm PS}(n_f,\frac{Q^2}{\mu^2})&=&
n_f \,\tilde {\cal C}_{i,q}^{\rm PS}(n_f,\frac{Q^2}{\mu^2})\,, \,
{\cal C}_{i,g}^{\rm S}(n_f,\frac{Q^2}{\mu^2})=
n_f \,\tilde {\cal C}_{i,g}^{\rm S}(n_f,\frac{Q^2}{\mu^2})\,,
%\nonumber\\[2ex]
\end{eqnarray}
where PS represents the purely singlet component.
Hence the remaining $n_f$ in the argument of the coefficient functions
marked with a tilde denotes the number of internal flavors.  
The same holds for the $n_f$ in the parton densities. However the 
argument $n_f$ in the structure functions is external and it refers to
the number of parton densities appearing in their expressions.
The parton densities satisfy the renormalization group equations.
If we define
\begin{eqnarray}
\label{eqn2.4}
D \equiv \mu \frac{\partial}{\partial \mu} + \beta(n_f,g) \frac{\partial}
{\partial g}\,, \qquad g \equiv g(n_f,\mu^2)
\end{eqnarray}
then
\begin{eqnarray}
\label{eqn2.5}
D f_q^{\rm NS}(n_f,\mu^2)&=& - \gamma_{qq}^{\rm NS}(n_f,g)f_q^{\rm NS}
(n_f,\mu^2)
\nonumber\\[2ex]
 D f_k^{\rm S}(n_f,\mu^2) &=& - \gamma_{kl}^{\rm S}(n_f,g)f_l^{\rm S}
(n_f,\mu^2)
\qquad k,l=q,g
\end{eqnarray}
where $\gamma_{kl}$ represent the anomalous dimensions of the operators
in the operator product expansion (OPE). 

The heavy flavor components 
($Q = {\rm c,b,t}, \bar Q = \bar{\rm c},\bar{\rm b},\bar{\rm t}$)
of the structure functions $F_2$ and $F_L$ 
arise from Feynman graphs with heavy flavors ($Q$ and $\bar Q$ with mass $m$)
in the final state and are given by
\begin{eqnarray}
\label{eqn2.6}
&&F_{i,Q}^{\rm EXACT}(n_f,Q^2,m^2) = \sum_{k=1}^{n_f} e_k^2 \left [
f_q^{\rm S}(n_f,\mu^2) \otimes 
L_{i,q}^{\rm PS}\Big(n_f,\frac{Q^2}{m^2},\frac{Q^2}{\mu^2}\Big) \right.
\nonumber\\[2ex]
&& \left. + f_g^{\rm S} (n_f,\mu^2) \otimes
L_{i,g}^{\rm S}\Big(n_f,\frac{Q^2}{m^2},\frac{Q^2}{\mu^2}\Big) 
+ f_{k+\bar k}(n_f,\mu^2) \otimes
L_{i,q}^{\rm NS}\Big(n_f,\frac{Q^2}{m^2}, \frac{Q^2}{\mu^2}\Big) \right ]
\nonumber\\[2ex]
&&+ e_Q^2 \left [ f_q^{\rm S} (n_f,\mu^2) \otimes 
H_{i,q}^{\rm PS}\Big(n_f,\frac{Q^2}{m^2},\frac{Q^2}{\mu^2}\Big)
 +  f_g^{\rm S}(n_f,\mu^2) \otimes 
H_{i,g}^{\rm S}\Big(n_f,\frac{Q^2}{m^2},\frac{Q^2}{\mu^2}\Big) \right ] \,,
\nonumber\\[2ex]
\end{eqnarray}
where $e_Q$ represents the charge of the heavy quark.
%--------------------------------
%fig1
\begin{figure}
\begin{center}
  \begin{picture}(130,65)
  \Gluon(0,0)(30,15){3}{7}
  \ArrowLine(60,15)(30,15)
  \ArrowLine(30,15)(30,45)
  \ArrowLine(30,45)(60,45)
  \Photon(0,60)(30,45){3}{7}
  \end{picture}
\hspace*{1cm}
  \begin{picture}(130,65)
  \Gluon(0,0)(30,15){3}{7}
  \ArrowLine(30,15)(60,15)
  \ArrowLine(30,45)(30,15)
  \ArrowLine(60,45)(30,45)
  \Photon(0,60)(30,45){3}{7}
  \end{picture}
\vspace*{5mm}
  \caption{Lowest-order photon-gluon fusion process 
   $\gamma^* + g \rightarrow Q + \bar Q$
  contributing to the coefficient functions $H_{i,g}^{\rm S,(1)}$.} 
  \label{fig:1}
\end{center}
\end{figure}
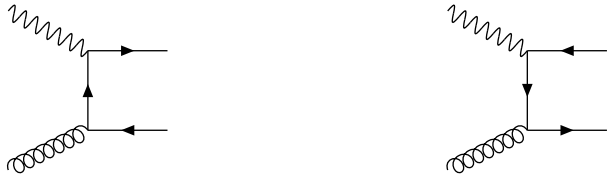    
Furthermore $L_{i,k}$ and $H_{i,k}$ $(i=2,L ; k=q,g)$ 
represent the heavy-quark coefficient
functions which are exactly calculated order by order in perturbation theory. 
In Figs. \ref{fig:1}-\ref{fig:5} we have shown some of the Feynman
diagrams contributing to the coefficient functions up to order $\alpha_s^2$.
Like in the case of the light-parton coefficient functions
${\cal C}_{i,k}$ they can be split into (purely)-singlet and non-singlet parts. 
The distinction between $L_{i,k}$
and $H_{i,k}$ can be traced back to the different (virtual) photon-parton
heavy-quark production mechanisms from which they originate. 
The functions $L_{i,k}$, $H_{i,k}$
are attributed to the reactions where the virtual photon couples to the light
quarks and the heavy quark respectively. Hence $L_{i,k}$ 
and $H_{i,k}$ in Eq. (\ref{eqn2.6})
are multiplied by $e_k^2$ and $e_Q^2$ respectively. 
As has been mentioned in the introduction the heavy quark coefficient
functions contain large logarithms of the type $\ln^i(Q^2/m^2)$ when
$Q^2 \gg m^2$ which can be removed from the former by using mass factorization.
%---------------------------
%fig2
\begin{figure}
\begin{center}
  \begin{picture}(60,60)
  \Gluon(0,0)(30,15){3}{7}
  \ArrowLine(60,15)(30,15)
  \ArrowLine(30,15)(30,45)
  \ArrowLine(30,45)(60,45)
  \Gluon(45,45)(45,15){3}{7}
  \Photon(0,60)(30,45){3}{7}
  \end{picture}
\hspace*{1cm}
%  \hfill
  \begin{picture}(60,60)
  \Gluon(0,0)(30,15){3}{7}
  \ArrowLine(60,15)(30,15)
  \ArrowLine(30,15)(30,45)
  \ArrowLine(30,45)(60,45)
  \Gluon(30,30)(45,15){3}{7}
  \Photon(0,60)(30,45){3}{7}
\end{picture}
\hspace*{1cm}
%  \hfill
  \begin{picture}(60,60)
  \Gluon(0,0)(30,15){3}{7}
  \ArrowLine(60,15)(30,15)
  \ArrowLine(30,15)(30,45)
  \ArrowLine(30,45)(60,45)
  \Gluon(30,30)(45,45){3}{7}
  \Photon(0,60)(30,45){3}{7}
\end{picture}
\hspace*{1cm}
%  \hfill
  \begin{picture}(60,60)
  \Gluon(0,0)(30,15){3}{7}
  \ArrowLine(60,15)(30,15)
  \ArrowLine(30,15)(30,45)
  \ArrowLine(30,45)(60,45)
  \Gluon(15,7)(45,45){3}{7}
  \Photon(0,60)(30,45){3}{7}
    \Text(9,0)[t]{$k_1$}
    \Text(45,35)[t]{$k$}
\end{picture}
\vspace*{5mm}
  \caption{Virtual gluon corrections to the process $\gamma^* + g \rightarrow
           Q + \bar Q$ contributing to the coefficient functions
           $H_{i,g}^{\rm S,(2)}$.}
  \label{fig:2}
\end{center}
\end{figure}
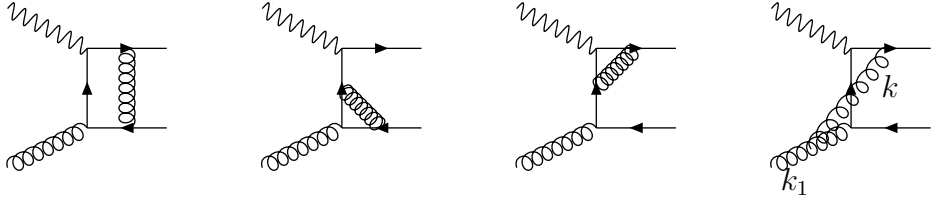
%----------------------------------------
To do this we first have to split the heavy quark coefficient
functions $L_{i,k}$ into soft and hard parts
\begin{eqnarray}
\label{eqn2.7}
L_{i,k}(n_f,\frac{Q^2}{m^2},\frac{Q^2}{\mu^2})
=L_{i,k}^{\rm HARD}(n_f,\Delta,\frac{Q^2}{m^2},\frac{Q^2}{\mu^2})
+L_{i,k}^{\rm SOFT}(n_f,\Delta,\frac{Q^2}{m^2},\frac{Q^2}{\mu^2})\,,
\end{eqnarray}
where $\Delta$ is a cut on the invariant mass $s_{Q\bar Q}$
of the heavy quark pair. The cut can be determined by experiment. 
It is chosen such that in the limit $m \rightarrow 0$
all mass singularities reside in the soft parts so that the hard parts
are collinearly finite. Taking Fig. \ref{fig:5} as an example we mean by
hard that one detects a $Q\bar Q$-pair with a large invariant mass which
is experimentally observable if $s_{Q\bar Q}>\Delta$. In the case 
$s_{Q\bar Q}<\Delta$ the $Q\bar Q$-pair is soft and becomes indistinguishable
from other light parton final states which contain 
contributions from virtual heavy quark loops. 
Next we add the soft parts to the other
contributions to $F_i^{\rm LIGHT}$ in Eq. (\ref{eqn2.1}) and the mass
factorization proceeds like
\begin{eqnarray}
\label{eqn2.8}
&& \tilde {\cal C}_{i,k}(n_f,\frac{Q^2}{\mu^2})+
{\cal C}_{i,k}^{\rm VIRT}(n_f,\frac{Q^2}{m^2},\frac{Q^2}{\mu^2})
+  L_{i,k}^{\rm SOFT}(n_f,\Delta,\frac{Q^2}{m^2},\frac{Q^2}{\mu^2}) = 
\nonumber\\[2ex]
&& A_{lk,Q}(n_f,\frac{\mu^2}{m^2}) 
\otimes \tilde {\cal C}_{i,l}(n_f,\frac{Q^2}{\mu^2})
\nonumber\\[2ex]
&& + A_{lk}(n_f,\frac{\mu^2}{m^2})
\otimes {\cal C}_{i,l,Q}^{\rm CSN,SOFT}
(n_f,\Delta,\frac{Q^2}{m^2},\frac{Q^2}{\mu^2}) \,,
\quad k,l=q,g \,.
\end{eqnarray}
Here ${\cal C}_{i,l,Q}$ are those parts of the light parton coefficient
functions ${\cal C}_{i,l}$ which contain the heavy quark loops.
The hard parts of $L_{i,k}$ are left in $F_{i,Q}^{\rm EXACT}$ in 
Eq. (\ref{eqn2.6}) and do not need any mass factorization.
Furthermore we have the condition that the dependence on the parameter
$\Delta$ cancels in the sums so
\begin{eqnarray}
\label{eqn2.9}
{\cal C}_{i,k,Q}^{\rm CSN}\Big(n_f,\frac{Q^2}{m^2},\frac{Q^2}{\mu^2}\Big)
&=& {\cal C}_{i,k,Q}^{\rm CSN,SOFT}
\Big(n_f,\Delta,\frac{Q^2}{m^2},\frac{Q^2}{\mu^2} \Big) 
\nonumber\\[2ex]
&& +L_{i,k}^{\rm HARD}\Big(n_f,\Delta,\frac{Q^2}{m^2},\frac{Q^2}{\mu^2}
\Big) \,,
\end{eqnarray}
where $\mu$ in the hard parts only represents the renormalization
scale. The coefficient functions $H_{i,k}$ satisfy the relations
\footnote{ In order to help the reader, who is more familiar with the notation 
in \cite{acot}, one can make the following comparison.
For instance Eq. (7) in the latter reference $f_g^{Q(1)}$
is equal to our $A_{Qg}^{(1)}$. Similarly in Eq. (8)
$\sum_{\lambda} \omega^{\lambda (1)}_{Bg}=H_{i,g}^{(1)}$ and
$\sum_{\lambda} \omega^{\lambda (0)}_{BQ}=H_{i,Q}^{(0)}$.} 
\begin{eqnarray}
\label{eqn2.10}
H_{i,k}(n_f,\frac{Q^2}{m^2},\frac{Q^2}{\mu^2}) = 
A_{lk}(n_f,\frac{\mu^2}{m^2}) 
\otimes {\cal C}_{i,l}^{\rm CSN}(n_f,\frac{Q^2}{m^2},\frac{Q^2}{\mu^2})
\quad k,l=Q,q,g \,.
\end{eqnarray}
Notice that mass factorization applied to the functions
$H_{i,q}$ and $H_{i,g}$ occurring in $F_{i,Q}^{\rm EXACT}$ (\ref{eqn2.6})
leads to the coefficient functions ${\cal C}_{i,Q}^{\rm CSN}$. The latter
also follow from mass factorization of the functions $H_{i,Q}$ which
represent processes with a heavy quark in the initial state. The
quantities $H_{i,Q}$, which do not appear in $F_{i,Q}^{\rm EXACT}$, 
together  with the corresponding operator matrix elements (OME's) $A_{QQ}$ 
are characteristic of variable flavor number schemes.  
The procedure above transfers the logarithms $\ln^i(\mu^2/m^2)$, appearing in 
$L_{i,k}^{\rm SOFT}$ and $H_{i,k}$, to the heavy quark operator
matrix elements $A_{lk,Q}$ and $A_{Qk}$. The latter are defined by
(see \cite{bmsn1} for details of renormalization and mass factorization)
\begin{eqnarray}
\label{eqn2.11}
A_{lk,Q}&=&\langle k \mid O_l(0)\mid k \rangle \qquad k,l=q,g \,,
\nonumber\\[2ex]
A_{Qk} &=& \langle k \mid O_Q(0)\mid k \rangle \qquad k=Q,q,g \,.
\end{eqnarray}
Note that the $O_l$ are the light quark and gluon operators and in
$A_{lk,Q}$ we only retain contributions from subgraphs which contain heavy
quark (Q) loops. The quantity $O_Q$ represents the heavy quark operator.
Here we want to stress that the operators are sandwiched between quark and 
gluon states. This will cause mass singularities of the type
$\ln^i(\mu^2/m^2)$ to appear in a similar way as they appear in partonic
cross sections.

The heavy quark coefficient functions defined in the 
CSN scheme in Eqs. (\ref{eqn2.8})-(\ref{eqn2.10})
are collinearly finite and tend asymptotically to the massless 
parton coefficient functions presented in Eq. (\ref{eqn2.1}) i.e.
%-------------------------------
%fig3
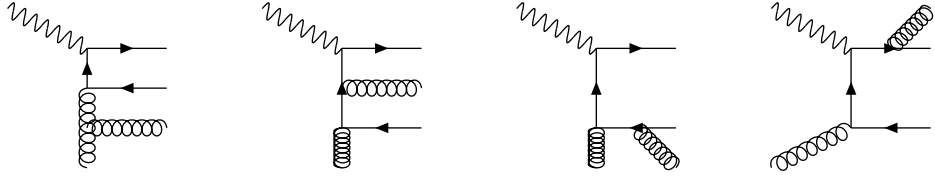
\begin{figure}
\begin{center}
  \begin{picture}(60,60)
  \Gluon(30,0)(30,30){3}{7}
  \ArrowLine(60,30)(30,30)
  \ArrowLine(30,30)(30,45)
  \ArrowLine(30,45)(60,45)
  \Gluon(30,15)(60,15){3}{7}
  \Photon(0,60)(30,45){3}{7}
  \end{picture}
\hspace*{1cm}
%  \hfill
  \begin{picture}(60,60)
  \Gluon(30,0)(30,15){3}{7}
  \ArrowLine(60,15)(30,15)
  \ArrowLine(30,15)(30,45)
  \ArrowLine(30,45)(60,45)
  \Gluon(30,30)(60,30){3}{7}
  \Photon(0,60)(30,45){3}{7}
\end{picture}
\hspace*{1cm}
%  \hfill
  \begin{picture}(60,60)
  \Gluon(30,0)(30,15){3}{7}
  \ArrowLine(60,15)(30,15)
  \ArrowLine(30,15)(30,45)
  \ArrowLine(30,45)(60,45)
  \Gluon(45,15)(60,0){3}{7}
  \Photon(0,60)(30,45){3}{7}
\end{picture}
\hspace*{1cm}
%  \hfill
  \begin{picture}(60,60)
  \Gluon(0,0)(30,15){3}{7}
  \ArrowLine(60,15)(30,15)
  \ArrowLine(30,15)(30,45)
  \ArrowLine(30,45)(60,45)
  \Gluon(45,45)(60,60){3}{7}
  \Photon(0,60)(30,45){3}{7}
\end{picture}
\vspace*{5mm}
  \caption{The  bremsstrahlung process $\gamma^* + g \rightarrow
           Q + \bar Q + g$ contributing to the coefficient functions
           $H_{i,g}^{\rm S,(2)}$.}
  \label{fig:3}
\end{center}
\end{figure}
%----------------------------------------
\begin{eqnarray}
\label{eqn2.12}
\lim_{Q^2 \rightarrow \infty}\,
{\cal C}_{i,k}^{\rm CSN}(n_f,\frac{Q^2}{m^2},\frac{Q^2}{\mu^2})
= {\cal C}_{i,k}(n_f,\frac{Q^2}{\mu^2})\,, \qquad k=Q,q,g \,.
\end{eqnarray}
In particular we have
\begin{eqnarray}
\label{eqn2.13}
{\cal C}_{i,k} (n_f,\frac{Q^2}{\mu^2})+ \lim_{Q^2 \gg m^2}
 {\cal C}_{i,k,Q}^{\rm CSN} (n_f,\frac{Q^2}{m^2},\frac{\mu^2}{m^2})
={\cal C}_{i,k} (n_f+1,\frac{Q^2}{\mu^2}) \,,
\end{eqnarray}
so that the number of internal flavors is enhanced by one unit.

The CSN scheme above has similarities with the VFNS schemes proposed 
in \cite{acot}, \cite{thro}. The decomposition of 
the $L_{i,k}$ into soft and hard parts has not been discussed previously. 
However one must address this issue because the mass singularities in
$L_{i,k}$ and ${\cal C}_{i,k}^{\rm VIRT}$ separately have such high powers
in $\ln(Q^2/m^2)$ that they cannot be removed via mass factorization. 
Another feature
is that in the limit $m \rightarrow 0$ the final state invariant energies
in the reactions which contribute to these two types of coefficient 
functions become equal.
Hence $L_{i,k}$ and ${\cal C}_{i,k}^{\rm VIRT}$ have to be added so that
the leading singularities cancel and the remaining ones are then removed by
mass factorization according to Eq. (\ref{eqn2.8}). Notice that the
the total coefficient functions $L_{i,k}$ cannot be moved to 
$F_i^{\rm LIGHT}$ because this would contradict the definitions of the latter 
structure functions where only light partons are observed in the final states.
Therefore it is sufficient to transfer the $L_{i,k}^{\rm SOFT}$ to the
$F_i^{\rm LIGHT}$ since they contain the same mass singularities as the 
$L_{i,k}$. If $\Delta$ is chosen small enough, the heavy quarks are 
unobservable in a measurement of $F_i^{\rm LIGHT}$. Note that the problem
of the separation of $L_{i,k}$ into soft and hard parts is not needed for 
the total structure function $F_i^{\rm LIGHT}+F_{i,Q}^{\rm EXACT}$. The
all order mass factorization of the latter is shown in \cite{col}.

To illustrate the procedure above we carry it out up to order
$\alpha_s^2$. The coefficients in the series expansion are defined as
follows
\begin{eqnarray}
\label{eqn2.14}
{\cal C}_{i,k} &=& \sum_{n=0}^{\infty} a_s^n {\cal C}_{i,k}^{(n)}
\quad, \quad
H_{i,k} = \sum_{n=0}^{\infty} a_s^n H_{i,k}^{(n)} 
\quad, \quad
L_{i,k} = \sum_{n=2}^{\infty} a_s^n L_{i,k}^{(n)} \,,
\nonumber\\[2ex]
A_{kl} &=& \sum_{n=0}^{\infty} a_s^n A_{kl}^{(n)} \quad , \quad \mbox{with}
\quad a_s \equiv \frac{\alpha_s}{4\pi} \,.
\end{eqnarray}
Up to second order the mass factorization relations become
\begin{eqnarray}
\label{eqn2.15}
&&{\cal C}_{i,q}^{\rm VIRT,NS,(2)}(\frac{Q^2}{m^2})
 + L_{i,q}^{\rm SOFT,NS,(2)}\Big 
(\Delta,\frac{Q^2}{m^2},\frac{Q^2}{\mu^2}\Big) = 
A_{qq,Q}^{\rm NS,(2)}\Big(\frac{\mu^2}{m^2}\Big) 
{\cal C}_{i,q}^{\rm NS,(0)}
\nonumber\\[2ex]
&& -\beta_{0,Q} \ln \left ( \frac{\mu^2}{m^2} \right ) 
{\cal C}_{i,q}^{\rm NS,(1)} \Big(\frac{Q^2}{\mu^2}\Big)
%\nonumber\\[2ex] && 
+ {\cal C}_{i,q,Q}^{\rm CSN,SOFT,NS,(2)}
\Big (\Delta,\frac{Q^2}{m^2},\frac{Q^2}{\mu^2}\Big) \,,
\end{eqnarray}
with
\begin{eqnarray}
\label{eqn2.16}
{\cal C}_{i,q}^{\rm VIRT,NS,(2)}(\frac{Q^2}{m^2})
=F^{(2)}(\frac{Q^2}{m^2})\, {\cal C}_{i,q}^{(0)}\,.
\end{eqnarray}

%---------------------------------
%fig4
\begin{figure}
\begin{center}
  \begin{picture}(60,60)
  \DashArrowLine(0,0)(30,0){3}
  \DashArrowLine(30,0)(60,0){3}
  \Gluon(30,0)(30,25){3}{7}
  \ArrowLine(60,25)(30,25)
  \ArrowLine(30,25)(30,50)
  \ArrowLine(30,50)(60,50)
  \Photon(0,60)(30,50){3}{7}
  \end{picture}
\hspace*{2cm}
%  \hfill
  \begin{picture}(60,60)
  \DashArrowLine(0,0)(30,0){3}
  \DashArrowLine(30,0)(60,0){3}
  \Gluon(30,0)(30,25){3}{7}
  \ArrowLine(30,25)(60,25)
  \ArrowLine(30,50)(30,25)
  \ArrowLine(60,50)(30,50)
  \Photon(0,60)(30,50){3}{7}
\end{picture}
\vspace*{5mm}
  \caption{Bethe-Heitler process $\gamma^* + q(\bar q) \rightarrow
           Q + \bar Q + q(\bar q)$ contributing to the coefficient functions
           $H_{i,q}^{\rm PS,(2)}$. The light quarks $q$ and the
  heavy quarks $Q$ are indicated by dashed and solid lines
  respectively.}
  \label{fig:4}
\end{center}
\end{figure}
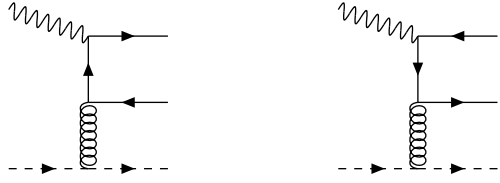
\noindent 
Here $F^{(2)}(Q^2/m^2)$ denotes the two-loop vertex correction in 
Fig. \ref{fig:6}. 
This function satisfies the decoupling theorem 
which implies that it vanishes in the 
limit $m \rightarrow \infty$. 
The heavy quark coefficient functions $L_{i,k}^{\rm NS,(2)}$ have been
calculated in \cite{bmsmn} and, after their convolution with the
partonic densities, yield contributions to the structure
functions which behave asymptotically like $\ln^3(Q^2/m^2)$.
These logarithms are canceled after adding $F^{(2)}(Q^2/m^2)$ 
in Ref. \cite{rijk}  
to $L_{i,k}^{\rm NS,SOFT,(2)}$ which contains the same mass singularities\
as $L_{i,k}^{\rm NS,(2)}$ (see the remark below Eq. (\ref{eqn2.7})).
To obtain the hard and soft parts
we divide the integral over $s_{Q\bar Q}$ in the graphs of Fig. \ref{fig:5}
in two regions i.e. $s > s_{Q\bar Q} > \Delta$ and 
$\Delta > s_{Q\bar Q} > 4~m^2$ which we denote by HARD and SOFT 
respectively.
Here $s_{Q\bar Q}$ and $s$ denote the CM energies squared of the $Q \bar Q$
system and the incoming photon-parton state respectively. The hard
and soft parts are presented in the Appendix. Finally $\beta_{0,Q} = -2/3$
denotes the heavy quark contribution to the lowest order coefficient
of the $\beta$-function in Eq. (2.4). We must change the running 
coupling constant when we change schemes.
 
The mass factorization of the heavy quark coefficient 
functions $H_{i,k}$ is simpler.  Here we get
\begin{eqnarray}
\label{eqn2.17}
H_{i,q}^{\rm PS,(2)}\Big(\frac{Q^2}{m^2},\frac{Q^2}{\mu^2}\Big)
 & = &  A_{Qq}^{\rm PS,(2)}\Big(\frac{\mu^2}{m^2}\Big)
{\cal C}_{i,Q}^{\rm CSN,NS,(0)}\Big (\frac{Q^2}{m^2} \Big )
%\nonumber\\[2ex]
%&& 
+{\cal C}_{i,q}^{\rm CSN,PS,(2)}
\Big(\frac{Q^2}{m^2},\frac{Q^2}{\mu^2}\Big) \,, 
\nonumber\\[2ex]
\end{eqnarray}
\begin{eqnarray}
\label{eqn2.18}
H_{i,Q}^{\rm NS,(1)}\Big(\frac{Q^2}{m^2}\Big)
 & = &  A_{QQ}^{\rm NS,(1)}\Big(\frac{\mu^2}{m^2}\Big)
{\cal C}_{i,Q}^{\rm CSN,NS,(0)}\Big(\frac{Q^2}{m^2}\Big)
%\nonumber\\[2ex]
%&& 
+ {\cal C}_{i,Q}^{\rm CSN,NS,(1)}
\Big(\frac{Q^2}{m^2},\frac{Q^2}{\mu^2}\Big) \,,
\nonumber\\[2ex]
\end{eqnarray}
\begin{eqnarray}
\label{eqn2.19}
H_{i,g}^{\rm S,(1)}\Big(\frac{Q^2}{m^2}\Big)
 & = &   A_{Qg}^{\rm S,(1)}\Big(\frac{\mu^2}{m^2}\Big)
{\cal C}_{i,Q}^{\rm CSN,NS,(0)}\Big (\frac{Q^2}{m^2}\Big )
%\nonumber\\[2ex]
%&& 
+ {\cal C}_{i,g}^{\rm CSN,S,(1)}
\Big(\frac{Q^2}{m^2},\frac{Q^2}{\mu^2}\Big)
\,,
\nonumber\\[2ex]
\end{eqnarray}
\begin{eqnarray}
\label{eqn2.20}
H_{i,g}^{\rm S,(2)}\Big(\frac{Q^2}{m^2},\frac{Q^2}{\mu^2}\Big)
 & = &   A_{Qg}^{\rm S,(2)}\Big(\frac{\mu^2}{m^2}\Big)
{\cal C}_{i,Q}^{\rm CSN,NS,(0)}\Big (\frac{Q^2}{m^2}\Big)
+{\cal C}_{i,g}^{\rm CSN,S,(2)}
\Big(\frac{Q^2}{m^2},\frac{Q^2}{\mu^2}\Big) 
\nonumber\\[2ex]
&& +  A_{Qg}^{\rm S,(1)}\Big(\frac{\mu^2}{m^2}\Big)
\otimes
{\cal C}_{i,Q}^{\rm CSN,NS,(1)}\Big(\frac{Q^2}{m^2},\frac{Q^2}{\mu^2}\Big)\,.
%\nonumber\\[2ex]
%&& 
%+{\cal C}_{i,g}^{\rm CSN,S,(2)}
%\Big(\frac{Q^2}{m^2},\frac{Q^2}{\mu^2}\Big) \,.
%\nonumber\\[2ex]
\end{eqnarray}
%--------------------------------------------------------
%fig5
\begin{figure}
\begin{center}
  \begin{picture}(150,100)(0,0)
  \DashArrowLine(10,30)(40,30){4}
  \DashArrowLine(40,30)(70,30){4}
  \DashArrowLine(70,30)(120,30){4}
  \Photon(40,0)(70,30){2}{5}
  \GlueArc(95,20)(55,170,120){4}{10}
%  \ArrowArc(95,60)(30,161,90)
  \ArrowArc(95,60)(30,90,161)
  \ArrowArc(95,75)(30,199,270)
    \Text(73,95)[t]{$p_2$}
    \Text(73,47)[t]{$p_1$}
    \Text(25,43)[t]{$p$}
    \Text(95,25)[t]{$p'$}
    \Text(30,0)[t]{$q$}
\end{picture}
\hspace*{1cm}
  \begin{picture}(150,100)(0,0)
  \DashArrowLine(10,30)(40,30){4}
  \DashArrowLine(40,30)(70,30){4}
  \DashArrowLine(70,30)(120,30){4}
  \Photon(10,0)(40,30){2}{5}
  \GlueArc(125,20)(55,170,120){4}{10}
%  \ArrowArc(95,60)(30,161,90)
  \ArrowArc(125,60)(30,90,161)
  \ArrowArc(125,75)(30,199,270)
    \Text(110,98)[t]{$p_2$}
    \Text(110,45)[t]{$p_1$}
    \Text(25,43)[t]{$p$}
    \Text(95,25)[t]{$p'$}
    \Text(0,0)[t]{$q$}
\end{picture}
\caption[]{Compton process 
 $\gamma^*(q) + q(p) \rightarrow Q(p_1) + \bar Q(p_2) + q(p')$ 
 contributing to the coefficient functions $L_{i,q}^{\rm NS,(2)}$.
  The light quarks $q$ and the
  heavy quarks $Q$ are indicated by dashed and solid lines
  respectively ($s=(p+q)^2$, $s_{Q\bar Q}=(p_1+p_2)^2$ see text).}
\label{fig:5}
\end{center}
\end{figure}
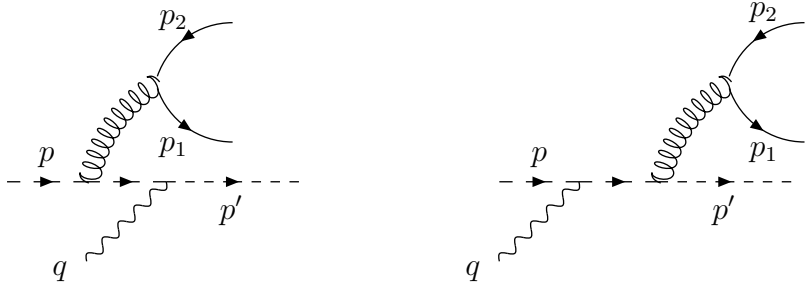
\noindent In the expressions above we have only given the arguments on which 
the coefficient functions and operator matrix elements depend, like
$n_f$, $Q^2/m^2$, or $Q^2/\mu^2$ (at least up to that order in
perturbation theory). Furthermore
we have dropped the convolution symbol when the corresponding coefficient
function behaves as a $\delta$-function of the type $\delta(1-z)$. 
The heavy quark coefficient functions correspond to the following
processes
\begin{eqnarray}
\label{eqn2.21}
H_{i,g}^{\rm S,(1)} &:& \gamma^* + g \rightarrow Q + \bar Q
\quad \mbox{Fig. \ref{fig:1}}
\nonumber\\[2ex]
H_{i,g}^{\rm S,(2)} &:& \gamma^* + g \rightarrow Q + \bar Q + g
\quad \mbox{Figs. \ref{fig:2}, \ref{fig:3}}
\nonumber\\[2ex]
H_{i,q}^{\rm PS,(2)} &:& \gamma^* + q(\bar q) \rightarrow Q
+ \bar Q + q(\bar q)
\quad \mbox{Fig. \ref{fig:4}}
\nonumber\\[2ex]
L_{i,q}^{\rm NS,(2)} &:& \gamma^*
+ q(\bar q) \rightarrow Q + \bar Q + q(\bar q)
\quad \mbox{Fig. \ref{fig:5}}
\nonumber\\[2ex]
H_{i,Q}^{\rm NS,(0)} &:& \gamma^* + Q \rightarrow Q
\nonumber\\[2ex]
H_{i,Q}^{\rm NS,(1)} &:&\gamma^* + Q \rightarrow Q + g
\quad \mbox{Fig. \ref{fig:7}} \,.
\end{eqnarray}
In the reactions above the virtual corrections to the lowest order
processes are implicitly understood. The coefficient functions 
$L_{i,q}^{\rm NS,(2)}$, $H_{i,q}^{\rm PS,(2)}$ and $H_{i,g}^{\rm S,(2)}$,
computed in the ${\overline {MS}}$-scheme,  
can be found in \cite{lrsn}
whereas the $H_{i,Q}^{\rm NS,(1)}$ are computed in the context 
of QED in \cite{neve}. The ${\overline {\rm MS}}$-scheme is also chosen for
the OME's in  Eqs. (\ref{eqn2.15})-(\ref{eqn2.20}) which are computed
up to order $\alpha_s^2$ in \cite{bmsn1} and \cite{bmsmn}. This also holds
for $A_{QQ}^{\rm NS,(1)}$ in Eq. (\ref{eqn2.18}) which
is presented in the context of QED in \cite{bnb}. Furthermore the 
running coupling constant appearing in the quantities above contains $n_f$
active flavors.

The mass singular logarithms of the
type $\ln(\mu^2/m^2)$, appearing in the OME's above, are
absorbed by the light parton densities. This procedure leads to parton 
densities which are represented in the $n_f+1$ light flavor scheme.  
For the light parton densities one obtains
%-----------------------------------
%fig6
\begin{figure}
\begin{center}
  \begin{picture}(120,120)
    \DashArrowLine(0,20)(25,20){5}
    \DashArrowLine(105,20)(130,20){5}
    \DashArrowLine(25,20)(65,90){5}
    \DashArrowLine(65,90)(105,20){5}
   \ArrowArc(65,20)(15,0,180)
   \ArrowArc(65,20)(15,180,360)
  \Gluon(25,20)(50,20){3}{4}
  \Gluon(80,20)(105,20){3}{4}
  \Photon(65,120)(65,90){3}{4}
   \Text(75,120)[t]{$\gamma^*$}
    \Text(0,32)[t]{$q$}
    \Text(65,2)[t]{$Q$}
    \Text(65,52)[t]{$\bar Q$}
  \end{picture}
  \caption{Two-loop vertex correction to the process
   $\gamma^* + q \rightarrow q$ containing a heavy quark (Q) loop.
          It contributes to ${\cal C}_{i,q}^{\rm VIRT,NS,(2)}(Q^2/m^2)
    =F^{(2)}(Q^2/m^2)~ {\cal C}_{i,q}^{(0)}$.}
  \label{fig:6}
\end{center}
\end{figure}
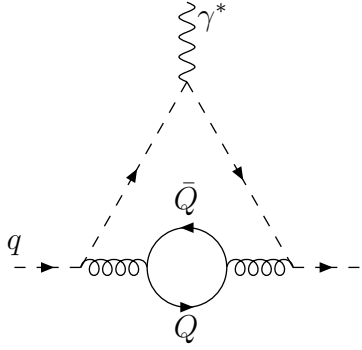
%--------------------------------------------------------
\begin{eqnarray}
\label{eqn2.22}
&& f_{k+\bar k}(n_f+1,\mu^2)  = 
 A_{qq,Q}^{\rm NS}\Big(n_f, \frac{\mu^2}{m^2}\Big)\otimes
 f_{k+\bar k}(n_f, \mu^2)
+ \tilde A_{qq,Q}^{\rm PS}\Big(n_f, \frac{\mu^2}{m^2}\Big) 
\nonumber \\  &&
\otimes f_q^{\rm S}(n_f, \mu^2)
%\nonumber \\ &&
+ \tilde A_{qg,Q}^{\rm S}\Big(n_f, \frac{\mu^2}{m^2}\Big) \otimes
f_g^{\rm S}(n_f, \mu^2)\,, \, k=1 \cdots n_f\,. 
%\nonumber \\ &&
\end{eqnarray}
The parton density representing the heavy quark in the $n_f+1$ flavor 
scheme is 
\begin{eqnarray}
\label{eqn2.23}
f_{Q+\bar Q}(n_f+1, \mu^2) 
&=& \tilde A_{Qq}^{\rm PS}\Big(n_f, \frac{\mu^2}{m^2}\Big)\otimes
f_q^{\rm S}(n_f, \mu^2)
\nonumber\\[2ex]
&& + \tilde A_{Qg}^{\rm S}\Big(n_f, \frac{\mu^2}{m^2}\Big) \otimes
f_g^{\rm S}(n_f, \mu^2) \,.
%\nonumber\\[2ex]
\end{eqnarray}
Finally the gluon density in the $n_f+1$ flavor scheme is
\begin{eqnarray}
\label{eqn2.24}
f_g^{\rm S}(n_f+1, \mu^2) &=&  A_{gq,Q}^{\rm S}(n_f,\frac{\mu^2}{m^2})
\otimes f_q^{\rm S}(n_f, \mu^2)
\nonumber\\[2ex]
&& + A_{gg,Q}^{\rm S}(n_f,\frac{\mu^2}{m^2})\otimes f_g^{\rm S}(n_f, \mu^2)\,. 
%\nonumber\\[2ex]
\end{eqnarray}
One can check (see \cite{bmsn1} ) that the new parton densities satisfy the 
renormalization
group equations in Eq. (\ref{eqn2.5}) wherein all quantities $n_f$ are 
replaced by $n_f+1$.
Up to order $a_s^2$ the above relations become
\begin{eqnarray}
\label{eqn2.25}
f_{k+\bar k}(n_f+1,\mu^2) &=& f_{k+\bar k}(n_f,\mu^2) 
\nonumber\\[2ex]
&& + a_s^2(n_f,\mu^2) A_{qq,Q}^{\rm NS,(2)}\Big(\frac{\mu^2}{m^2}\Big)
\otimes f_{k+\bar k}(n_f, \mu^2)\,, 
\end{eqnarray}
\begin{eqnarray}
\label{eqn2.26}
&& f_{Q+\bar Q}(n_f+1, \mu^2)=
a_s(n_f,\mu^2) \tilde A_{Qg}^{\rm S,(1)}\Big(\frac{\mu^2}{m^2}\Big) \otimes
f_g^{\rm S}(n_f, \mu^2) 
\nonumber\\[2ex]  
&& + a_s^2 (n_f,\mu^2)\Big [\tilde A_{Qq}^{\rm PS,(2)}
\Big(\frac{\mu^2}{m^2}\Big)\otimes f_q^{\rm S}(n_f, \mu^2)
+ \tilde A_{Qg}^{\rm S,(2)}\Big(\frac{\mu^2}{m^2}\Big) 
\otimes f_g^{\rm S}(n_f, \mu^2)\Big ] \,,
\nonumber\\
\end{eqnarray}
\begin{eqnarray}
\label{eqn2.27}
&& f_g^{\rm S}(n_f+1, \mu^2) = f_g^{\rm S}(n_f, \mu^2)
+a_s(n_f,\mu^2)A_{gg,Q}^{\rm S,(1)}(\frac{\mu^2}{m^2}) \otimes 
f_g^{\rm S}(n_f, \mu^2)
\nonumber\\[2ex]  
&& + a_s^2(n_f,\mu^2) \Big [ A_{gq,Q}^{\rm S,(2)}(\frac{\mu^2}{m^2})
\otimes f_q^{\rm S}(n_f, \mu^2) 
 + A_{gg,Q}^{\rm S,(2)}(\frac{\mu^2}{m^2})
\otimes f_g^{\rm S}(n_f, \mu^2) \Big ]\,.
\nonumber\\
\end{eqnarray}
Notice that in passing from an $n_f$-flavor to an $n_f+1$-flavor scheme
the running coupling constant becomes
\begin{eqnarray}
\label{eqn2.28}
a_s(n_f+1,\mu^2)&=&a_s(n_f,\mu^2) \Big [ 1 - a_s(n_f,\mu^2)\beta_{0,Q}
\ln(\frac{\mu^2}{m^2}) \Big ]\,, 
\end{eqnarray}
which has to be used in all expressions for the structure functions
in the CSN scheme. 
%--------------------------
%fig7
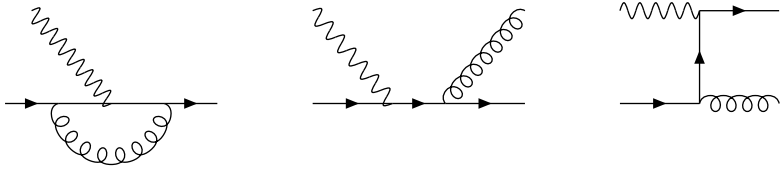
\begin{figure}
\begin{center}
  \begin{picture}(80,50)
  \ArrowLine(0,15)(20,15)
  \Line(20,15)(60,15)
  \ArrowLine(60,15)(80,15)
  \GlueArc(40,15)(20,180,360){3}{8}
  \Photon(10,50)(40,15){3}{9}
  \end{picture}
\hspace*{1cm}
  \begin{picture}(80,50)
  \ArrowLine(0,15)(30,15)
  \ArrowLine(30,15)(50,15)
  \ArrowLine(50,15)(80,15)
  \Gluon(50,15)(80,50){3}{7}
  \Photon(0,50)(30,15){3}{7}
  \end{picture}
\hspace*{1cm}
  \begin{picture}(60,50)
  \ArrowLine(0,15)(30,15)
  \ArrowLine(30,15)(30,50)
  \ArrowLine(30,50)(60,50)
  \Gluon(30,15)(60,15){3}{4}
  \Photon(0,50)(30,50){3}{5}
\end{picture}
\vspace*{5mm}
  \caption{Order $\alpha_s$ corrections to the process
  $\gamma^* + Q \rightarrow Q$ and the reaction
  $\gamma^* + Q \rightarrow Q + g$
 contributing to the coefficient functions $H_{i,Q}^{\rm NS,(1)}$.}
  \label{fig:7}
\end{center}
\end{figure}
%---------------------------------------------
Using the mass factorization relations in Eqs. (\ref{eqn2.8}), (\ref{eqn2.10})
and the redefinitions of the parton densities in 
Eqs. (\ref{eqn2.22})-(2.24) we obtain from Eq. (\ref{eqn2.6}) the heavy 
quark components of the structure functions in the CSN scheme are
\begin{eqnarray}
\label{eqn2.29}
&&F_{i,Q}^{\rm CSN} (n_f+1,\Delta,Q^2,m^2)  =
\nonumber\\[2ex]
&& e_Q^2 \left [ f_{Q +\bar Q}(n_f+1,\mu^2) \otimes \left \{
{\cal C}_{i,Q}^{\rm CSN,NS}\Big(n_f+1,\frac{Q^2}{m^2},\frac{Q^2}{\mu^2}\Big)
\right. \right.
\nonumber\\[2ex]
&& \left. \left. + {\cal C}_{i,Q}^{\rm CSN,PS}
\Big(n_f+1,\frac{Q^2}{m^2},\frac{Q^2}{\mu^2}\Big) \right \} \right. 
\nonumber\\[2ex]
&& \left. + \sum_{l=1}^{n_f} f_{l+\bar l}(n_f+1,\mu^2)
\otimes {\cal C}_{i,q}^{\rm CSN,PS}
\Big(n_f+1,\frac{Q^2}{m^2},\frac{Q^2}{\mu^2}\Big) \right.
\nonumber\\[2ex]
&&\left. + f_g^{\rm S}(n_f+1, \mu^2)\otimes  {\cal C}_{i,g}^{\rm CSN,S}
\Big(n_f+1,\frac{Q^2}{m^2},\frac{Q^2}{\mu^2}\Big) \right ]
\nonumber\\[2ex]
&&+ \sum_{k=1}^{n_f} e_k^2 \left [ \sum_{l=1}^{n_f} 
f_{l+\bar l}(n_f+1,\mu^2) \otimes
L_{i,q}^{\rm HARD,PS}\Big(n_f,\Delta,\frac{Q^2}{m^2},
\frac{Q^2}{\mu^2}\Big) \right.
\nonumber\\[2ex]
&& \left. + f_g^{\rm S} (n_f+1,\mu^2) \otimes
L_{i,g}^{\rm HARD,S}\Big(n_f,\Delta,\frac{Q^2}{m^2},
\frac{Q^2}{\mu^2}\Big) \right.
\nonumber\\[2ex]
&& \left. + f_{k+\bar k}(n_f+1,\mu^2) \otimes
L_{i,q}^{\rm HARD,NS}\Big(n_f,\Delta,\frac{Q^2}{m^2},\frac{Q^2}{\mu^2}
 \Big) \right ] \,.
\end{eqnarray}
In a similar way one obtains from Eq. (\ref{eqn2.1}) the light parton 
components of the structure functions in the CSN scheme are
\begin{eqnarray}
\label{eqn2.30}
&& F_i^{\rm CSN,LIGHT}(n_f,\Delta,Q^2,m^2) =
\sum\limits_{k=1}^{n_f} e_k^2 \Biggl [ 
\nonumber\\[2ex]
&& \sum_{l=1}^{n_f} f_{l+\bar l}(n_f+1,\mu^2) \otimes
\left (\tilde{\cal C}_{i,q}^{\rm PS}\Big(n_f,\frac{Q^2}{\mu^2}\Big)
+{\cal C}_{i,q,Q}^{\rm CSN,SOFT,PS}
\Big(n_f,\Delta,\frac{Q^2}{m^2},\frac{Q^2}{\mu^2}\Big) \right ) 
\nonumber\\[2ex]
&& + f_g^{\rm S}(n_f+1,\mu^2) \otimes
\left (\tilde{\cal C}_{i,g}^{\rm S}\Big(n_f,\frac{Q^2}{\mu^2}\Big)
+ {\cal C}_{i,g,Q}^{\rm CSN,SOFT,S}
\Big(n_f,\Delta,\frac{Q^2}{m^2},\frac{Q^2}{\mu^2}\Big) \right ) 
\nonumber\\[2ex]
&& + f_{k+\bar k}(n_f+1,\mu^2) \otimes
\left ({\cal C}_{i,q}^{\rm NS}\Big(n_f,\frac{Q^2}{\mu^2}\Big)
+{\cal C}_{i,q,Q}^{\rm CSN,SOFT,NS}
\Big(n_f,\Delta,\frac{Q^2}{m^2},\frac{Q^2}{\mu^2}\Big) \right ) \Biggr ] \,.
\nonumber\\
\end{eqnarray}
Up to a given order Eqs. (\ref{eqn2.29}) and 
(\ref{eqn2.30}) do not differ from the structure functions presented  
in Eqs. (\ref{eqn2.6}) and (\ref{eqn2.1}) respectively as long as one
uses fixed order perturbation theory. This can be
checked up to second order when the coefficient functions in 
Eqs. (\ref{eqn2.29}) and (\ref{eqn2.30}) are substituted using the mass 
factorization relations in (\ref{eqn2.15})-(\ref{eqn2.20}). The difference
arises if the logarithms of the type $\ln^i(\mu^2/m^2)$, which show up in the 
parton densities, are resummed using the renormalization group equations
in Eq. (\ref{eqn2.5}). This resummation induces corrections beyond fixed
order perturbation theory which become noticeable for $\mu^2 \gg m^2$.
On the other hand we do not want that the resummation bedevils the
threshold behavior of the structure functions. In this region the
best representation is still given by Eqs. (\ref{eqn2.1}) and (\ref{eqn2.6}).
Therefore one has to look for a scale at which expressions
(\ref{eqn2.29}) and (\ref{eqn2.30}) coincide with those given by
fixed order perturbation theory in (\ref{eqn2.6}) and (\ref{eqn2.1})
respectively. Finding this scale is the most important issue in CSN as
we will show below.
Both expressions $F_{i,Q}^{\rm CSN}$ and
$F_i^{\rm CSN,LIGHT}$ are renormalization group invariants. Hence they
satisfy the renormalization group equations (see Eq. (\ref{eqn2.4}) )
\begin{eqnarray}
\label{eqn2.31}
D\,F_{i,Q}^{\rm CSN}=0\,, \qquad D\,F_i^{\rm CSN,LIGHT}=0 \,.
\end{eqnarray}
The same holds for the total structure function in the variable
flavor number scheme which is defined as
\begin{eqnarray}
\label{eqn2.32}
&& F_i^{\rm CSN}(n_f+1,Q^2,m^2) = F_i^{\rm CSN,LIGHT}(n_f,\Delta,Q^2,m^2)
\nonumber\\[2ex] && \qquad\qquad 
+ F_{i,Q}^{\rm CSN}(n_f+1,\Delta,Q^2,m^2)
\,.
\end{eqnarray}
One can now show that for large $Q^2$, $F_i^{\rm CSN}(n_f+1,Q^2,m^2)$
turns into the same expression as Eq. (\ref{eqn2.1}) where $n_f$ in
the light parton coefficient functions ${\cal C}_{i,k}$ is replaced
by $n_f+1$ and ${\cal C}_{i,k}^{\rm VIRT}=0$ for the 
$n_f+1$ heavy flavor piece.

After having discussed the general procedure to construct CSN scheme
structure functions we now turn to the practical issues. 
For asymptotic values of $Q^2$, far above the heavy $Q\bar Q$ threshold at
$(1-x)Q^2/x = 4m^2$, all coefficient functions 
${\cal C}_{i,k}^{\rm CSN}$ in Eq. (\ref{eqn2.29}) can be replaced by the 
light parton coefficient functions so that, after having removed
$L_{i,k}^{\rm HARD}$, one gets the heavy quark
components of the structure functions in the so-called zero mass 
variable flavor number scheme (ZM-CSN). 
However near threshold at low $Q^2$ and large $x$ there is a problem, which 
has not been solved satisfactorily in the literature. 
In this region one would like the $F_{i,Q}^{\rm CSN}$ to vanish 
in the same way that the $F_{i,Q}^{\rm EXACT}$ vanish. Unfortunately
the coefficient functions ${\cal C}_{i,Q}^{\rm CSN}$ do not vanish
in the threshold region due to the presence of the OME's $A_{Qk}^{\rm S}$ and
the functions $H_{i,Q}^{\rm NS}$ which describe processes with 
ONE heavy quark in the final state (see Eq. (\ref{eqn2.21}) ) 
contrary to $H_{i,g}$ and $H_{i,q}$ which 
originate from reactions with at least TWO heavy quarks (Q and $\bar {\rm Q}$) 
in the final state. Only the latter functions 
have the correct threshold behavior. In the literature
two ways have been proposed to obtain reasonable threshold behaviors.
The first one was given in a paper by Aivasis, Collins, Olness
and Tung \cite{acot}, which will be denoted 
as the ACOT boundary conditions. 
The second one was proposed in a paper by Thorne and Roberts
\cite{thro}, which we shall call the TR
boundary conditions. In both approaches the structure functions have
the properties
\begin{eqnarray}
\label{eqn2.33}
%Q^2 < m^2 \quad \rightarrow \quad 
F_{i,Q}(Q^2,m^2) =
F_{i,Q}^{\rm EXACT}(n_f,Q^2,m^2)\,
&\mbox{for}& Q^2 < m^2\,,
\nonumber\\[2ex]
%Q^2 \ge m^2 \quad \rightarrow \quad 
F_{i,Q}(Q^2,m^2) =
F_{i,Q}^{\rm CSN}(n_f+1,Q^2,m^2) 
&\mbox{for}& Q^2 \ge m^2\,,
\end{eqnarray}
where the parton densities satisfy the boundary conditions at $\mu^2= m^2$
\begin{eqnarray}
\label{eqn2.34}
%\mu^2\le m^2 \quad \rightarrow 
f_{k + \bar k}(n_f+1,x,\mu^2)&=&
f_{k + \bar k}(n_f,x,\mu^2)\,,
\nonumber\\[2ex]
f_{Q + \bar Q}(n_f+1, x, \mu^2)&=&0 \quad \protect\footnotemark \,,
\nonumber\\[2ex]
f_g(n_f+1,x,\mu^2)&=&f_g(n_f,x,\mu^2)\,.
\end{eqnarray}
\addtocounter{footnote}{-1}
\footnotetext{Note that in the
existing parton density sets $f_{Q + \bar Q}(n_f+1, x, \mu^2)$ also vanishes
for $\mu^2<m^2$.}
Notice that there is no relation between the scale $\mu$, chosen in 
these boundary conditions, and the production threshold of heavy quarks
$(1-x)~Q^2/x \ge 4~m^2$. If one e.g. takes $\mu^2=Q^2$ and $Q^2 \le m^2$ all
terms in Eq. (\ref{eqn2.29}), where the heavy quark density is multiplied with 
${\cal C}_{i,Q}^{\rm CSN}$, vanish in spite of the fact that heavy flavors
are still produced as long as $x < Q^2/(Q^2 + 4~m^2)$. On the other hand
it is possible that $Q^2 > m^2$ and $x \ge Q^2/(Q^2 + 4~m^2)$ which implies
a non-vanishing heavy quark density with no heavy quark pair production.
Therefore the value for $Q^2$ chosen in Eq. (\ref{eqn2.33}) is a little
arbitrary and no condition is imposed on $x$. Since we do not have any 
alternative we shall choose the same value of $Q^2$ as in  Eq. (\ref{eqn2.33})
above which $F_{i,Q}^{\rm EXACT}$ turns into $F_{i,Q}^{\rm CSN}$. 
Another problem is is that the boundary conditions in Eq. (\ref{eqn2.34}) 
do not agree with the relations in Eqs. (\ref{eqn2.22})-(\ref{eqn2.24}) 
if the computations are extended beyond order $a_s^2$ which happens
in the ${\overline {\rm MS}}$-scheme. In this case the continuity
at $\mu^2=m^2$ is changed into a discontinuity.
Finally there is a problem with the heavy quark density if we choose
an arbitrary value for $\mu^2$ in $f_{Q + \bar Q}(n_f+1, x, \mu^2)$.
In deep inelastic scattering one very often chooses $\mu^2=Q^2$. If
$Q^2<m^2$ one has to know $f_{Q + \bar Q}(n_f+1, x, \mu^2)$ for values 
$\mu^2<m^2$ which are not specified in  Eq. (\ref{eqn2.34}). Furthermore the 
OME's do not vanish for $\mu^2<m^2$ so that $C_{i,k}^{\rm CSN} \not = H_{i,k}$.
(see Eqs. (\ref{eqn2.17})-(\ref{eqn2.20})). Hence in Eq. (\ref{eqn2.33})
$F_{i,Q}\not = F_{i,Q}^{\rm EXACT}$ for $Q^2<m^2$ so that the threshold 
behavior will be spoiled. 
Therefore one has to avoid chosing $\mu^2<m^2$ which can be achieved
by the prescription given by ACOT in \cite{acot}
\begin{eqnarray}
\label{eqn2.35}
\mu^2=m^2+kQ^2 \left (1 - \frac{m^2}{Q^2} \right )^n 
&\mbox{for}& Q^2 > m^2\,,
\nonumber\\[2ex]
\mu^2=m^2 \hspace*{34mm} 
& \mbox{for}& Q^2 \le m^2\,,
\end{eqnarray}
with $k=1/2$ and $n=2$.
In this way one gets $C_{i,k}^{\rm CSN} = H_{i,k}$ for $Q^2 < m^2$
at least up to order $a_s$. For higher orders one has to use the relations
in Eqs. (\ref{eqn2.22})-(\ref{eqn2.24}) instead of those given in 
Eq. (\ref{eqn2.34}) as the latter only hold up to order $a_s$.
The new conditions are presented up to order $a_s^2$ in 
Eqs. (\ref{eqn2.25})-(2.27). 

In the TR prescription one first chooses $\mu^2=Q^2$ and then requires
\begin{eqnarray}
\label{eqn2.36}
F_i^{\rm CSN}(n_f+1,Q^2,m^2)\mid_{Q^2=m^2}&=&
F_i^{\rm EXACT}(n_f,Q^2,m^2)\mid_{Q^2=m^2} \,,
\nonumber\\[2ex]
\frac{d\, F_i^{\rm CSN}(n_f+1,Q^2,m^2)}{d\,\ln(Q^2/m^2)} \mid_{Q^2=m^2}
&=& \frac{d\, F_i^{\rm EXACT}(n_f,Q^2,m^2)}{d\,\ln(Q^2/m^2)}\mid_{Q^2=m^2}\,.
\nonumber\\
\end{eqnarray}
Using the mass factorization relations in Eqs. (\ref{eqn2.17})-(\ref{eqn2.20}) 
the TR boundary conditions
lead to new heavy quark coefficient functions ${\cal C}_{i,Q}^{\rm CSN}$
which have nothing to do with the reactions in Eq. (\ref{eqn2.21}). 
For instance the lowest order coefficient functions in the TR prescription
corresponding to the reaction
$\gamma^* + Q \rightarrow Q$ in Eq. (\ref{eqn2.21}) vanish 
at the $Q \bar Q$ threshold although this process only contains one
heavy quark in the final state.
Moreover, as already admitted by the authors in
\cite{thro}, this procedure breaks down beyond order $a_s$ 
because there are more coefficient functions than relations between them.
Another problem is
that it is unclear in which subtraction scheme one is working since
the subtraction terms in \cite{thro} have nothing to do with the usual OME's
except in the limit $Q^2 \gg m^2$. For example in order $a_s$ the 
subtraction term needed for $H_{i,Q}^{\rm NS,(1)}$ in Eq. (\ref{eqn2.18}) 
is not given by $A_{QQ}^{\rm NS,(1)}$. The same holds for 
$H_{i,g}^{\rm S,(1)}$ in Eq. (\ref{eqn2.19}),
which gets another subtraction than given by $A_{Qg}^{\rm S,(1)}$.
From the theoretical viewpoint the boundary conditions in Eq. (\ref{eqn2.36}) 
seem very unattractive to us because the relations between the coefficient
functions ${\cal C}_{i,k}^{\rm CSN}$ ($k=Q,q,g$) and the actual parton 
reactions are broken and the scheme is unknown. 
Notice that the parton densities
in \cite{thro} are still presented in the ${\overline {\rm MS}}$-scheme.

A different VFNS from that discussed above has been proposed
by Buza, Matiounine, Smith and van Neerven in \cite{bmsn1}, \cite{bmsn2}, 
which we call the BMSN scheme.
In the latter it was advocated that only when the large logarithms dominate 
the heavy quark coefficient functions do they have to be removed via 
mass factorization so that they can be resummed via the renormalization group 
equations. 
In the BMSN scheme we need the asymptotic heavy quark coefficient functions 
defined by
\begin{eqnarray}
\label{eqn2.37}
\lim_{Q^2 \gg m^2} H_{i,k}(n_f,\frac{Q^2}{m^2},\frac{Q^2}{\mu^2})
=H_{i,k}^{\rm ASYMP}(n_f,\frac{Q^2}{m^2},\frac{Q^2}{\mu^2})\,,
\end{eqnarray}
which behave like
\begin{eqnarray}
\label{eqn2.38}
H^{{\rm ASYMP},(l)}(n_f,\frac{Q^2}{m^2},\frac{Q^2}{\mu^2})
\sim a_s^l \sum_{n+j\leq l} a_{nj}
\ln^n\Big(\frac{Q^2}{m^2}\Big)\ln^j\Big(\frac{Q^2}{\mu^2}\Big)\,,
\end{eqnarray}
with a similar behavior for $L_{i,k}^{\rm ASYMP}$. 
In the BMSN scheme $F_{i,Q}^{\rm EXACT}$ is given by Eq. (\ref{eqn2.6})
except that $L_{i,k}\rightarrow L_{i,k}^{\rm HARD}$ analogous to the CSN.
The asymptotic heavy quark structure functions, 
denoted by $F_{i,Q}^{\rm ASYMP}$, are given by the same
expressions as presented for $F_{i,Q}^{\rm EXACT}$ where now all
exact heavy quark functions are replaced by their asymptotic ones. 
Up to second order the latter can be found in \cite{bmsmn}. The functions
$L_{i,q}^{\rm SOFT,ASYMP,NS,(2)}$ and $L_{i,q}^{\rm HARD,ASYMP,NS,(2)}$
are given in Appendix A.
In the BMSN scheme the charm components of the structure functions are
defined as
\begin{eqnarray}
\label{eqn2.39}
&&F_{i,Q}^{\rm BMSN}(n_f+1,\Delta,Q^2,m^2)=
F_{i,Q}^{\rm EXACT}(n_f,\Delta,Q^2,m^2) 
\nonumber\\[2ex]
&& - F_{i,Q}^{\rm ASYMP}(n_f,\Delta,Q^2,m^2)
+ F_{i,Q}^{\rm PDF}(n_f+1,\Delta,Q^2,m^2)\,,
\end{eqnarray}
with
\begin{eqnarray}
\label{eqn2.40}
&& F_{i,Q}^{\rm PDF} (n_f+1,\Delta,Q^2,m^2) =  
\nonumber\\[2ex]
&& e_Q^2 \left [ f_q^{\rm S}(n_f+1,\mu^2)\otimes\tilde{\cal C}_{i,q}^{\rm PS}
\Big(n_f+1,\frac{Q^2}{\mu^2}\Big)
 + f_g^{\rm S}(n_f+1, \mu^2) \right.
\nonumber\\[2ex]
&& \left. 
\otimes \tilde{\cal C}_{i,g}^{\rm S}\Big(n_f+1,\frac{Q^2}{\mu^2}\Big)
+ f_{Q +\bar Q}(n_f+1,\mu^2) \otimes
{\cal C}_{i,q}^{\rm NS}\Big(n_f+1,\frac{Q^2}{\mu^2}\Big) \right] 
\nonumber\\[2ex]
&& + \sum_{k=1}^{n_f} e_k^2 \left [ \sum_{l=1}^{n_f} f_{l+\bar l}(n_f+1,\mu^2)
\otimes L_{i,q}^{\rm HARD,ASYMP,PS}
\Big(n_f,\Delta,\frac{Q^2}{m^2},\frac{Q^2}{\mu^2}\Big) \right.
\nonumber\\[2ex]
&& \left. + f_g^{\rm S}(n_f+1,\mu^2)\otimes L_{i,g}^{\rm HARD,ASYMP,S}
\Big(n_f,\Delta,\frac{Q^2}{m^2},\frac{Q^2}{\mu^2}\Big)  \right.
\nonumber\\[2ex]
&& \left. + f_{k +\bar k}(n_f+1,\mu^2) \otimes L_{i,q}^{\rm HARD,ASYMP,NS}
\Big(n_f,\Delta,\frac{Q^2}{m^2},\frac{Q^2}{\mu^2}\Big) \right ] \,.
\end{eqnarray}
The structure functions $F_{i,Q}^{\rm PDF}$ are obtained from 
the $F_{i,Q}^{\rm ASYMP}$
via the mass factorization relations in Eqs. (\ref{eqn2.7}), (\ref{eqn2.10})
by making the replacements $H_{i,k} \rightarrow H_{i,k}^{\rm ASYMP}$,
$L_{i,k} \rightarrow L_{i,k}^{\rm ASYMP}$ and
${\cal C}_{i,k}^{\rm CSN} \rightarrow \tilde {\cal C}_{i,k}$ on the
left and right-hand sides respectively. Furthermore we have
used the definitions for the parton densities in Eqs. (\ref{eqn2.22})-
(\ref{eqn2.24}). Notice that if the coefficient functions, indicated
by $L_{i,k}^{\rm HARD,ASYMP}$, are removed from $F_{i,Q}^{\rm PDF}$ 
one obtains the structure functions in the ZM-CSN.
The light parton components of the structure functions become
\begin{eqnarray}
\label{eqn2.41}
&& F_i^{\rm BMSN,LIGHT}(n_f,\Delta,Q^2,m^2) =
\nonumber\\[2ex]
&& \sum\limits_{k=1}^{n_f} e_k^2  \left [
\sum_{l=1}^{n_f} f_{l+\bar l}(n_f+1,\mu^2) \otimes  
\delta\, {\cal C}_{i,q}^{\rm PS}\Big(n_f,\Delta,\frac{Q^2}{m^2},
\frac{Q^2}{\mu^2} \Big) \right.
\nonumber\\[2ex]
&& \left.  +  f_g^{\rm S}(n_f+1,\mu^2) \otimes 
\delta \, {\cal C}_{i,g}^{\rm S}\Big(n_f,\Delta,\frac{Q^2}{m^2},
\frac{Q^2}{\mu^2} \Big) \right.
\nonumber\\[2ex]
&& \left. + f_{k+\bar k}(n_f+1, \mu^2) \otimes 
\delta \, {\cal C}_{i,q}^{\rm NS}\Big(n_f,\Delta,\frac{Q^2}{m^2},
\frac{Q^2}{\mu^2} \Big) \right ] \,,
\end{eqnarray}
with
\begin{eqnarray}
\label{eqn2.42}
&& \delta \, {\cal C}_{i,k}\Big(n_f,\Delta,\frac{Q^2}{m^2},
\frac{Q^2}{\mu^2}\Big)= \tilde {\cal C}_{i,k}\Big(n_f+1,\frac{Q^2}{\mu^2}\Big)
+\tilde {\cal C}_{i,k}^{\rm VIRT}\Big(n_f,\frac{Q^2}{m^2},\frac{Q^2}
{\mu^2}\Big)
\nonumber\\[2ex]
&& +L_{i,k}^{\rm SOFT}\Big(n_f,\Delta,\frac{Q^2}{m^2},\frac{Q^2}{\mu^2}\Big)
-\tilde {\cal C}_{i,k}^{\rm VIRT,ASYMP}\Big(n_f, \frac{Q^2}{m^2},
\frac{Q^2}{\mu^2}\Big)
\nonumber\\[2ex]
&& -L_{i,k}^{\rm ASYMP}\Big(n_f,\frac{Q^2}{m^2},\frac{Q^2}{\mu^2} \Big) \,.
\end{eqnarray}
The coefficient functions above satisfy the property that
\begin{eqnarray}
\label{eqn2.43}
&& \mathop{\mbox{lim}}\limits_{\vphantom{\frac{A}{A}} Q^2 \gg m^2} 
\delta \, {\cal C}_{i,k}\Big(n_f,\Delta,\frac{Q^2}{m^2},\frac{Q^2}{\mu^2}\Big)
= 
\nonumber\\[2ex]
&& \tilde {\cal C}_{i,k}\Big(n_f+1,\frac{Q^2}{\mu^2}\Big)
 -L_{i,k}^{\rm HARD,ASYMP}
\Big(n_f,\Delta,\frac{Q^2}{m^2},\frac{Q^2}{\mu^2} \Big)\,.
\end{eqnarray}
Using the relations in Eqs. (\ref{eqn2.9}), (\ref{eqn2.12}) one can make 
a comparison between the CSN and the BMSN schemes. In the asymptotic limit
the heavy quark components satisfy (see Eqs. (\ref{eqn2.29}), 
(\ref{eqn2.39}), (\ref{eqn2.40}))
\begin{eqnarray}
\label{eqn2.44}
&& \mathop{\mbox{lim}}\limits_{\vphantom{
\frac{A}{A}} Q^2 \gg m^2}
F_{i,Q}^{\rm BMSN}(n_f+1,\Delta,Q^2,m^2)
= \mathop{\mbox{lim}}\limits_{\vphantom{
\frac{A}{A}} Q^2 \gg m^2} F_{i,Q}^{\rm CSN}(n_f+1,\Delta,Q^2,m^2) 
\nonumber\\[2ex]
&& = \mathop{\mbox{lim}}\limits_{\vphantom{\frac{A}{A}} Q^2 \gg m^2}
F_{i,Q}^{\rm PDF}(n_f+1,\Delta,Q^2,m^2) \,,
\end{eqnarray}
and the light parton components satisfy (see Eqs. (\ref{eqn2.30}),
(\ref{eqn2.41}))
\begin{eqnarray}
\label{eqn2.45}
&& \mathop{\mbox{lim}}\limits_{\vphantom{
\frac{A}{A}} Q^2 \gg m^2}
F_i^{\rm BMSN,LIGHT}(n_f,\Delta,Q^2,m^2)
= \mathop{\mbox{lim}}\limits_{\vphantom{
\frac{A}{A}} Q^2 \gg m^2} F_i^{\rm CSN,LIGHT}(n_f,\Delta,Q^2,m^2) \,,
\nonumber\\
\end{eqnarray}
provided we impose the same boundary conditions on both schemes. From 
the discussion above we infer that the only difference between the CSN 
and the BMSN schemes can be attributed to the $m^2/Q^2$-terms which are 
present in ${\cal C}_{i,Q}^{\rm CSN}$. Such terms 
do not occur in the ${\cal C}_{i,q}$
appearing in $F_{i,Q}^{\rm PDF}$ if one chooses the BMSN-scheme.
In the next Section we will make a study where in $Q^2$
these differences are noticeable.
%\end{document}

%%%%%%%%%%%%%%%%%%%%%%%%%%%%%%%%%%%%%%%%%%%%%%%%%%%%%%%%%%%%%%%%%%%%%%%%
%\format=latex
%\documentstyle[12pt]{article}
%\pagestyle{myheadings}  
%\begin{document}
%------------------This is Section 3---------------------------------
\mysection{Comparison between the CSN and the BMSN scheme}
%----------------------------------------------------------
%\topmargin=0in
%\headheight=0in
%\headsep=0in
%\oddsidemargin=7.2pt
%\evensidemargin=7.2pt
%\footheight=1in
%\marginparwidth=0in
%\marginparsep=0in
%\textheight=9in
%\textwidth=6in 26 
%\newcommand{\be}{\begin{eqnarray}}
%\newcommand{\ee}{\end{eqnarray}}
\newcommand{\ssim}{\scriptstyle \sim}

In this Section we will make a comparison in the case of charm production
between the above CSN scheme and the BMSN scheme proposed in
\cite{bmsn1}, \cite{bmsn2}. For that purpose we construct a parton
density set with four active flavors from an existing three flavor set
in the literature following Eqs. (2.22)-(2.24).
The charm quark density of our set will be compared with those in
other sets with four active flavors presented in \cite{mrst98} (MRST98, central
gluon) and \cite{cteq5} (CTEQ5HQ). Using our set we
will study the differences between the charm components of the structure
functions $F_{i,c}^{\rm CSN}(n_f+1)$ in Eq. (\ref{eqn2.29}) 
and $F_{i,c}^{\rm BMSN}(n_f+1)$ 
in Eq. (\ref{eqn2.39}) in particular in the threshold region.

Since all coefficient functions are computed in the 
${\overline{\rm MS}}$-scheme we choose the leading order (LO) and 
next-to-leading order (NLO) parton density sets presented in
\cite{grv98} which contain three active flavors only (i.e. u,d,s).
This implies that one has chosen $n_f=3$ for the anomalous dimensions.
In order to make this paper self contained 
we give some details here. In LO where the input scale $\mu_0$
is chosen to be $\mu_0^2=\mu_{\rm LO}^2=0.26~({\rm GeV/c})^2$ the input parton
densities are
\begin{eqnarray}
\label{eqn3.1}
xu_v(3,x,\mu_{\rm LO}^2) & = & 1.239\,\,x^{0.48}\,(1-x)^{2.72}\, 
    (1-1.8\sqrt{x} + 9.5x)\nonumber\\
xd_v(3,x,\mu_{\rm LO}^2) & = & 0.614\,\, (1-x)^{0.9}\,\, 
     xu_v(3,x,\mu_{\rm LO}^2)\nonumber\\
x\Delta(3,x,\mu_{\rm LO}^2) & = & 0.23 \,\, x^{0.48}\,(1-x)^{11.3}\,
    (1-12.0\sqrt{x} + 50.9x)\nonumber\\
x(\bar{u}+\bar{d})(3,x,\mu_{\rm LO}^2) & = & 1.52\,\, x^{0.15}\, (1-x)^{9.1}\,
    (1-3.6\sqrt{x} + 7.8x)\nonumber\\
xg(3,x,\mu_{\rm LO}^2) & = & 17.47\,\, x^{1.6}\, (1-x)^{3.8}\nonumber\\
xs(3,x,\mu_{\rm LO}^2) & = & x\bar{s}(x,\mu_{\rm LO}^2) = 0
\end{eqnarray}
In NLO where the input scale equals 
$\mu_{\rm NLO}^2=0.40~({\rm GeV/c})^2$ we have
\begin{eqnarray}
\label{eqn3.2}
xu_v(3,x,\mu_{\rm NLO}^2) & = & 0.632\,\,x^{0.43}\,(1-x)^{3.09}\, 
    (1+18.2x)\nonumber\\
xd_v(3,x,\mu_{\rm NLO}^2) & = & 0.624\,\, (1-x)^{1.0}\,\, 
     xu_v(3,x,\mu_{\rm NLO}^2)\nonumber\\
x\Delta(3,x,\mu_{\rm NLO}^2) & = & 0.20 \,\, x^{0.43}\,(1-x)^{12.4}\,
    (1-13.3\sqrt{x} + 60.0x)\nonumber\\
x(\bar{u}+\bar{d})(3,x,\mu_{\rm NLO}^2) & = & 1.24\,\, x^{0.20}\, (1-x)^{8.5}\,
    (1-2.3\sqrt{x} + 5.7x)\nonumber\\
xg(3,x,\mu_{\rm NLO}^2) & = & 20.80\,\, x^{1.6}\, (1-x)^{4.1}\nonumber\\
xs(3,x,\mu_{\rm NLO}^2) & = & x\bar{s}(x,\mu_{\rm NLO}^2) = 0.
\end{eqnarray}
where $\Delta\equiv\bar{d}-\bar{u}$. Furthermore in \cite{grv98} the 
heavy quark masses are $m_c=1.4~{\rm GeV/c^2}$, $m_b=4.5~{\rm GeV/c}$.
In both sets the densities are evolved from a very low
starting scale, where it is necessary to use the exact numerical solution
for the running coupling constant $\alpha_s(\mu^2)$. The latter follows
from the implicit equation 
\begin{eqnarray}
\label{eqn3.3}
&& \ln\frac{Q^2}{\Lambda_{n_f}^{\overline{\rm MS}}} = \frac{4\pi}
{\beta_0\alpha_s(\mu^2)}\,-\, \frac{\beta_1}{\beta_0^2} \, \ln
\left[ \frac{4\pi}{\beta_0\alpha_s(\mu^2)}\, +\, \frac{\beta_1}{\beta_0^2}
\right] \,,
\nonumber\\[2ex]
&& \beta_0=11-\frac{2}{3}n_f \,, \qquad \beta_1=102-\frac{38}{3}n_f \,,
\end{eqnarray}
and will be used in all the following formulae.
Furthermore we adopt the values 
$\Lambda_{3,4,5,6}^{\overline{\rm MS}}=299.4, 246, 167.7, 67.8~{\rm MeV}$   
which yield $\alpha_s(5,M_Z^2)=0.114$. Notice that the values for 
$\Lambda_{n_f}$ follow from the matching conditions
\begin{eqnarray}
\label{eqn3.4}
\alpha_s(n_f,\Lambda_{n_f},m^2)=\alpha_s(n_f+1,\Lambda_{n_f+1},m^2) \,.
\end{eqnarray}
where for $n_f=3$ and $n_f=4$ one has to choose $m=m_c$ and $m=m_b$ 
respectively. For the computation of $F_{i,c}^{\rm EXACT}$ (\ref{eqn2.6})
and $F_{i,c}^{\rm ASYMP}$ ((\ref{eqn2.39}) we take $n_f=3$ for the parton 
densities and the running coupling constant in Eq. (\ref{eqn3.3}). 
However for $F_{i,c}^{\rm CSN}$
(\ref{eqn2.29}), $F_{i,c}^{\rm BMSN}$ (\ref{eqn2.39}) and $F_{i,c}^{\rm PDF}$ 
(\ref{eqn2.40}) we need $n_f=4$ for the coupling constant in Eq. (\ref{eqn3.3})
and a parton density set with four active flavors (i.e. u,d,s,c) when the 
scale $\mu$
becomes larger or equal to the heavy flavor mass $m$. For reasons which will 
be explained below our computations are performed with parton densities 
represented in LO, NLO and NNLO (next-to-next-to leading order). 
The LO densities are convoluted
with the order $\alpha_s^2$ contributions to the coefficient functions. The NLO
densities are convoluted with the order $\alpha_s$ parts of the coefficient 
functions. The zeroth order contributions to the latter have to be multiplied
with the NNLO densities. Notice that for $n_f=3$ we only need LO and NLO
densities here since the heavy quark coefficient functions in 
$F_{i,c}^{\rm EXACT}$ 
and $F_{i,c}^{\rm ASYMP}$ start in order $\alpha_s$. 
Our sets with four active flavors are derived from the 
ones with three active flavors by putting $n_f=3$ in 
Eqs. (\ref{eqn2.25})-(\ref{eqn2.27}) starting at
a specific scale which we choose as $\mu^2 = m^2$. Hence it follows that
in LO or in zeroth order $\alpha_s$ one gets
\begin{eqnarray}
\label{eqn3.5}
f_{k+{\bar k}}^{\rm LO}(4,x,m^2)& = &f_{k+{\bar k}}^{\rm LO}(3,x,m^2)\,,
\nonumber\\[2ex]
f_{c + {\bar c}}^{\rm LO}(4,x,m^2)& = & 0\,,
\nonumber\\[2ex]
f_g^{\rm S,LO}(4,x,m^2)& = &f_g^{\rm S,LO}(3,x,m^2)\,,
%\nonumber\\[2ex]
\end{eqnarray}
whereas in NLO or in first order $\alpha_s$
one obtains
\begin{eqnarray}
\label{eqn3.6}
f_{k+{\bar k}}^{\rm NLO}(4,x,m^2)& =& f_{k+{\bar k}}^{\rm NLO}(3,x,m^2)\,,
\nonumber\\[2ex]
f_{c + {\bar c}}^{\rm NLO}(4,x,m^2)& = & 0\,,
\nonumber\\[2ex]
f_g^{\rm S, NLO}(4,x,m^2)& = &f_g^{\rm S, NLO}(3,x,m^2)\,.
%\nonumber\\[2ex]
\end{eqnarray}
Since the two-loop OME's $A_{Qk}^{(2)}$, $A_{kl,Q}^{(2)}$ in Eqs. 
(\ref{eqn2.25})-(\ref{eqn2.27}) do not vanish at $\mu^2=m^2$ in the 
${\overline {\rm MS}}$-scheme
we find that in NNLO or in order $\alpha_s^2$ 
the parton densities are discontinuous at $\mu^2=m^2$ while going from
three to four flavors i.e.
\begin{eqnarray}
\label{eqn3.7}
f_{k+{\bar k}}^{\rm NNLO}(4,x,m^2)& \not= &f_{k+{\bar k}}^{\rm NNLO}
(3,x,m^2)\,,
\nonumber\\[2ex]
f_{c + {\bar c}}^{\rm NNLO}(4,x,m^2)& \not= & 0\,,
\nonumber\\[2ex]
f_g^{\rm S, NNLO}(4,x,m^2)& \not= &f_g^{\rm S, NNLO}(3,x,m^2)\,.
%\nonumber\\[2ex]
\end{eqnarray}
Note that if we would drop the terms independent of 
$\ln \mu^2/m^2$ in the two-loop
operator matrix elements the inequalities in Eq. (\ref{eqn3.7}) would
become equalities. Above $\mu=m_c$ all four flavor number densities evolve with 
$n_f=4$ (we have not yet included a bottom quark density above $\mu=m_b$).
The evolution of the parton densities, given by the renormalization group
equations in Eq. (\ref{eqn2.5}), is determined by the anomalous dimensions 
$\gamma_{ij}^{(0)}$ (LO), $\gamma_{ij}^{(1)}$ (NLO) and $\gamma_{ij}^{(2)}$ 
(NNLO). Unfortunately the three-loop anomalous dimensions 
$\gamma_{ij}^{(2)}$ are
not known yet except for the moments $N=2,4,6,8$ (see \cite{mom}).
However an analysis of the light parton structure 
function in Eq. (\ref{eqn2.1}) \cite{nevo}
reveals that the contribution from $\gamma_{ij}^{(2)}$ is less 
important numerically than the contribution
due to the two-loop coefficient functions computed in \cite{zn}.
Hence our ignorance about the three-loop anomalous dimension will
not appreciably alter our results.
Therefore our NNLO analysis is only determined by the boundary
conditions in Eq. (\ref{eqn3.7}) which only affects the charm density
appearing in $F_{i,c}^{\rm CSN}$, $F_{i,c}^{\rm BMSN}$ and 
$F_{i,c}^{\rm PDF}$.
The evolution of the parton densities above $\mu^2=m_c^2$, presented in 
\cite{chsm}, was performed 
using a computer program which implements the direct $x$-space method
(similar to that of \cite{Botje}). The code is written in C++ and has
the options to evolve densities in LO and NLO
whereas the NNLO option presently only uses the NLO anomalous dimensions.
We have checked that the evolution of the parton densities
is in agreement with the results in \cite{brnv}. 

Before substituting the parton densities into the structure functions
we encounter a problem caused by the inequalities in Eq. (\ref{eqn3.7}). 
This happens in the threshold region which, according to the 
ACOT boundary conditions in Eq. (\ref{eqn2.35}), is defined by $Q^2<m^2$. 
In this region one has to choose $\mu^2=m^2$ so 
that $F_{i,c}^{\rm CSN}(n_f=4)$ and $F_{i,c}^{\rm BMSN}(n_f=4)$ are equal
to $F_{i,c}^{\rm EXACT}(n_f=3)$. Notice that the latter has to be understood
in the sense mentioned below Eq. (\ref{eqn2.38}) where $L_{i,k} \rightarrow
L_{i,k}^{\rm HARD}$. However since $\alpha_s(m^2)$
is rather large we have to truncate the perturbation series for the structure
functions to the desired order otherwise the threshold behavior of all the
$F_{i,c}^{\rm CSN}(n_f=4)$ and $F_{i,c}^{\rm BMSN}(n_f=4)$ will be destroyed.
Let us explain this for the BMSN scheme in Eq. (\ref{eqn2.39}). The arguments
for the CSN scheme proceed in an analogous way. The conditions that the
$F_{i,c}^{\rm BMSN}(n_f=4)=F_{i,c}^{\rm EXACT}(n_f=3)$ for $\mu^2=m^2$ implies
that the $F_{i,c}^{\rm ASYMP}(n_f=3)$ in Eq. (\ref{eqn2.39}) are canceled by
the $F_{i,c}^{\rm PDF}(n_f=4)$. 
Further we have to bear in mind that the $F_{i,c}^{\rm EXACT}(n_f=3)$
and $F_{i,c}^{\rm ASYMP}(n_f=3)$ are determined by the parton densities in the
three flavor number scheme whereas the $F_{i,c}^{\rm PDF}(n_f=4)$ 
are constructed out of
the four flavor number scheme parton densities. The latter have the form 
\begin{eqnarray}
\label{eqn3.8}
f_k(4,x,m^2)^{\rm NNLO}&=&f_k(3,x,m^2)^{\rm NNLO} [ 1 + O(\alpha_s^2)]
\nonumber\\[1ex]
f_g^{\rm S}(4,x,m^2)^{\rm NNLO}&=&f_g^{\rm S}(3,x,m^2)^{\rm NNLO} 
[ 1 + O(\alpha_s^2)] 
\nonumber\\[1ex]
c(4,x,m^2)^{\rm NNLO}&=&f_g^{\rm S}(3,x,m^2)^{\rm NNLO}O(\alpha_s^2)
+f_q^{\rm S}(3,x,m^2)^{\rm NNLO} O(\alpha_s^2)\,.
\nonumber\\
\end{eqnarray}
If these densities are substituted in $F_{i,c}^{\rm PDF}(n_f=4)$ 
in Eq. (\ref{eqn2.40}) one obtains additional terms of order 
$\alpha_s^3$ and $\alpha_s^4$ in the perturbation series which are not canceled
by $F_{i,c}^{\rm ASYMP}(n_f=3)$. 
Notice that the latter are only computed up to order 
$\alpha_s^2$. This effect is caused by multiplying the four flavor number 
densities 
with the coefficient functions ${\cal C}_{i,k}$ corrected beyond zeroth order 
in $\alpha_s$. To avoid these higher order terms we propose the 
following formulae for the structure functions in the CSN scheme 
\begin{eqnarray}
\label{eqn3.9}
&& F_{i,Q}^{\rm CSN} (n_f+1,\Delta,Q^2,m^2)  =  e_Q^2 \left [
f_{Q +\bar Q}^{\rm NNLO}(n_f+1,\mu^2) {\cal C}_{i,Q}^{\rm CSN,NS,(0)}
(\frac{Q^2}{m^2}) \right.
\nonumber\\[2ex]
&& \left. +a_s(n_f+1,\mu^2) \left \{ f_{Q +\bar Q}^{\rm NLO}(n_f+1,\mu^2) 
\otimes
{\cal C}_{i,Q}^{\rm CSN,NS,(1)}\Big (\frac{Q^2}{m^2},\frac{Q^2}{\mu^2}\Big )
\right. \right.
\nonumber\\[2ex]
&& \left. \left. +f_g^{\rm S,NLO}(n_f+1,\mu^2) \otimes
{\cal C}_{i,g}^{\rm CSN,S,(1)} \Big (\frac{Q^2}{m^2},\frac{Q^2}{\mu^2}
\Big ) \right \}
\right.
\nonumber\\[2ex]
&& \left. +a_s^2(n_f+1,\mu^2) \left \{
f_{Q +\bar Q}^{\rm LO}(n_f+1,\mu^2)\otimes \left (
{\cal C}_{i,Q}^{\rm CSN,NS,(2)}
\Big (n_f+1,\frac{Q^2}{m^2}, \frac{Q^2}{\mu^2}\Big ) \right. \right. \right.
\nonumber\\[2ex]
&& \left. \left. \left. + {\cal C}_{i,Q}^{\rm CSN,PS,(2)}
\Big(\frac{Q^2}{m^2}, \frac{Q^2}{\mu^2}\Big) \right )
 + \sum_{l=1}^{n_f}f_{l + \bar l}^{\rm LO} (n_f+1,\mu^2) \otimes 
{\cal C}_{i,q}^{\rm CSN,PS,(2)}\Big(\frac{Q^2}{m^2}, \frac{Q^2}{\mu^2}\Big)
\right. \right.
\nonumber\\[2ex]
&& \left. \left. + f_g^{\rm S,LO}(n_f+1,\mu^2)\otimes 
{\cal C}_{i,g}^{\rm CSN,S,(2)}
\Big (\frac{Q^2}{m^2},\frac{Q^2}{\mu^2} \Big )\right \}\right ]
\nonumber\\[2ex]
&& + a_s^2(n_f+1,\mu^2) \sum_{k=1}^{n_f}\,e_k^2 \,
f_{k + \bar k}^{\rm LO}(n_f+1,\mu^2)
\otimes L_{i,q}^{\rm HARD,NS,(2)} \Big (\Delta,\frac{Q^2}{m^2}\Big ) \,.
\end{eqnarray}
Notice that from the perturbative point of view the heavy quark density
$f_{Q +\bar Q}^{\rm LO}$ starts in order $\alpha_s(\mu^2)$ so that
after multiplication with  
${\cal C}_{i,Q}^{\rm CSN,NS,(2)}$ and ${\cal C}_{i,Q}^{\rm CSN,PS,(2)}$
the product becomes of order $\alpha_s^3(\mu^2)$. Hence these 
coefficient functions did not appear in the mass factorization relations 
(\ref{eqn2.17})-(\ref{eqn2.20}) since the latter are carried out up to order
$\alpha_s^2(\mu^2)$. Since the heavy quark density in CSN has to be treated
on the same footing as the light flavor densities, in particular after
resummation of the terms in $\ln^i(\mu^2/m^2)$, all densities are considered to
start in zeroth order in perturbation theory. This explains the form of
of the above expression. Furthermore the coefficient functions
${\cal C}_{i,Q}^{\rm CSN,NS,(2)}$ and ${\cal C}_{i,Q}^{\rm CSN,PS,(2)}$
which originate from parton processes with an heavy
quark in the initial state have not been calculated yet. Therefore we 
have to approximate them by the replacements
\begin{eqnarray}
\label{eqn3.10}
&&{\cal C}_{i,Q}^{\rm CSN,NS,(2)}
\Big (n_f+1,\frac{Q^2}{m^2}, \frac{Q^2}{\mu^2} \Big ) 
\rightarrow 
{\cal C}_{i,q}^{\rm NS,(2)}\Big (n_f+1,\frac{Q^2}{\mu^2}
\Big ) \,,
\nonumber\\[2ex]
&& {\cal C}_{i,Q}^{\rm CSN,PS,(2)}
\Big (\frac{Q^2}{m^2}, \frac{Q^2}{\mu^2} \Big ) 
\rightarrow {\cal C}_{i,q}^{\rm CSN,PS,(2)}
\Big (\frac{Q^2}{m^2}, \frac{Q^2}{\mu^2} \Big ) \,,
\end{eqnarray}
respectively. The remaining CSN scheme coefficient functions
can be computed via the relations in Eqs. (\ref{eqn2.17})-(\ref{eqn2.20}),
which are defined in terms of the known $H$'s and $A$'s. The light partonic
parts of the CNS structure functions up to second order are given by
\begin{eqnarray}
\label{eqn3.11}
&& F_i^{\rm CSN,LIGHT}(n_f,\Delta,Q^2) =
\sum_{k=1}^{n_f} e_k^2 \Biggl [ f_{k + \bar k}^{\rm NNLO}(n_f+1,\mu^2)
{\cal C}_{i,q}^{\rm NS,(0)}  
\nonumber\\[2ex]
&& +a_s(n_f+1,\mu^2) \left \{ f_{k + \bar k}^{\rm NLO}(n_f+1,\mu^2) \otimes
{\cal C}_{i,q}^{\rm NS,(1)} (\frac{Q^2}{\mu^2})
+ f_g^{\rm S,NLO}(n_f+1,\mu^2) \right. 
\nonumber\\[2ex]
&& \left. \otimes \tilde {\cal C}_{i,g}^{\rm S,(1)} (\frac{Q^2}{\mu^2})
\right \}
+ a_s^2(n_f+1,\mu^2) \left \{ f_{k + \bar k}^{\rm LO}(n_f+1,\mu^2) 
\otimes
\left ( {\cal C}_{i,q}^{\rm NS,(2)} \Big (n_f,\frac{Q^2}{\mu^2}\Big )
\right. \right. 
\nonumber\\[2ex]
&& \left. \left. + {\cal C}_{i,q,Q}^{\rm SOFT,NS,(2)}
\Big (\Delta,\frac{Q^2}{m^2},\frac{Q^2}{\mu^2} \Big ) \right )
 + \sum_{l=1}^{n_f}f_{l + \bar l}^{\rm LO}(n_f+1,\mu^2)\otimes
\tilde {\cal C}_{i,q}^{\rm PS,(2)} \Big (\frac{Q^2}{\mu^2} \Big )
\right. 
\nonumber\\[2ex]
&& \left. + f_g^{\rm S,LO}(n_f+1,\mu^2)\otimes \tilde 
{\cal C}_{i,g}^{\rm S,(2)} \Big (\frac{Q^2}{\mu^2}\Big ) \right \} \Biggr ] \,,
\end{eqnarray}
with
\begin{eqnarray}
\label{eqn3.12}
&& {\cal C}_{i,q,Q}^{\rm SOFT,NS,(2)}\Big (\Delta,\frac{Q^2}{m^2},
\frac{Q^2}{\mu^2}\Big )
=L_{i,q,Q}^{\rm SOFT,NS,(2)}\Big (\Delta,\frac{Q^2}{m^2}\Big )
+\left ( F^{(2)}(Q^2,m^2) \right.
\nonumber\\[2ex]
&& \left. -A_{qq,Q}^{\rm NS,(2)} \Big (\frac{\mu^2}{m^2})\right )
 {\cal C}_{i,q}^{\rm NS,(0)}
 +\beta_{0,Q} \ln \left (\frac{\mu^2}{m^2} \right ) {\cal C}_{i,q}^{\rm NS,(1)}
\Big (\frac{Q^2}{\mu^2} \Big )
\end{eqnarray}
For the BMSN scheme we need the expression for $F_{i,Q}^{\rm EXACT}$ as 
defined above Eq. (\ref{eqn2.39})
\begin{eqnarray}
\label{eqn3.13}
&&F_{i,Q}^{\rm EXACT}(n_f,\Delta,Q^2,m^2) = 
 e_Q^2 \left [ a_s(n_f,\mu^2) f_g^{\rm S,NLO}(n_f,\mu^2) \otimes 
H_{i,g}^{\rm S,(1)}\Big(\frac{Q^2}{m^2}\Big) \right.
\nonumber\\[2ex]
&&\left. + a_s^2(n_f,\mu^2)\left \{ \sum_{k=1}^{n_f} f_{k + \bar k}^{\rm LO} 
(n_f,\mu^2) \otimes
H_{i,q}^{\rm PS,(2)}\Big(\frac{Q^2}{m^2},\frac{Q^2}{\mu^2}\Big) \right. \right.
\nonumber\\[2ex] 
&& \left. \left. +  f_g^{\rm S,LO}(n_f,\mu^2) \otimes
H_{i,g}^{\rm S,(2)}\Big(\frac{Q^2}{m^2},\frac{Q^2}{\mu^2}\Big) \right \}
\right ] 
\nonumber\\[2ex]
&& + a_s^2(n_f,\mu^2)\sum_{k=1}^{n_f} e_k^2 f_{k+\bar k}^{\rm LO}(n_f,\mu^2) 
\otimes L_{i,q}^{\rm HARD,NS,(2)} \Big(\Delta,\frac{Q^2}{m^2}\Big )\,.
\end{eqnarray}
If we choose the maximum $\Delta=s$ (defined in 
the figure caption for Fig. \ref{fig:5}), we get
$L_{i,q}^{\rm HARD,NS,(2)}=0$. On the other hand if the minimum value
is adopted i.e. $\Delta=4m^2$ one gets $L_{i,q}^{\rm HARD,NS,(2)}
=L_{i,q}^{\rm NS,(2)}$
and $F_{i,Q}^{\rm EXACT}$ becomes equal to the conventional expression
given in Eq. (\ref{eqn2.6}).
The structure function $F_{i,Q}^{\rm ASYMP}$ is obtained from
the expression above by replacing the exact coefficient functions by their 
asymptotic analogues. Furthermore to calculate $F_{i,Q}^{\rm BMSN}$ in 
Eq. (\ref{eqn2.39}) we need
\begin{eqnarray}
\label{eqn3.14}
&& F_{i,Q}^{\rm PDF} (n_f+1,\Delta,Q^2,m^2)  =  e_Q^2 \Biggl [
f_{Q +\bar Q}^{\rm NNLO}(n_f+1,\mu^2) {\cal C}_{i,q}^{\rm NS,(0)}
\nonumber\\[2ex]
&& +a_s(n_f+1,\mu^2) \left \{ f_{Q +\bar Q}^{\rm NLO}(n_f+1,\mu^2) 
\otimes {\cal C}_{i,q}^{\rm NS,(1)}(\frac{Q^2}{\mu^2}) \right. 
\nonumber\\[2ex] 
&& \left. +f_g^{\rm S,NLO}(n_f+1,\mu^2)\otimes \tilde 
{\cal C}_{i,g}^{\rm S,(1)} (\frac{Q^2}{\mu^2}) \right \} 
\nonumber\\[2ex]
&& +a_s^2(n_f+1,\mu^2) \left \{
f_{Q +\bar Q}^{\rm LO}(n_f+1,\mu^2)\otimes \left ({\cal C}_{i,q}^{\rm NS,(2)}
(n_f+1,\frac{Q^2}{\mu^2}) \right. \right.
\nonumber\\[2ex] 
&& \left. \left. 
+ \tilde {\cal C}_{i,q}^{\rm PS,(2)}\Big(\frac{Q^2}{\mu^2}\Big)\right) 
+ \sum_{l=1}^{n_f}f_{l + \bar l}^{\rm LO} (n_f+1,\mu^2) \Big )\otimes
\tilde {\cal C}_{i,q}^{\rm PS,(2)}\Big(\frac{Q^2}{\mu^2}\Big) 
\right. 
\nonumber\\[2ex] 
&&\left. + f_g^{\rm S,LO}(n_f+1,\mu^2)\otimes \tilde 
{\cal C}_{i,g}^{\rm S,(2)} (\frac{Q^2}{\mu^2})\right \}\Biggr ]
\nonumber\\[2ex]
&& + a_s^2(n_f+1,\mu^2)\sum_{k=1}^{n_f} e_k^2 f_{k+\bar k}^{\rm LO}(n_f,\mu^2)
\otimes L_{i,q}^{\rm HARD,ASYMP,NS,(2)} \Big(\Delta,\frac{Q^2}{m^2}\Big )\,.
\nonumber\\[2ex]
\end{eqnarray}
Finally up to order $\alpha_s^2$, Eq. (\ref{eqn2.41}) becomes
\begin{eqnarray}
\label{eqn3.15}
&& F_i^{\rm BMSN,LIGHT}(n_f+1,\Delta,Q^2) =
\sum_{k=1}^{n_f} e_k^2 \Biggl [ f_{k + \bar k}^{\rm NNLO}(n_f+1,\mu^2)
{\cal C}_{i,q}^{\rm NS,(0)}  
\nonumber\\[2ex]
&& +a_s(n_f+1,\mu^2)
\left \{ f_{k + \bar k}^{\rm NLO}(n_f+1,\mu^2) \otimes
{\cal C}_{i,q}^{\rm NS,(1)} (\frac{Q^2}{\mu^2})
+ f_g^{\rm S,NLO}(n_f+1,\mu^2) \right.  
\nonumber\\[2ex]
&& \left. \otimes \tilde {\cal C}_{i,g}^{\rm S,(1)}(\frac{Q^2}{\mu^2}) \right \}
+ a_s^2(n_f+1,\mu^2) \left \{ f_{k + \bar k}^{\rm LO}
(n_f+1,\mu^2) \otimes
 {\cal C}_{i,q}^{\rm NS,(2)} (n_f+1,\frac{Q^2}{\mu^2}) \right. 
\nonumber\\[2ex]
&& \left. + \sum_{l=1}^{n_f}f_{l + \bar l}^{\rm LO}(n_f+1,\mu^2)\otimes
\tilde {\cal C}_{i,q}^{\rm PS,(2)} (\frac{Q^2}{\mu^2})
 + f_g^{\rm S,LO}(n_f+1,\mu^2)\otimes \tilde {\cal C}_{i,g}^{\rm S,(2)}
(\frac{Q^2}{\mu^2}) \right \} \Biggr ] 
\nonumber\\[2ex]
&& + a_s^2(n_f+1,\mu^2)\sum_{k=1}^{n_f} e_k^2 f_{k+\bar k}^{\rm LO}(n_f,\mu^2)
\otimes \left [ 
L_{i,q}^{\rm SOFT,NS,(2)} \Big(\Delta,\frac{Q^2}{m^2}\Big ) \right.
\nonumber\\[2ex]
&& \left. -L_{i,q}^{\rm ASYMP,NS,(2)} \Big(\frac{Q^2}{m^2}\Big )
+ \Big (F^{(2)}(Q^2,m^2)
-  F^{\rm ASYMP,(2)}(Q^2,m^2) \Big ){\cal C}_{i,q}^{\rm NS,(0)} \right ]\,.
\nonumber\\
\end{eqnarray}
Notice that we have the relations
\begin{eqnarray}
\label{eqn3.16}
{\cal C}_{i,q}^{\rm NS,(2)}(n_f+1,\frac{Q^2}{\mu^2})=
{\cal C}_{i,q}^{\rm NS,(2)}(n_f,\frac{Q^2}{\mu^2})+
{\cal C}_{i,q,Q}^{\rm NS,(2)}(\frac{Q^2}{\mu^2})\,,
\end{eqnarray}
with
\begin{eqnarray}
\label{eqn3.17}
&& {\cal C}_{i,q,Q}^{\rm NS,(2)}(\frac{Q^2}{\mu^2})
=L_{i,q}^{\rm ASYMP,NS,(2)}(\frac{Q^2}{m^2})
+ \Big (F^{\rm ASYMP,(2)}(Q^2,m^2)
\nonumber\\[2ex]
&&  -A_{qq,Q}^{\rm NS,(2)} (\frac{\mu^2}{m^2})\Big )\,
 {\cal C}_{i,q}^{\rm NS,(0)}
+\beta_{0,Q}\, \ln (\frac{\mu^2}{m^2}) {\cal C}_{i,q}^{\rm NS,(1)}
(\frac{Q^2}{\mu^2})\,.
\end{eqnarray}
The first and second terms in the expression above cancel the third last and 
final term in Eq. (\ref{eqn3.15}). The result is then equal to 
${\cal C}_{i,q,Q}^{\rm SOFT,NS,(2)}$ in Eq. (\ref{eqn3.11}) in the limit
$Q^2 \gg m^2$.
The form of the above structure functions also suppresses higher order
terms beyond $\alpha_s^2$ arising from the three flavor number parton densities 
since the latter also contain terms proportional to $\alpha_s$ and higher. 
This becomes apparent if one takes the Nth moments of the 
densities. For instance we observe the 
following behavior up to NNLO in the non-singlet case
\begin{eqnarray}
\label{eqn3.18}
f_q^{\rm LO,(N)}(\mu^2)&\sim&
\left [\frac{ \alpha_s(\mu^2)}{\alpha_s(\mu_0^2)}
\right ]^{\gamma_{qq}^{(0)}/2 \beta_0}
f_q^{\rm LO,(N)}(\mu_0^2)\,,
\nonumber\\[1ex]
f_q^{\rm NLO,(N)}(\mu^2)&\sim&\Big [ 1 + \alpha_s(\mu^2) A_q^{(1)} \Big ]
\left [\frac{ \alpha_s(\mu^2)}{\alpha_s(\mu_0^2)}
\right ]^{\gamma_{qq}^{(0)}/2 \beta_0} f_q^{\rm NLO,(N)}(\mu_0^2)\,,
\nonumber\\[1ex]
f_q^{\rm NNLO,(N)}(\mu^2)&\sim&\Big [1 + \alpha_s(\mu^2) A_q^{(1)}
+ \alpha_s^2(\mu^2) A_q^{(2)}\Big ]
\left [\frac{ \alpha_s(\mu^2)}{\alpha_s(\mu_0^2)}
\right ]^{\gamma_{qq}^{(0)}/2 \beta_0}
\nonumber\\[1ex]
&& \times f_q^{\rm NNLO,(N)}(\mu_0^2)\,.
\end{eqnarray}
The choice of the multiplication rules above avoids the appearance of scheme
dependent terms beyond the order in which we
want to compute the structure functions.
The above prescription guarantees that
for $Q^2<m^2$ we satisfy the condition $F_{i,c}=F_{i,c}^{\rm EXACT}(n_f)$ in
both schemes.

In the subsequent part of this section we will only discuss the case
where the heavy quark is the charm quark, i.e. $Q=c$. Hence
in all expressions above we have to choose $n_f=3$. Further we have
to make a choice for the cut off $\Delta$ appearing in the coefficient
functions $L_{k,q}^{\rm SOFT,NS,(2)}$ ($k=2,L$).
The $\delta = (\Delta-4m^2)/(s-4m^2)$ dependence of
$xL_{2,q}^{\rm SOFT,NS,(2)}$ is shown in Fig. 8, where one notes that it peaks
at large $x$, i.e., near threshold. (The plot for $xL_{L,q}^{\rm SOFT,NS,(2)}$
has a similar shape). After convoluting this function with the parton densities
its contribution to the structure function $F_2^{\rm LIGHT}$ only amounts 
to a few percent at $Q^2 = 10^3$  $({\rm GeV/c})^2$. At decreasing $Q^2$ 
the contribution becomes even smaller. The same holds for $L_{2,q}^{\rm HARD}$
contributing to $F_{2,c}$. Hence the dependence of the structure functions on 
the value of $\Delta$ will be very small. Therefore in the subsequent analysis
we choose $\Delta = 10$ ${\rm GeV}^2$. Other choices hardly affect
the plots so that our conclusions will be unaltered.

Next we present the $x$-dependence of the NNLO charm density 
(see Eq. (\ref{eqn3.7})) for various values of $\mu^2$ in Figs. 9a,b. 
The latter plot emphasizes the region $0.01<x<1$.
At $\mu=m$ it becomes negative for $x<0.007$ which is due to the boundary
condition in Eq. (\ref{eqn2.23}) and the momentum sum rule. When $\mu>m$ the 
density becomes positive
over the whole $x$ range. In Figs. 9c and 9d we have shown the charm densities
in NLO which are obtained from \cite{mrst98} and \cite{cteq5} respectively,
with an offset scale so that they can easily be compared with Fig. 9a.
The former is constructed in the TR scheme whereas the latter follows the
prescription of ACOT. Both are positive over the whole $x$ range. 
Our LO and NLO parameterizations, which are not shown in the figures, are 
also positive for all values of $x$.
This property can be traced back to the boundary conditions which yield in
LO and NLO $c(x,m^2)=0$.
Note that a direct comparison between the charm densities from different 
groups is not meaningful because each group fits
different data to determine their respective input three flavor number
gluon densities.
However it seems that our charm density, at small $\mu^2$, does not rise as 
steeply as that of the CTEQ5HQ \cite{cteq5} at small $x$. 
It is more similar to the  MRST98 (central gluon) \cite{mrst98} 
density. In Fig. 10 we make a comparison between our charm density which 
evolves according to the renormalization group equation and the one computed
in fixed order perturbation theory (FOPT) via Eq. (\ref{eqn2.23}). 
To that order we have plotted
\begin{eqnarray}
\label{eqn3.19}
R(x,\mu^2)=\frac{c^{\rm EVOLVED}(x,\mu^2)}{c^{\rm FOPT}(x,\mu^2)}\,.
\end{eqnarray}
The density $c^{\rm FOPT}$ is computed up to order $\alpha_s^2$ since
the OME's in Eq. (\ref{eqn2.26}) are only known up
to that order. In Fig. 10a we have shown the ratio in LO. The latter implies
that we have only kept the leading logarithms in $\ln \mu^2/m^2$ in the OME's
which are resummed in all orders in $c^{\rm EVOLVED}$. The deviation of
$R$ from unity shows the effect of the resummation. 
The same ratio is shown
in NLO in Fig. 10b where we also included the subleading terms in the
OME's. Finally if we take into account the non-logarithmic terms in the 
two-loop OME's $A_{Qg}^{\rm S,(2)}$ and $A_{Qq}^{\rm PS,(2)}$ (\ref{eqn2.26})
one obtains the NNLO ratio (see Fig. 10c). The figures reveal that 
in LO and NLO the effect of the resummation is very small except near $x=1$. 
This picture changes if we go to NNLO where the deviation of $R$ from one 
becomes appreciable when $x$ tends to zero. Here $R$ can even become negative
which happens for $\mu^2\approx 3~({\rm GeV/c})^2$. This effect is wholly due to
the boundary condition $c(x,m^2)\not =0$ which occurs beyond NLO. Furthermore
the figures reveal that $R>1$ at large $x$ whereas $R<1$ at small $x$. Notice
that in Fig. 10c $c^{\rm FOPT}(x,\mu^2)=0$ for $x=0.007$ at 
$\mu^2=2~({\rm GeV/c})^2$ so that $R=\infty$ which explains the bump in the 
figure. Figure 10c is important because it shows that
$c^{\rm EVOLVED}(x,\mu^2)<c^{\rm FOPT}(x,\mu^2)$ at small $x$. The consequence
is that $F_{i,c}^{\rm BMSN}(x,Q^2)$ and $F_{i,c}^{\rm CSN}(x,Q^2)$ will 
become smaller than $F_{i,c}^{\rm EXACT}(x,Q^2)$ when $Q^2$ becomes slightly
larger than $m^2$ due to the choice made for the 
factorization scale in Eq. (\ref{eqn2.35}). This can even lead to a negative
structure function as will happen for $F_{L,c}^{\rm CSN}$ which we will see 
later on.

Now we present results for the various structure functions. In Fig.11
we show the charm quark structure functions in NNLO given by
$F_{2,c}^{\rm CSN}(n_f=4)$, 
$F_{2,c}^{\rm BMSN}(n_f=4)$, 
$F_{2,c}^{\rm PDF}(n_f=4)$ and
$F_{2,c}^{\rm EXACT}(n_f=3)$  
plotted in the region $1<Q^2<10^3~({\rm GeV/c})^2$ for $x=0.05$. 
The figure reveals that there is hardly
any difference between the BMSN and CSN prescriptions. The curves in both
prescriptions lie between the ones representing $F_{2,c}^{\rm PDF}(n_f=4)$
and $F_{2,c}^{\rm EXACT}(n_f=3)$ except for low $Q^2$. In this region the
latter is a little bit larger than the other ones which is expected from the 
discussion of the charm density given above. Notice that in the low $Q^2$ 
region $F_{2,c}^{\rm PDF}(n_f=4)$ becomes negative which means that charm
quark electroproduction cannot be described by this quantity anymore.
In Fig. 12 we present the same plots for $x=0.005$. Again one cannot
distinguish between $F_{2,c}^{\rm BMSN}(n_f=4)$ and $F_{2,c}^{\rm CSN}(n_f=4)$
but now both are smaller than $F_{2,c}^{\rm EXACT}(n_f=3)$ over the whole $Q^2$
range. The latter is even larger than $F_{2,c}^{\rm PDF}(n_f=4)$ in particular
for $Q^2>50~({\rm GeV/c})^2$.
Further we want to emphasize that due to our careful treatment of the
threshold region there is an excellent cancellation (to three significant
places) between $F_{2,c}^{\rm PDF}(n_f=4)$ and
and $F_{2,c}^{\rm ASYMP}(n_f=3)$ at very small $Q^2$ so that
both $F_{2,c}^{\rm CSN}(n_f=4)$ 
and $F_{2,c}^{\rm BMSN}(n_f=4)$ 
tend to 
$F_{2,c}^{\rm EXACT}(n_f=3)$. 
Also at large $Q^2$ we have
an excellent cancellation between 
$F_{2,c}^{\rm ASYMP}(n_f=3)$ 
and 
$F_{2,c}^{\rm EXACT}(n_f=3)$  
so that both 
$F_{2,c}^{\rm CSN}(n_f=4)$ 
and
$F_{2,c}^{\rm BMSN}(n_f=4)$ 
tend to 
$F_{2,c}^{\rm PDF}(n_f=4)$ (see Eq. (\ref{eqn2.44})). 

In Fig. 13 we show similar plots as in Fig. 11 but now for the 
charm quark longitudinal structure functions. 
Here we observe a difference between the plots for
$F_{L,c}^{\rm CSN}(n_f=4)$ and $F_{L,c}^{\rm BMSN}(n_f=4)$ in the
region $m_c^2<Q^2<40~({\rm GeV/c})^2$.
In particular the latter tends to
$F_{L,c}^{\rm EXACT}(n_f=3)$ while the former is larger. Furthermore
$F_{L,c}^{\rm PDF}(n_f=3)$ is considerably larger than the other three
structure functions, which differs from the behavior seen 
in Fig. 11. This can be mainly attributed to the gluon density which 
plays a more prominant role in $F_{L,c}$ than in $F_{2,c}$. For $x=0.005$ 
(see Fig. 14) the difference between the BMSN and 
the CSN descriptions becomes even more conspicuous. In this case
$F_{L,c}^{\rm CSN}(n_f=4)$ becomes negative in the region
$m_c^2<Q^2<7~({\rm GeV/c})^2$ which is unphysical. This effect can be 
attributed to the zeroth order longitudinal coefficient function 
in Eq. (\ref{eqn3.9}), which behaves
like ${\cal C}_{L,Q}^{\rm CSN,NS,(0)}=4m^2/Q^2$ (see \cite{neve}), and
the non-vanishing charm density at $\mu^2=m^2$. In the case of BMSN
the longitudinal coefficient function is equal to zero in lowest order
so that $F_{L,c}^{\rm BMSN}(n_f=4)$ does not become negative.

In Figs. 15 and 16 we make a comparison between the NLO and the NNLO
structure functions $F_{2,c}^{\rm CSN}(n_f=4)$ and 
$F_{2,c}^{\rm BMSN}(n_f=4)$. Both prescriptions i.e. CSN and BMSN lead to
the same result in NNLO. However while going from NLO to NNLO the
the structure function $F_{2,c}^{\rm CSN}(n_f=4)$ decreases whereas
$F_{2,c}^{\rm BMSN}(n_f=4)$ increases a little bit. The differences 
in the case of $x=0.005$ are even smaller than those observed for $x=0.05$.
The same comparison between NLO and NNLO is made for the longitudinal
structure functions in Figs. 17 and 18. Here the differences between
NLO and NNLO are much larger than in the case of $F_{2,c}$ in Figs. 15,16.
In NLO 
$F_{L,c}^{\rm BMSN}(n_f=4)$ is smaller than the one plotted for NNLO. However
for $F_{L,c}^{\rm CSN}(n_f=4)$ we see a decrease in the small $Q^2$-region
while going from NLO to NNLO whereas for large $Q^2$ we observe the opposite.
In particular the valley in the region $m_c^2<Q^2<7~({\rm GeV/c})^2$
observed for $F_{L,c}^{\rm CSN}(n_f=4)$ at $x=0.005$ in NNLO turns into a bump.
This is due the boundary condition on the charm density which in NLO
vanishes at $\mu=m_c$ whereas in NNLO it is negative at small $x$-values
(see Fig. 10c). From the observations above one can conclude that the CSN
prescription bedevils the threshold (low $Q^2$) behavior for $F_{L,c}$ due
to the non-vanishing zeroth order longitudinal coefficient function
${\cal C}_{L,Q}^{\rm CSN,NS,(0)}$. 
This problem is avoided by TR in \cite{thro} by imposing a condition
on the structure functions as indicated in Eq. (\ref{eqn2.36}).
Hence our results for $F_{i,c}^{\rm BMSN}$ agree reasonably well for $i=2$ 
and $i=L$ with 
those presented in NLO by TR in \cite{thro}. This is mainly due to the fact 
that there is only a small difference between the NLO and NNLO approximations 
in the BMSN scheme. It also reveals that the condition in Eq. (\ref{eqn2.36}) 
for $i=L$ can be mimicked by a vanishing zeroth order longitudinal coefficient 
function. Note that results for the $x$-values presented above are 
representative for the whole range $5\times 10^{-5}<x<0.5$.

To summarize the main points of this paper we have discussed 
two variable flavor number schemes for charm quark electroproduction
in NNLO. They are distinguished by the way mass factorization
is implemented. In the CSN scheme this is done with respect to the
full heavy and light quark structure functions at finite $Q^2$. 
In the BMSN scheme the mass factorization is only applied to the
coefficient functions in the large $Q^2$ limit. Both schemes require
three flavor and four flavor number parton densities which satisfy NNLO
matching conditions at a scale $\mu^2= m^2$. 
We have constructed these densities using our own evolution code.
The schemes also require matching conditions on the coefficient functions 
which are implemented in this paper.
We have also made a careful analysis of the removal of 
dangerous terms in $\ln(Q^2/m^2)$ from the Compton contributions so that both
$F_{i,c}^{\rm CSN}(n_f+1)$ and
$F_{i,c}^{\rm BMSN}(n_f+1)$ are collinearly safe. We have done this in a
way which is simplest from the theoretical point of view, by
implementing a cut $\Delta$ on the mass of the $c-\bar c$ pair which has
to be determined by experiment.
This cut is not required in the
fixed order perturbation theory approach given by $F_{i,c}^{\rm EXACT}$
in \cite{lrsn} for moderate $Q^2$-values.

Finally we made a careful analysis of the threshold behaviors of
$F_{i,c}^{\rm CSN}(n_f+1)$ and
$F_{i,c}^{\rm BMSN}(n_f+1)$. In order to
achieve the required cancellations near threshold so that they both become
equal to $F_{i,c}^{\rm EXACT}(n_f)$ one must be 
very careful to combine terms with the same order in the
expansion in the running coupling constant $\alpha_s$.
Therefore technically we require six sets of parton densities, namely
the LO, NLO and NNLO three flavor number sets and the LO, NLO and NNLO four
flavor number sets. However not all the necessary theoretical inputs
are available to us to finish this task. The approximations we made
in this paper were sufficient to provide very clear answers.
We successfully implemented the required cancellations near threshold and the
corresponding limits at large scales came out naturally. Inconsistent sets
of parton densities automatically spoil these cancelations. We did not have 
to use matching conditions on derivatives of structure functions as
proposed in \cite{thro}, which seem very artificial. The numerical 
results do however end up quite similar. We have also shown that the CSN
scheme defined above leads to an unnatural behavior of the longitudinal 
structure function in the threshold region which is due to a 
non-vanishing zeroth order coefficient function. 
Since there are no other differences between
the CSN and BMSN schemes we recommend the latter because it is less 
complicated than the former. In particular
it does not need additional coefficient functions other than
the existing heavy quark and light parton coefficient functions available
in the literature.

ACKNOWLEDGMENTS

The work of A. Chuvakin and J. Smith was partially supported
by the National Science Foundation grant PHY-9722101.
The work of W.L. van Neerven was supported
by the EC network `QCD and Particle Structure' under contract
No.~FMRX--CT98--0194.

%\end{document}

\appendix
%format=latex
%\documentstyle[12pt]{article}
%\begin{document}
%\bigskip
\mysection*{Appendix A}
\setcounter{section}{1}

In this appendix we present the exact expressions for the heavy quark
coefficient functions $L^{(2)}_{i,q}$ corresponding to the Compton process
in Fig. \ref{fig:5} when there is a cut $\Delta$ on the invariant mass 
$s_{Q\bar Q}$ of the heavy quark pair.
As explained in \cite{bmsmn} the calculation is straightforward because
one can first integrate over the heavy quark momenta in the final state without
affecting the momentum of the remaining light quark. The phase space
integrals are the same as the ones obtained for the process 
$\gamma(q) + q(k_1) \rightarrow g^* + q$  ($g^* \rightarrow Q + \bar Q$)
where the gluon becomes virtual. In the expressions for the complete
integration over the virtual mass $s_{Q\bar Q}$ of the gluon 
(see Fig. \ref{fig:5}) one integrates over the 
range $4\,m^2 \le s_{Q\bar Q} \le s$ with $s=(q+k_1)^2$. The 
resulting expressions, called $L_{i,q}^{\rm r,(2)}$ with $r={\rm NS}, {\rm S}$
and $i=2,L$\, are presented in appendix A of \cite{bmsmn}. Notice that 
up to order $\alpha_s^2$ there is no difference between singlet and non-singlet
so that $L_{i,q}^{\rm NS,(2)}=L_{i,q}^{\rm S,(2)}$. If we limit 
the range of integration to $4\,m^2 \le s_{Q\bar Q}\le \Delta$ one obtains
%(A.1)
\begin{eqnarray}
\label{eqnA.1}
&&L^{\rm SOFT,NS,(2)}_{L,q}(z,\Delta,\frac{Q^2}{m^2}) = 
C_F T_f \left \{ 96 a^2(s) z (1-z)^2  \Big [L_1(L_2 + L_4 + L_5) \right.
\nonumber \\ 
&& \left. \quad - 2(DIL_1 - DIL_2 - DIL_3 + DIL_4) \Big ]
- 32 a^2(s) [1 + 3(1-z) \right.
\nonumber \\
&& \left. \quad  - 6(1-z)^2] L_1
+ \frac{16}{3} z \Big [ 1 - 26 a(s)(1-z) + 88 a^2(s)(1-z)^2 \Big ]
\frac{L_3}{sq_2}
\right.
\nonumber \\ 
&& \left. \quad  - \frac{256}{3} b(\Delta) a(s) z (1-z) (1-d(\Delta)) L_6
+ \frac{64}{3} b(\Delta) a(s) \Big [ 2 + 10(1-z) 
\right.
\nonumber \\ 
&& \left. \quad -14(1-z)^2 - d(\Delta)\Big (2 -(1-z) - 3(1-z)^2 \Big ) \Big ]
\right.
\nonumber \\
&& \left. \quad - 128 b(\Delta) d(\Delta) a^2(s) z (1-z)^2 (1 + 2 d(\Delta)) 
L_6 \right.
\nonumber \\ 
&& \left. \quad  + \frac{64}{3} b(\Delta) d(\Delta) a^2(s)
 \Big (1+2d(\Delta) \Big )
\Big [ 1 + 3(1-z) - 6(1-z)^2 \Big ]  \right.
\nonumber \\
&& \left. \quad  - \frac{32}{3} b(\Delta) z  + \frac{16}{9} b^3(\Delta) z 
\right \} \,, 
\end{eqnarray}
%(A.2)
\begin{eqnarray}
\label{eqnA.2}
&&L^{\rm SOFT,NS,(2)}_{2,q}(z,\Delta,\frac{Q^2}{m^2}) = 
C_F T_f \left \{ \Big( \frac{4}{3} \frac{1+z^2}{1-z} 
- 16 a^2(s) (1-z) [1 \right.
\nonumber \\ 
&& \left. \qquad - 9(1-z) + 9(1-z)^2]\Big)
[ L_1(L_2 + L_4 + L_5) - 2( DIL_1 - DIL_2 \right.
\nonumber \\ && \left. \quad - DIL_3 + DIL_4)]
-\Big( \frac{8}{3} - \frac{4}{1-z} + 8 a^2(s)[2 + 18(1-z) \right.
\nonumber \\ && \left. \quad 
- 36 (1-z)^2 +  \frac{1}{1-z}]\Big) L_1 
+ \Big( \frac{8}{9}[ 28 - 17 (1-z) - \frac{19}{1-z}] \right.
\nonumber \\ && \left. \quad
+ \frac{16}{9} a(s) [61 - 160(1-z) + 128 (1-z)^2] \right.
\nonumber \\ && \left. \quad 
- \frac{64}{9} a^2(s) (1-z)[ 23 - 104(1-z) + 94 (1-z)^2 ]\Big)\frac{L_3}{sq_2}
\right.
\nonumber \\ && \left. \quad 
+ \frac{64}{3} b(\Delta) a(s) \Big [\Big [ 2 - 7(1-z) + 6 (1-z)^2 \Big ] 
( 1 - d(\Delta)) 
\right.
\nonumber \\ && \left. \quad 
+ d(\Delta) a(s) (1-z)[ 1 - 9(1-z) + 9(1-z)^2 ](1 + 2 d(\Delta))\Big ] 
L_6\right.
\nonumber \\ && \left. \quad 
+ \frac{8}{9} b(\Delta) \Big ( 6 - b^2(\Delta) \Big )
\Big [2 - (1-z) - \frac{2}{1-z} 
\Big ]L_6 \right.
\nonumber \\ && \left. \quad 
- \frac{32}{3} b(\Delta) d(\Delta) a(s) \Big [ 7 - 3(1-z) - 9(1-z)^2 
-\frac{1}{1-z} \Big ]
\right.
\nonumber \\ && \left.\quad 
- \frac{32}{9} b(\Delta) a(s) \Big [ 2 - 95(1-z) + 121 (1-z)^2 
+ \frac{3}{1-z} \Big ]
\right.
\nonumber \\ && \left. \quad 
+ \frac{16}{3} b(\Delta) d(\Delta) a^2(s) \Big (1 + 2d(\Delta) \Big )
\Big [ 2 + 18 (1-z)  \right.
\nonumber \\ && \left. \quad 
- 36(1-z)^2 +\frac{1}{1-z} \Big ]
- \frac{8}{9} b(\Delta) \Big [ 50 - 34 (1-z) - \frac{29}{1-z} \Big ] \right.
\nonumber\\ && \left. \quad
+ \frac{4}{27} b^3(\Delta) \Big [ 38 - 28(1-z) - \frac{17}{1-z} \Big ] 
\right \}\,,
\end{eqnarray}
where the partonic scaling variable is equal to
$ z = Q^2/(2 q\cdot k_1) = Q^2/(s + Q^2)$,  
Further we have defined
%(A.3)
\begin{eqnarray}
\label{eqnA.3}
&& \xi=\frac{Q^2}{m^2} \quad , 
\quad sq_1 = \sqrt{1 - 4 \frac{z}{(1-z) \xi}} \quad , \quad
\quad sq_2 = \sqrt{1 - 4 \frac{z}{\xi}}\,,
\nonumber \\ &&
a(s) = \frac{m^2}{s} = \frac{z}{(1-z)\xi} \quad  , \quad 
b(\Delta) = \sqrt{1-4\frac{m^2}{\Delta}} \quad  , \quad d(\Delta) = 
\frac{\Delta}{4~m^2}
\nonumber \\ &&  
L_1 = \ln\Big( \frac{ 1+b(\Delta)}{1- b(\Delta)}\Big) \quad , \quad
L_2 = \ln\Big( \frac{ 1+sq_2}{1- sq_2}\Big) \quad , \quad
L_3 = \ln\Big( \frac{ sq_2+b(\Delta)}{sq_2- b(\Delta)}\Big)\,,
\nonumber \\ &&  
L_4 = \ln\Big( \frac{ 1-z}{z^2}\Big) \quad , \quad
L_5 = \ln\Big( \frac{ 1-b^2(\Delta)}{1- sq_1^2}\Big) \quad , \quad
L_6 = \ln\Big( \frac{ sq_2^2- b^2(\Delta)}{z(1- sq_1^2)}\Big)\,,
\nonumber \\ &&  
DIL_1 = {\rm Li}_2\Big(\frac{(1-sq_2^2)(1+b(\Delta))}
{(1+sq_2)(1-b^2(\Delta))}\Big)
 \quad , \quad
DIL_2 = {\rm Li}_2\Big(\frac{1-sq_2}{1+b(\Delta)}\Big) \quad , \quad
\nonumber \\ &&  
DIL_3 = {\rm Li}_2\Big(\frac{1-b(\Delta)}{1+sq_2}\Big) \quad , \quad
DIL_4 = {\rm Li}_2\Big(\frac{1+b(\Delta)}{1+sq_2}\Big) \,.
\end{eqnarray}
The variable $\Delta $, which allows us to distinguish between
soft and hard (observable) heavy quark anti-quark pairs, is in the
range $4m^2 \le \Delta \le s$.
The variable $z$ is in the range $ 0 \le z \le \xi/(\xi + 4)$. 

Note that when $\Delta = s$ one obtains $L^{\rm SOFT,r,(2)}_{L,q}\rightarrow 
L_{i,q}^{\rm r,(2)}$ which are reported in \cite{bmsmn}.
When the integration range is given by $\Delta \le s_{Q\bar Q}\le s$
we get $L_{i,q}^{\rm HARD,r,(2)}$ which are given by
%(A.4)
\label{eqnA.4}
\begin{eqnarray}
L_{i,q}^{\rm HARD,NS,(2)}(z,\Delta,\frac{Q^2}{m^2})=
L_{i,q}^{\rm NS,(2)}(z,\frac{Q^2}{m^2})-
L_{i,q}^{\rm SOFT,NS,(2)}(z,\Delta,\frac{Q^2}{m^2})\,.
\end{eqnarray}
Notice that $L_{i,q}^{\rm HARD,NS,(2)}$ is finite in the limit $m \rightarrow 0$
so that it does not contain collinear divergences. The latter can be wholly
attributed to $L_{i,q}^{\rm SOFT,NS,(2)}$ as is revealed if one takes 
the limit $Q^2\rightarrow \infty$. In this case the expressions (A.1) and (A.2)
reduce to
%(A.5)
\begin{eqnarray}
\label{eqnA.5}
&&L^{\rm SOFT,ASYMP,NS,(2)}_{L,q}(z,\Delta,\frac{Q^2}{m^2}) = 
C_F T_f\left [\frac{16}{3}z\ln\frac{Q^2}{m^2}
-\frac{16}{3}z\ln\Big(\frac{Q^2}{\Delta}-z\Big)\right.
\nonumber\\ \ && \left. \quad
+\frac{32}{3}z\ln\frac{1+b(\Delta)}{2}
-\frac{80}{9}z -\frac{80}{9} (b(\Delta)-1)z 
+ \frac{16}{9} b(\Delta) \Big (b^2(\Delta)-1 \Big ) z  \right.
\nonumber \\ &&  \left. \quad 
+b(\Delta) \left \{\Big(\frac{64}{3} \frac{\Delta}{Q^2}z^2 
 - 16 \frac{\Delta^2}{Q^4} z^3\Big) \ln\Big[\Big(\frac{Q^2}{\Delta}-z\Big)
\frac{(1-z)}{z^2}\Big] 
\right. \right.
\nonumber \\ && \left. \left.\quad 
-  \frac{16}{3}\frac{\Delta}{Q^2}z \Big( \frac{2}{1-z} -4 +3z\Big)
+\frac{8}{3} \frac{\Delta^2}{Q^4}z^2 \Big( \frac{1}{(1-z)^2}
+\frac{3}{1-z} - 6\Big) \right \} \right ] \,,
\end{eqnarray}
and 
%(A.6)
\begin{eqnarray}
\label{eqnA.6}
&&L^{\rm SOFT,ASYMP,NS,(2)}_{2,q}(z,\Delta,\frac{Q^2}{m^2}) = 
C_F T_f \left [ \Big( \frac{1+z^2}{1-z} \Big) 
\left \{ \frac{8}{3} \ln\frac{Q^2}{m^2}
\ln \Big(\frac{1-z}{ z^2} \Big) \right. \right.
\nonumber \\ && \left. \left. \quad 
+\Big(\frac{8}{3} \ln\frac{Q^2}{m^2} 
+ \frac{16}{3} \ln\Big( \frac{1-z}{z^2}\Big)
+ \frac{8}{3} \ln \frac{Q^2}{\Delta} 
-\frac{116}{9} \Big)\ln \Big(\frac{1+b(\Delta)}{2}\Big) \right. \right.
\nonumber\\ && \left. \left. \quad
+\frac{4}{3} \ln^2\frac{Q^2}{m^2} -\frac{4}{3} \ln^2 \frac{Q^2}{\Delta}
- \frac{8}{3} \ln \frac{Q^2}{\Delta} \ln\Big( \frac{1-z}{z^2}\Big)
- \frac{8}{3}{\rm Li}_2 \Big(\frac{\Delta z(1+b(\Delta))}{2Q^2} \Big) 
 \right. \right.
\nonumber\\ && \left. \left. \quad
- \frac{8}{3}{\rm Li}_2 \Big(\frac{1+b(\Delta)}{2}\Big)
+ \frac{8}{3}{\rm Li}_2 \Big(\frac{2m^2}{\Delta(1+b(\Delta))} \Big) 
- \frac{58}{9} \ln\frac{Q^2}{m^2} - 2\ln\frac{Q^2}{\Delta} 
\right.\right.
\nonumber \\ && \left. \left. \quad 
+ 4 \ln\Big(\frac{Q^2}{\Delta} - z\Big)
- \frac{40}{9} \ln\Big( \frac{1-z}{z^2}\Big ) + \frac{314}{27} \right \}
+ \Big( \frac{2}{3}+\frac{26}{3}z \Big)\ln\frac{Q^2}{m^2}
\right.
\nonumber \\ && \left. \quad 
+\Big (\frac{2}{3}-2z \Big ) \ln\frac{Q^2}{\Delta}
-\Big( \frac{4}{3}+\frac{20}{3}z \Big)\ln(\frac{Q^2}{\Delta}-z) 
+\Big( \frac{4}{3}+\frac{52}{3}z \Big)  \ln \frac{1+b(\Delta)}{2} 
\right. 
\nonumber\\ && \left. \quad
-\frac{10}{9} - \frac{130}{9}z
+ b(\Delta) \left \{\Big[ -\frac{16}{3} 
\frac{\Delta}{Q^2} z \Big( \frac{2}{1-z} - 1  - 6z\Big ) 
+\frac{8}{3} \frac{\Delta^2}{Q^4}
z^2 \Big( \frac{1}{1-z} \right. \right.
\nonumber \\ && \left. \left. \quad 
- 9z\Big)\Big] \ln\Big[\Big(\frac{Q^2}{\Delta}-z\Big)\frac{(1-z)}{z^2}\Big] 
-\frac{8}{3} \frac{\Delta}{Q^2}z \Big( \frac{7}{1-z} -12 +9z
-\frac{1}{(1-z)^2}\Big) \right. \right.
\nonumber\\ && \left. \left. \quad
+\frac{2}{3} \frac{\Delta^2}{Q^4} z^2
\Big( \frac{2}{(1-z)^2} + \frac{18}{1-z}
-36 +\frac{1}{(1-z)^3} \Big) \right \}+
\Big ( \frac{8}{9}b(\Delta) (1-b^2(\Delta) ) \right.
\nonumber\\ && \left. \quad
+\frac{40}{9}\Big (b(\Delta)-1)\Big )
\Big (1+z-\frac{2}{1-z}\Big )
\ln \left (\Big (\frac{Q^2}{\Delta}-z \Big )\frac{(1-z)}{z^2} \right )
\right.
\nonumber\\ && \left. \quad
+\frac{4}{27}b(\Delta) \Big (b^2(\Delta)-1\Big ) 
\Big ( 10 +28z-\frac{17}{1-z}\Big )
+(b(\Delta)-1)\Big (-\frac{344}{27} \right.
\nonumber\\ && \left. \quad
-\frac{704}{27}z +\frac{628}{27}\frac{1}{1-z}\Big ) \right ]
\end{eqnarray}
respectively. In the limit $m \rightarrow 0$ the results above show the same 
logarithmic terms in $\ln^i(Q^2/m^2)$ $(i = 1,2)$
as the asymptotic expressions for $L_{i,q}^{\rm r,(2)}$ given in 
Eqs. (D.7) and (D.8) of \cite{bmsmn}. Hence the differences between the
results there and the asymptotic expressions above are free of any
terms in $\ln(Q^2/m^2)$. The asymptotic expressions of 
$L_{i,q}^{\rm HARD,r,(2)}$ for $Q^2 \rightarrow \infty$ are given by
%(A.7)
\begin{eqnarray}
\label{eqnA.7}
&&L^{\rm HARD,ASYMP,NS,(2)}_{L,q}(z,\Delta,\frac{Q^2}{m^2})  =
C_F T_f\left [\frac{16}{3}z\ln\frac{1-z}{z^2}
+\frac{16}{3}z\ln\Big(\frac{Q^2}{\Delta}-z\Big)\right.
\nonumber\\ \ && \left. \quad
-\frac{32}{3}z\ln\frac{1+b(\Delta)}{2}+\frac{16}{3}
-\frac{40}{3}z +\frac{80}{9} (b(\Delta)-1)z + \frac{16}{9} b(\Delta) 
\Big (1 \right.
\nonumber \\ && \left.  \quad
-b^2(\Delta) \Big ) z -b(\Delta) 
\left \{\Big(\frac{64}{3} \frac{\Delta}{Q^2}z^2 
 - 16 \frac{\Delta^2}{Q^4} z^3\Big) \ln\left (\Big(\frac{Q^2}{\Delta}-z\Big)
\frac{(1-z)}{z^2}\right ) \right. \right . 
\nonumber \\ && \left. \left.\quad
-  \frac{16}{3}\frac{\Delta}{Q^2}z  \Big( \frac{2}{1-z}
-4 +3z\Big)+\frac{8}{3} \frac{\Delta^2}{Q^4}z^2 \Big( \frac{1}{(1-z)^2}
+\frac{3}{1-z} - 6\Big) \right \} \right ] \,,
\nonumber\\
\end{eqnarray}
and
%(A.8)
\begin{eqnarray}
\label{eqnA.8}
&&L^{\rm HARD,ASYMP,,(2)}_{2,q}(z,\Delta,\frac{Q^2}{m^2})  =
C_F T_f \left [ \Big( \frac{1+z^2}{1-z} \Big)
\left \{ \frac{4}{3} \ln^2 (1-z) 
\right. \right.
\nonumber \\ && \left. \left. \quad
- 2\ln \Big(\frac{1-z}{z^2}\Big)-\frac{16}{3} \ln z \ln (1-z) 
+ 6\ln z + \frac{5}{3}
+\frac{4}{3} \ln^2 \frac{Q^2}{\Delta} 
\right. \right.
\nonumber \\ && \left. \left. \quad
+ \frac{8}{3} \ln \frac{Q^2}{\Delta} \ln\Big( \frac{1-z}{z^2}\Big)
+\Big(\frac{116}{9} -\frac{8}{3} \ln\frac{Q^2}{m^2} 
- \frac{16}{3} \ln\Big( \frac{1-z}{z^2}\Big)
- \frac{8}{3} \ln \frac{Q^2}{\Delta} \Big) \right. \right.
\nonumber\\ && \left. \left. \quad
\right. \right.
\nonumber \\ && \left. \left. \quad
\times \ln \Big(\frac{1+b(\Delta)}{2}\Big)
+ \frac{8}{3}{\rm Li}_2 \Big(\frac{\Delta z(1+b(\Delta))}{2Q^2} \Big)
+ \frac{8}{3}{\rm Li}_2 \Big(\frac{1+b(\Delta)}{2} \Big) 
- \frac{8}{3}\zeta(2) \right. \right.
\nonumber\\ && \left. \left. \quad
-\frac{8}{3}{\rm Li}_2 (1-z)
- \frac{8}{3}{\rm Li}_2 \Big(\frac{2m^2}{\Delta(1+b(\Delta))} \Big)
+ 2\ln\frac{Q^2}{\Delta}
- 4 \ln\Big(\frac{Q^2}{\Delta} - z\Big) 
\right \}
\right.
\nonumber \\ && \left. \quad
-\Big (\frac{2}{3}-2z \Big ) \ln\frac{Q^2}{\Delta} 
+\Big( \frac{4}{3}+\frac{20}{3}z \Big)\ln(\frac{Q^2}{\Delta}-z)
-\Big( \frac{4}{3}+\frac{52}{3}z \Big)  
\ln \Big(\frac{1+b(\Delta)}{2}\Big) \right.
\nonumber \\ && \left. \quad
+\Big( \frac{2}{3}+\frac{26}{3}z \Big)  \ln (1-z)
-\Big( 2+\frac{46}{3}z \Big)  \ln z +\frac{13}{3} - \frac{55}{3}z
+ b(\Delta) \left \{ \Big[\frac{16}{3} \frac{\Delta}{Q^2} z 
\right. \right.
\nonumber\\ && \left. \left. \quad
\times\Big( \frac{2}{1-z} - 1  - 6z\Big )
-\frac{8}{3} \frac{\Delta^2}{Q^4}
z^2 \Big( \frac{1}{1-z} - 9z)\Big)\Big]
\ln\Big[\Big(\frac{Q^2}{\Delta}-z\Big)\frac{(1-z)}{z^2}\Big] \right. \right.
\nonumber \\ && \left. \left. \quad
+\frac{8}{3} \frac{\Delta}{Q^2}z \Big( \frac{7}{1-z} -12 +9z
-\frac{1}{(1-z)^2}\Big)
-\frac{2}{3} \frac{\Delta^2}{Q^4} z^2
\Big( \frac{2}{(1-z)^2} + \frac{18}{1-z}  \right. \right.
\nonumber\\ && \left. \left. \quad
-36 +\frac{1}{(1-z)^3} \Big) \right \}+
\Big ( \frac{8}{9}b(\Delta)(b^2(\Delta)-1)+\frac{40}{9}(1-b)\Big )
\Big (1+z-\frac{2}{1-z}\Big )
\right.
\nonumber\\ && \left. \quad
\times \ln \Big [\Big (\frac{Q^2}{\Delta}-z \Big )\frac{(1-z)}{z^2} \Big ]
+\frac{4}{27}b(\Delta)(1-b^2(\Delta))\Big ( 10 +28z-\frac{17}{1-z} \Big ) 
\right.
\nonumber\\ && \left. \quad
+(1-b(\Delta))\Big (-\frac{344}{27}-\frac{704}{27}z
+\frac{628}{27}\frac{1}{1-z}\Big ) \right ]
\end{eqnarray}
respectively. As has been mentioned below Eq. (\ref{eqnA.4}) 
the expressions above
are finite in the limit $m \rightarrow 0$ ($b \rightarrow 1$) so that
they do not contain collinear divergences.
%\end{document}

%\include{vfsab}
%\documentstyle[12pt]{article}
%\pagestyle{myheadings}  
%\begin{document}
%----------------------------References-------------------------------------
%

%\documentstyle[12pt]{article}
%\begin{document}
\centerline{\bf \large{Figure Captions}}
\begin{description}
\item[Fig. 1.]
%--------------------------------
The lowest-order photon-gluon fusion process 
$\gamma^* + g \rightarrow Q + \bar Q$
contributing to the coefficient functions $H_{i,g}^{\rm S,(1)}$.
%----------------------
\item[Fig. 2.]
%---------------------------
%fig2
Some virtual gluon corrections to the process $\gamma^* + g \rightarrow
Q + \bar Q$ contributing to the coefficient functions
$H_{i,g}^{\rm S,(2)}$.
%--------------------------------------------
\item[Fig. 3.]
%-------------------------------
%fig3
The bremsstrahlung process $\gamma^* + g \rightarrow
Q + \bar Q + g$ contributing to the coefficient functions
$H_{i,g}^{\rm S,(2)}$.
%----------------------------------------
\item[Fig. 4.]
%---------------------------------
%fig4
The Bethe-Heitler process $\gamma^* + q(\bar q) \rightarrow
Q + \bar Q + q(\bar q)$ contributing to the coefficient functions
$H_{i,q}^{\rm PS,(2)}$. The light quarks $q$ and the
heavy quarks $Q$ are indicated by dashed and solid lines
respectively.
%---------------------------------------------
\item[Fig. 5.]
%-----------------------------------
%fig5
The Compton process 
$\gamma^* + q(\bar q) \rightarrow Q + \bar Q + q(\bar q)$ 
contributing to the coefficient functions $L_{i,q}^{\rm NS,(2)}$.
The light quarks $q$ and the
heavy quarks $Q$ are indicated by dashed and solid lines
respectively ($s=(p+q)^2$, $s_{Q\bar Q}=(p_1+p_2)^2$ see text).
%-------------------------------
\item[Fig. 6.]
%-----------------------------------
%fig6
The two-loop vertex correction to the process
$\gamma^* + q \rightarrow q$ containing a heavy quark (Q) loop.
It contributes to ${\cal C}_{i,q}^{\rm VIRT,NS,(2)}(Q^2/m^2)
=F^{(2)}(Q^2/m^2)~ {\cal C}_{i,q}^{(0)}$.
%----------------------------
\item[Fig. 7.]
%--------------------------
%fig7
Order $\alpha_s$ corrections to the process
$\gamma^* + Q \rightarrow Q$ and the reaction
$\gamma^* + Q \rightarrow Q + g$
contributing to the coefficient functions $H_{i,Q}^{\rm NS,(1)}$.
%--------------------------------
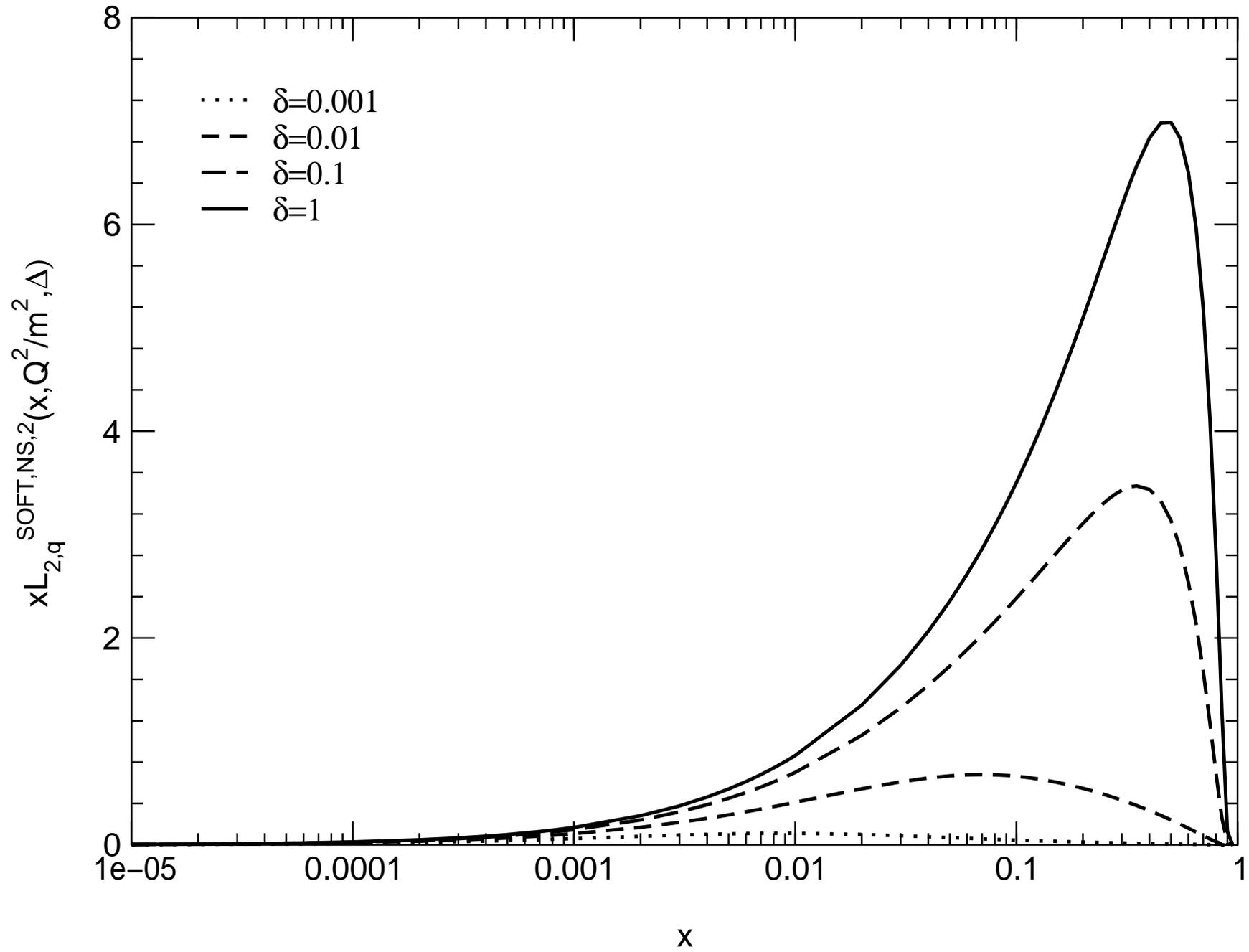
\item[Fig. 8.]
%----------------------
The $\delta = (\Delta-4m^2)/(s-4m^2)$ dependence of 
$xL_{2,q}^{\rm SOFT,NS,(2)}(x,Q^2/m^2,\Delta)$ at $Q^2/m^2=50$  (Eq. (A.2)) 
plotted as a function of $x$ for 
$\delta = $ 1, 0.1, 0.01 and 0.001 respectively.
%--------------------------------------------------
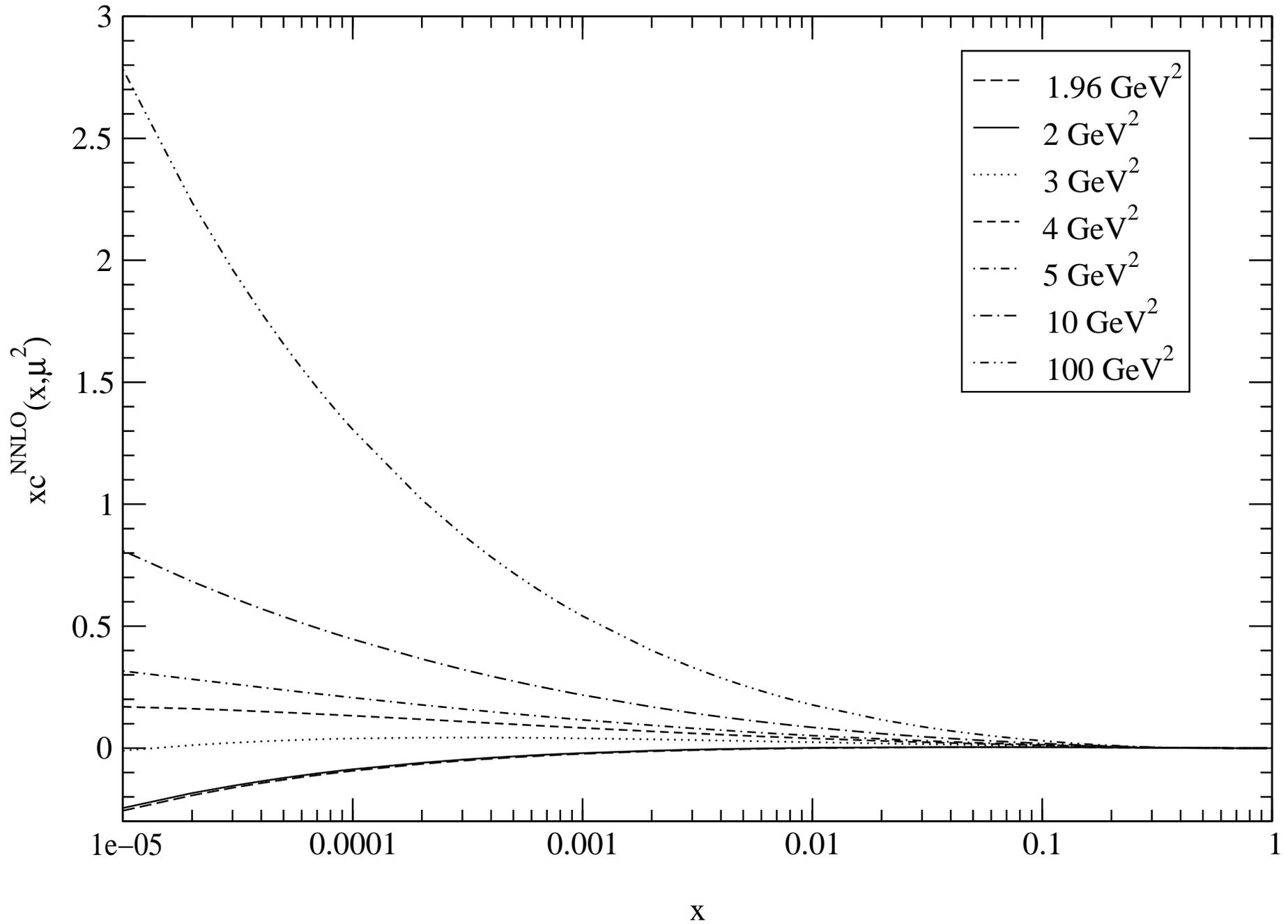
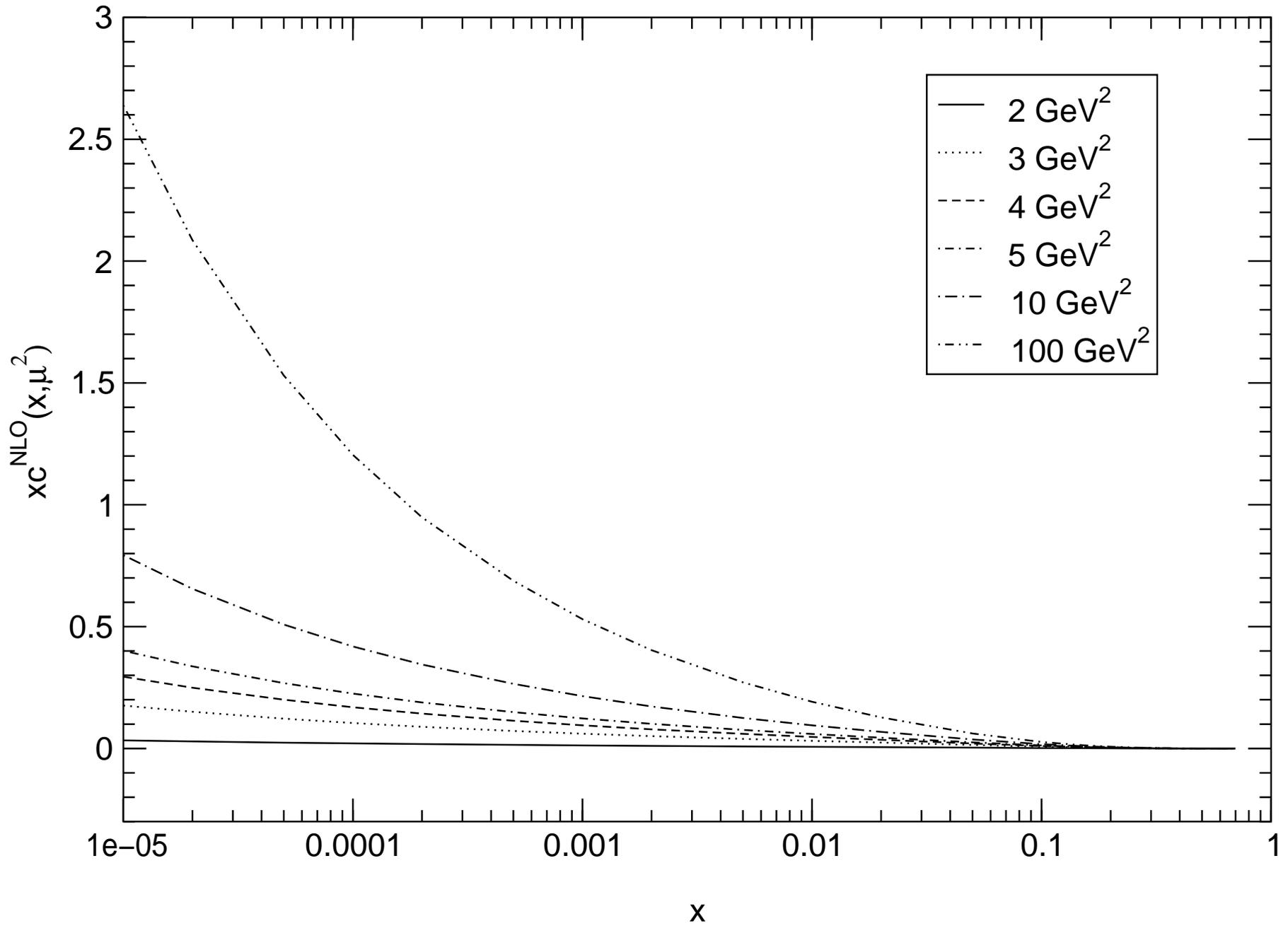
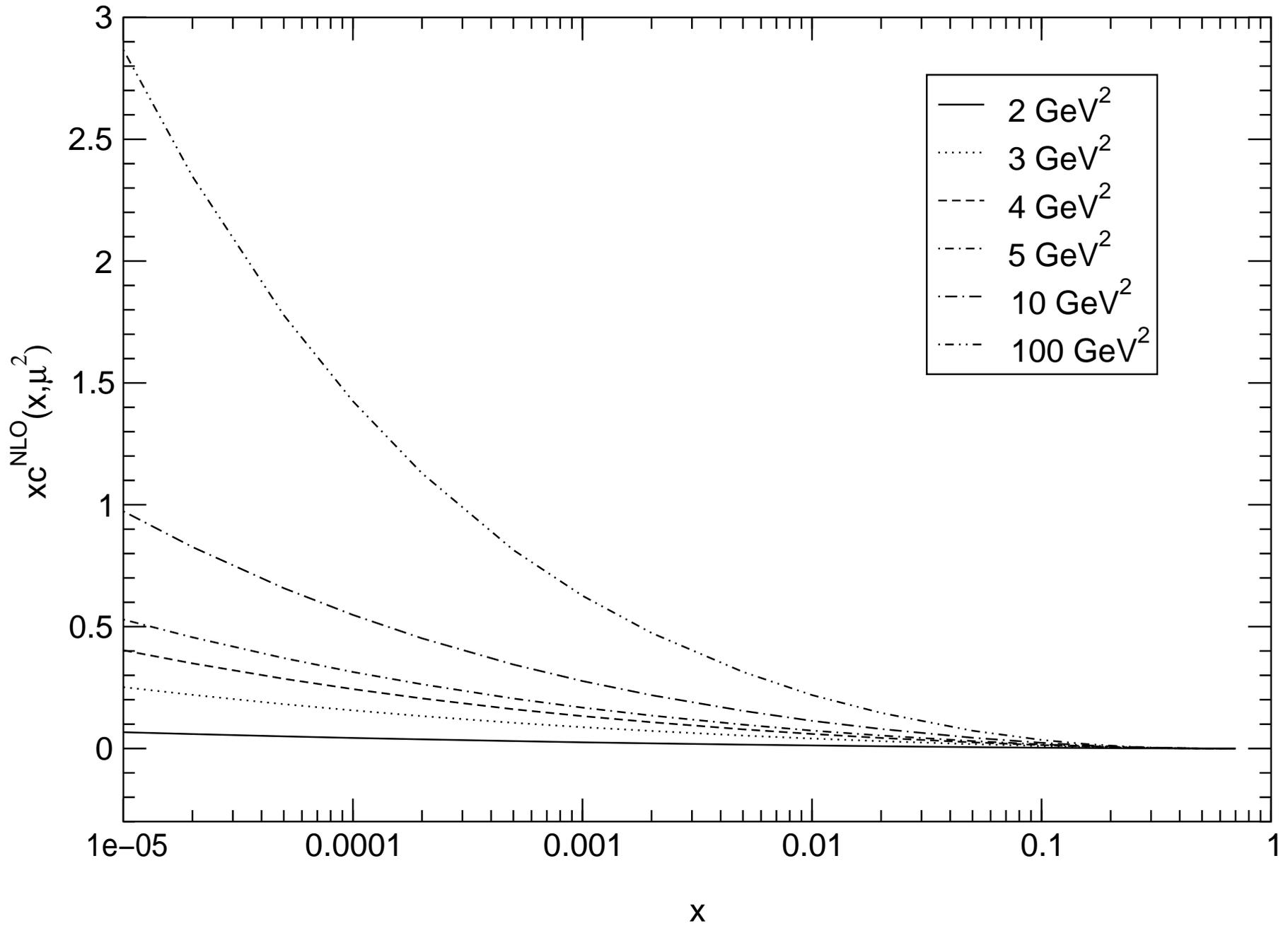
\item[Fig. 9.]
%----------------
(a) The charm density $xc^{\rm NNLO}(4,x,\mu^2)$ shown in the range
$10^{-5}<x<1$ for 
$\mu^2 =$ 1.96, 2, 3, 4, 5, 10 and 100 in units 
of $({\rm GeV/c})^2$. (b) similar plot as in (a) but now for $0.01<x<1$.
For a comparison we 
have also shown the NLO results obtained by MRST98 and CTEQ5HQ for the range
$10^{-5}<x<1$ in 
(c) and (d) respectively.
%------------------------------------------------------
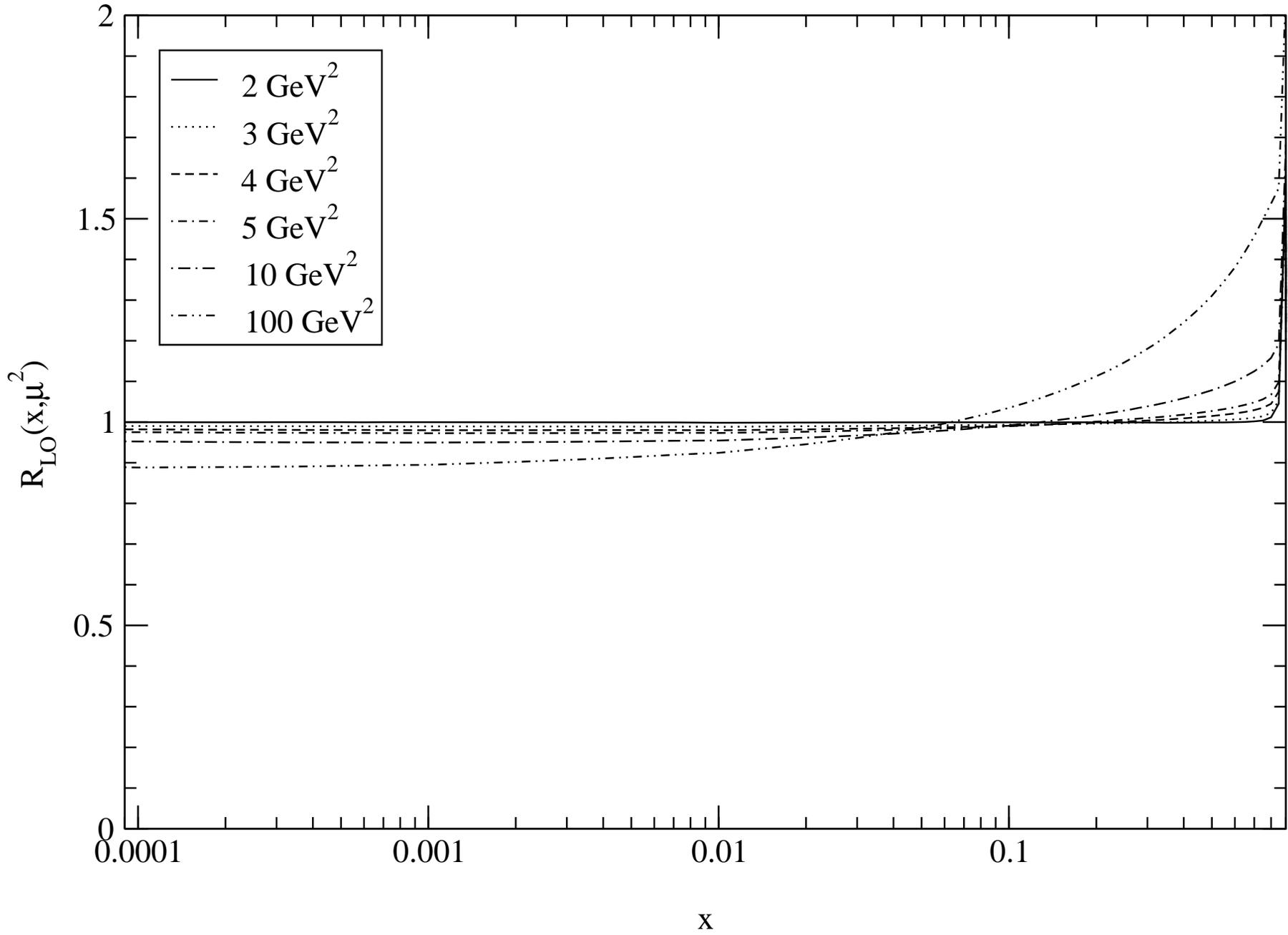
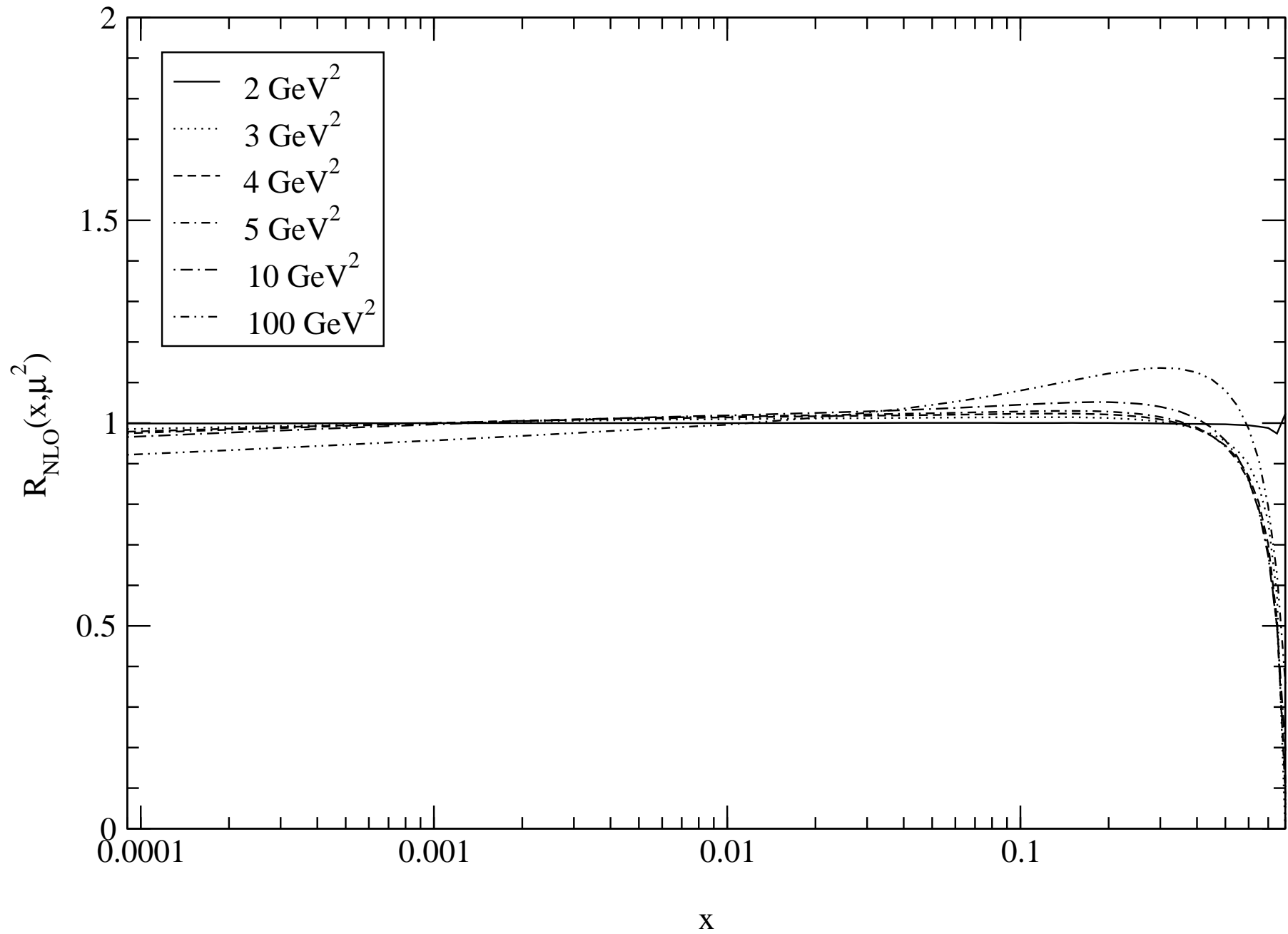
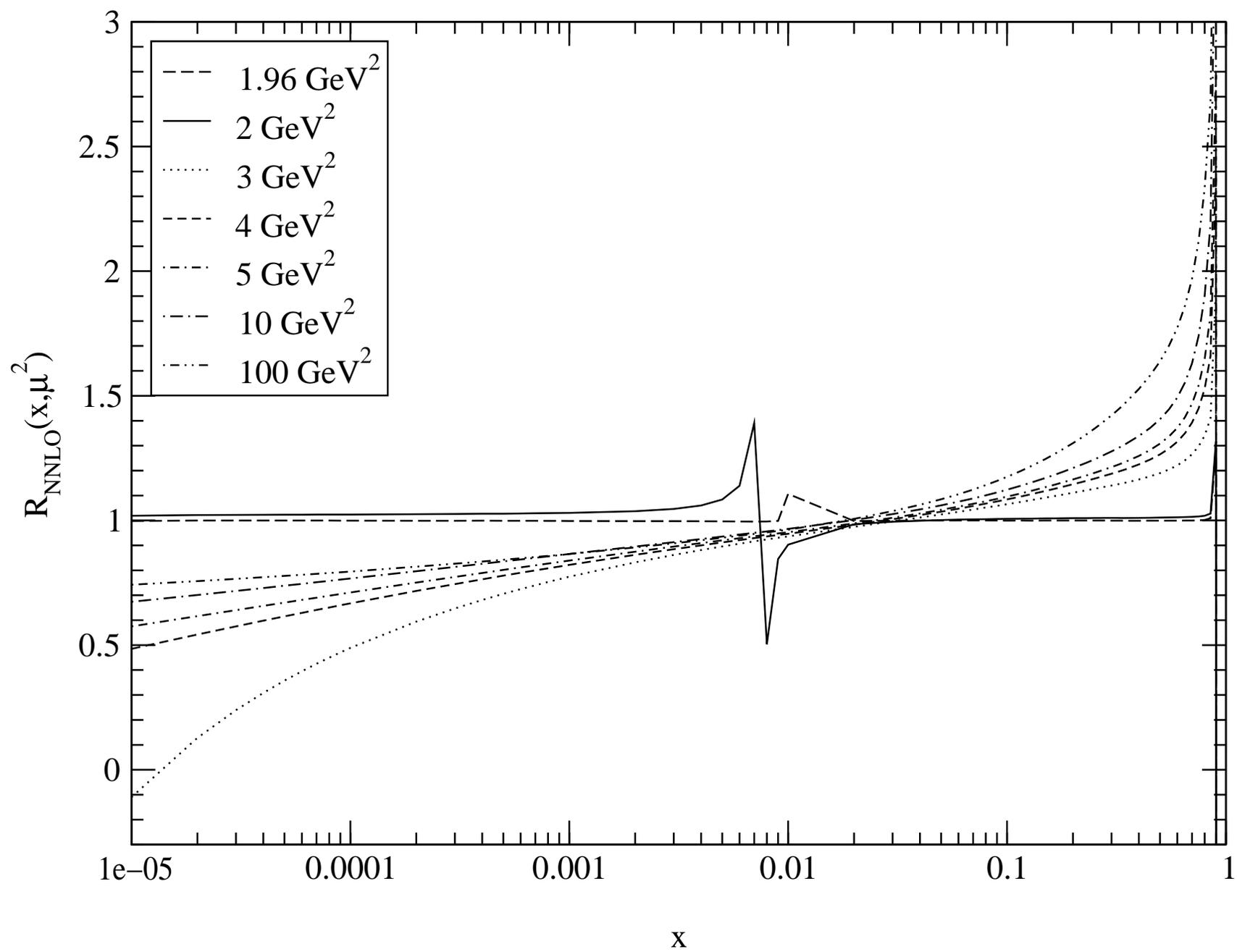
\item[Fig. 10.]
Ratios 
$R(x,\mu^2)=xc^{\rm EVOLVED}(4,x,\mu^2)$/$xc^{\rm FOPT}(4,x,\mu^2)$ 
for the scales $\mu^2 =$2, 3, 4, 5, 10, 100 in units of 
$({\rm GeV}/{\rm c})^2$.  (a) LO, (b) NLO, (c) NNLO.
%-----------------------------------------------
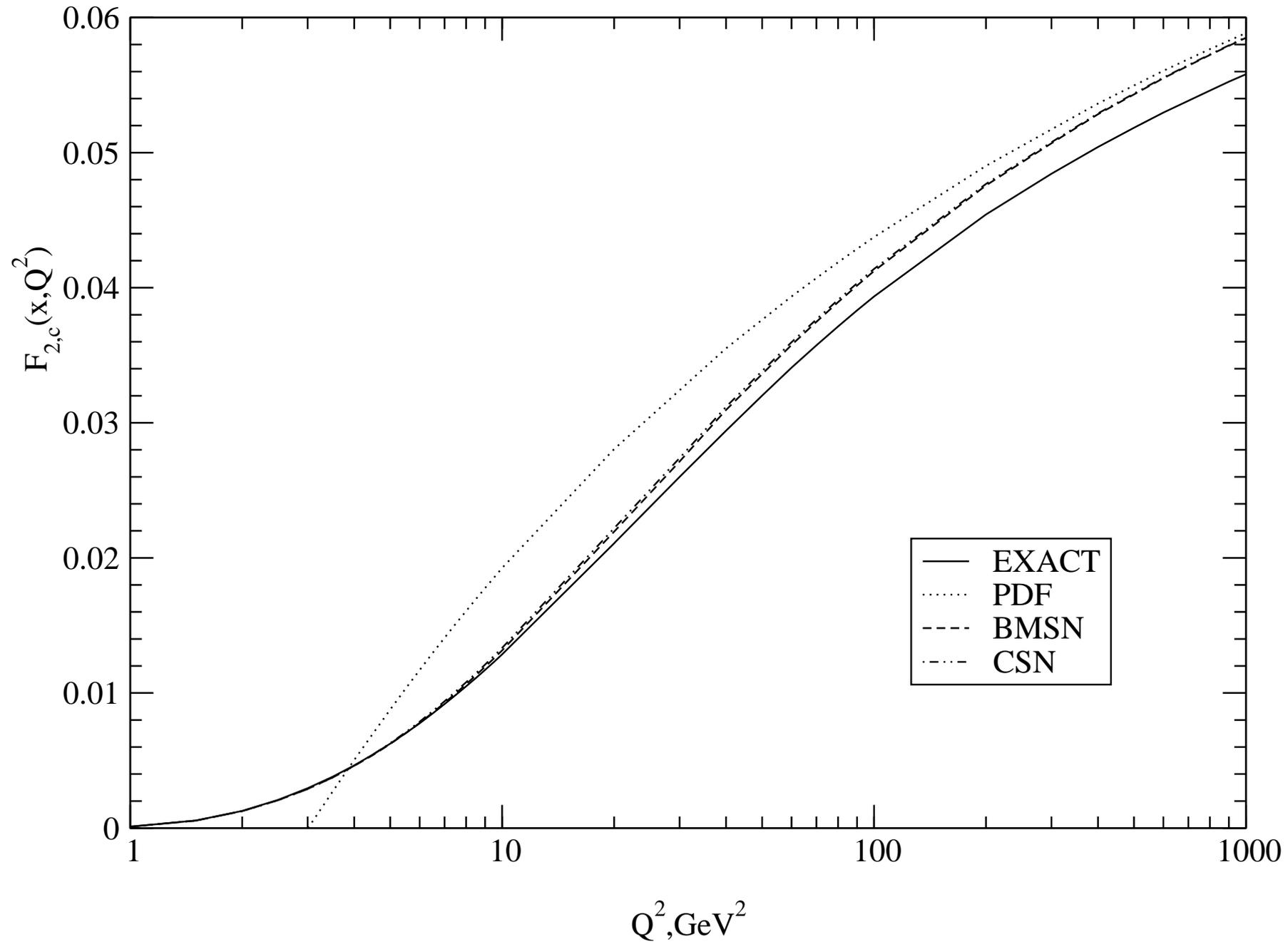
\item[Fig. 11.]
The charm quark structure functions 
$F_{2,c}^{\rm EXACT}(n_f=3)$ (solid line) 
$F_{2,c}^{\rm CSN}(n_f=4)$, (dot-dashed line)  
$F_{2,c}^{\rm BMSN}(n_f=4)$, (dashed line) and
$F_{2,c}^{\rm PDF}(n_f=4)$, (dotted line) 
in NNLO for $x=0.05$ 
plotted as functions of $Q^2$.
%-------------------------------------------------
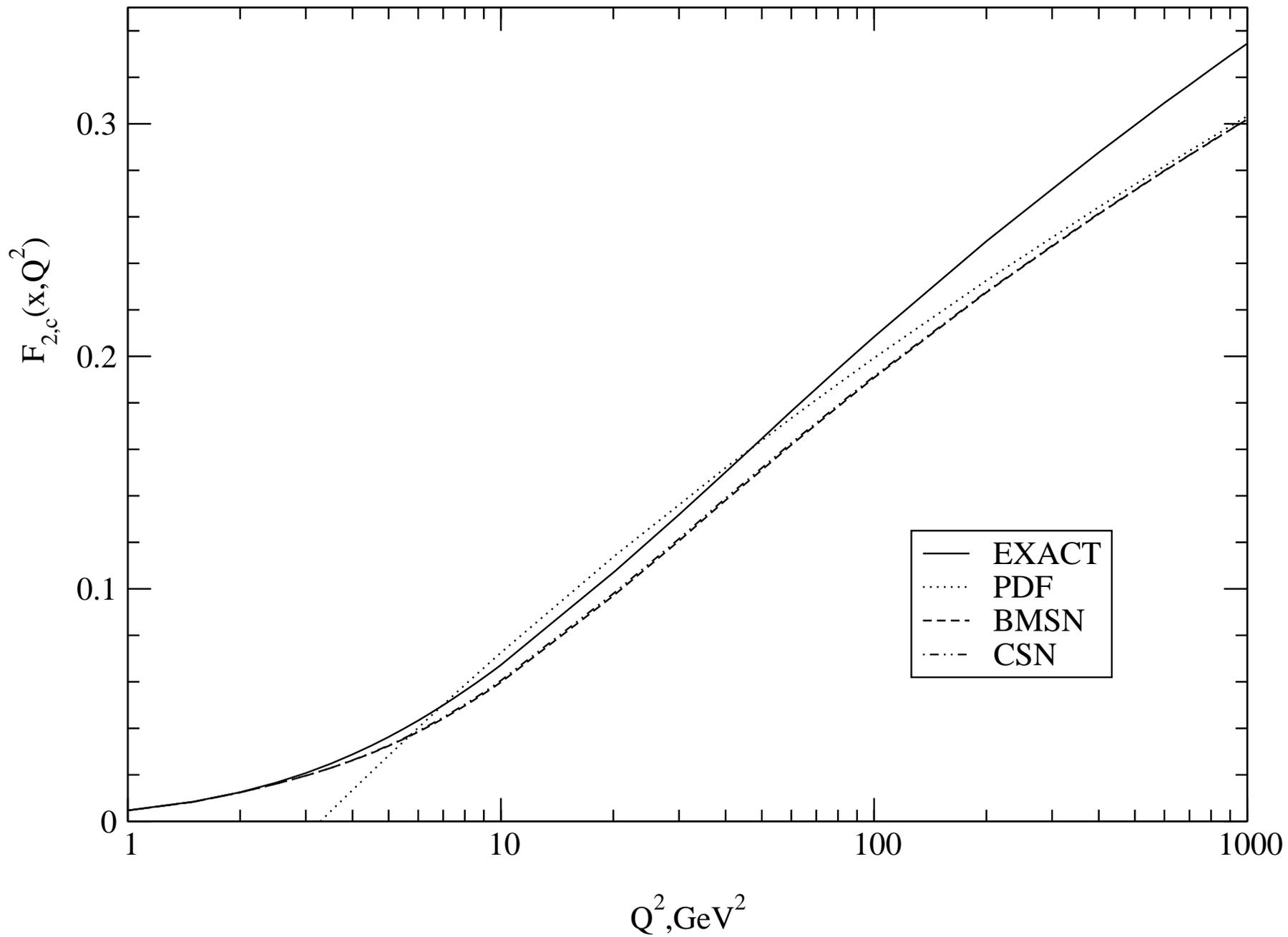
\item[Fig. 12.] Same as in Fig. 11 but now for $x=0.005$.
%-------------------------------------------------------
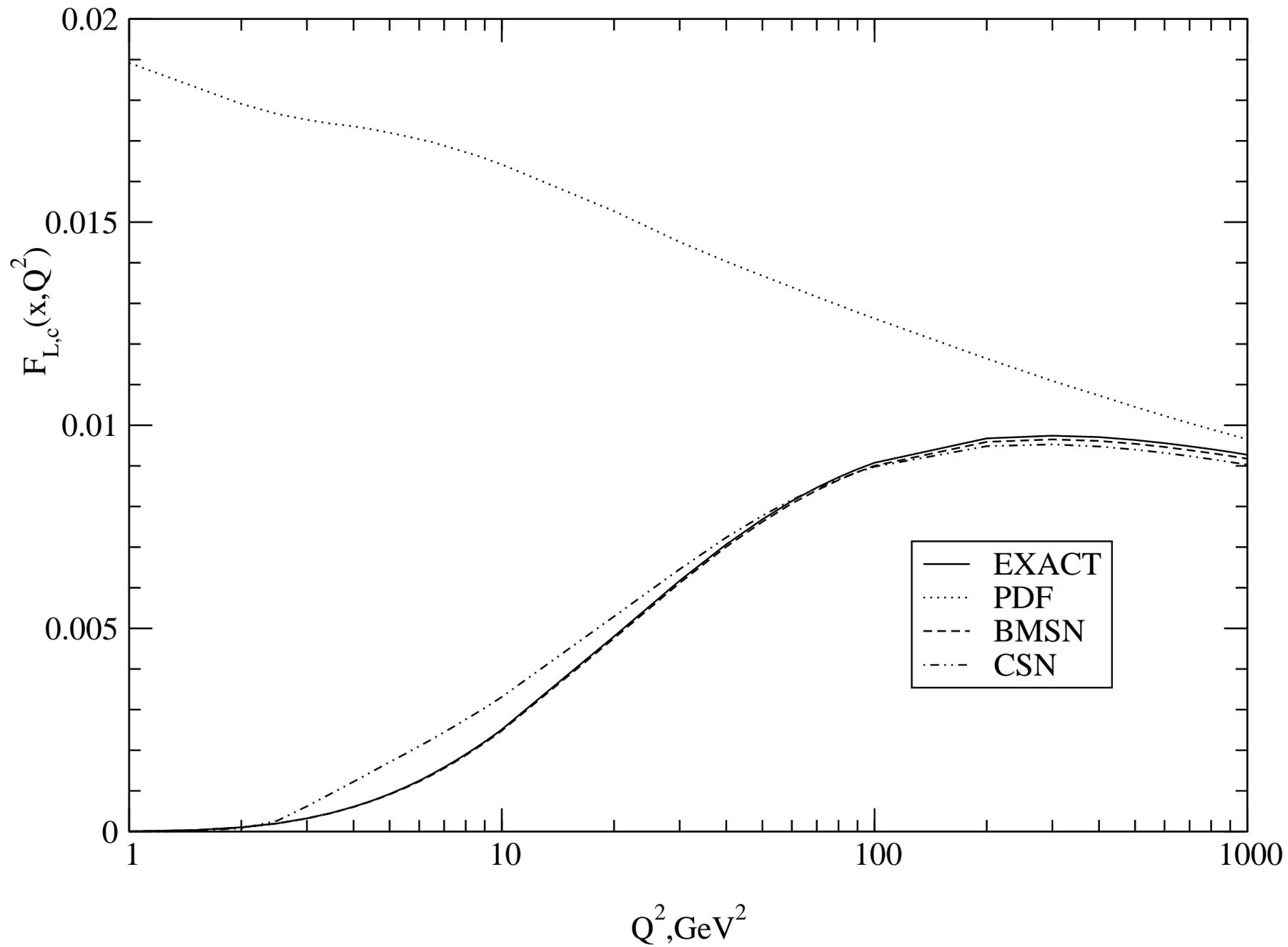
\item[Fig. 13.]
The charm quark structure functions 
$F_{L,c}^{\rm EXACT}(n_f=3)$ (solid line) 
$F_{L,c}^{\rm CSN}(n_f=4)$, (dot-dashed line)
$F_{L,c}^{\rm BMSN}(n_f=4)$, (dashed line) and
$F_{L,c}^{\rm PDF}(n_f=4)$, (dotted line) 
in NNLO for $x=0.05$
plotted as functions of $Q^2$.
%------------------------------------------------
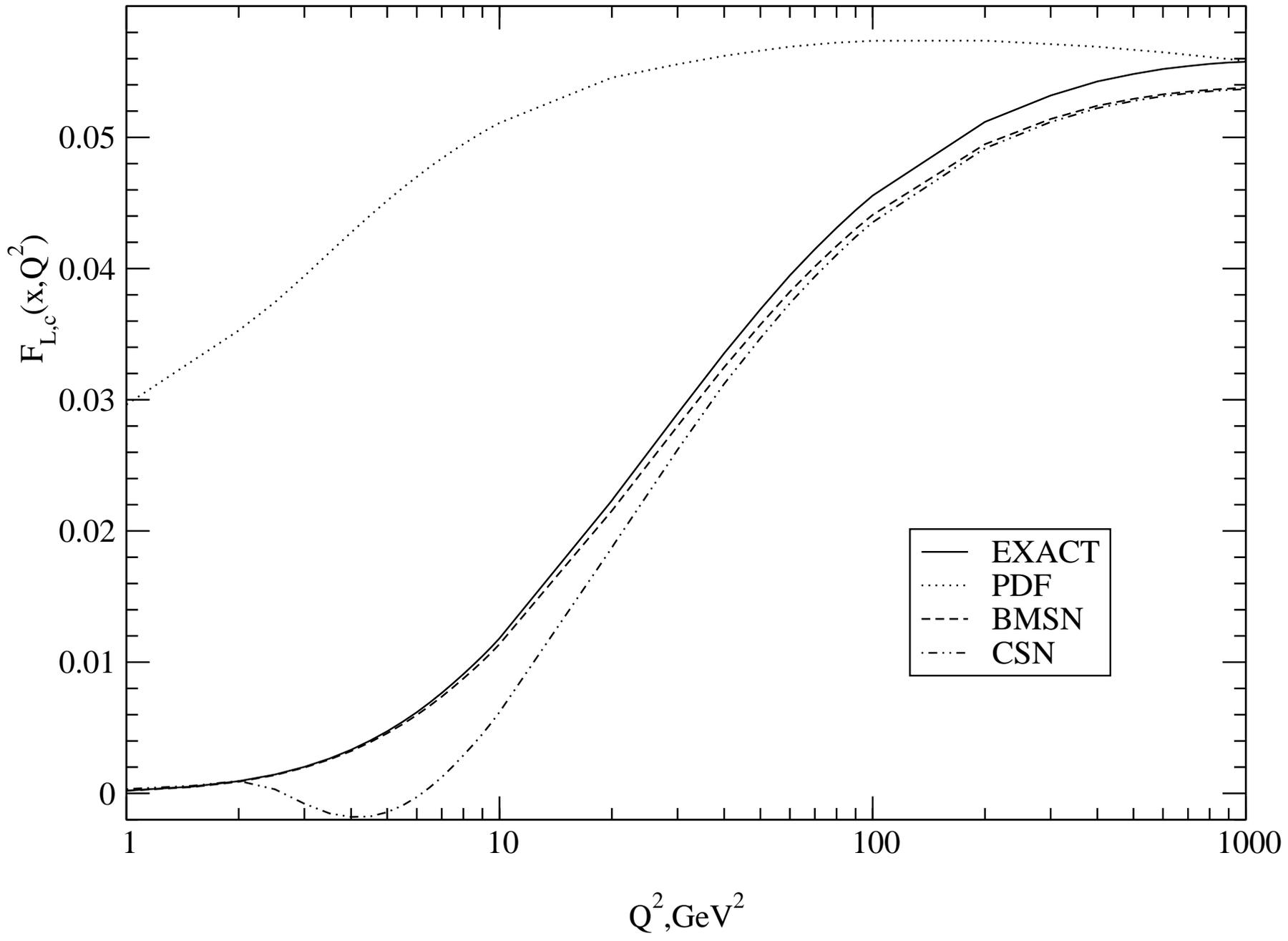
\item[Fig. 14.] Same as in Fig. 13 but now for $x=0.005$.
%---------------------------------------------------
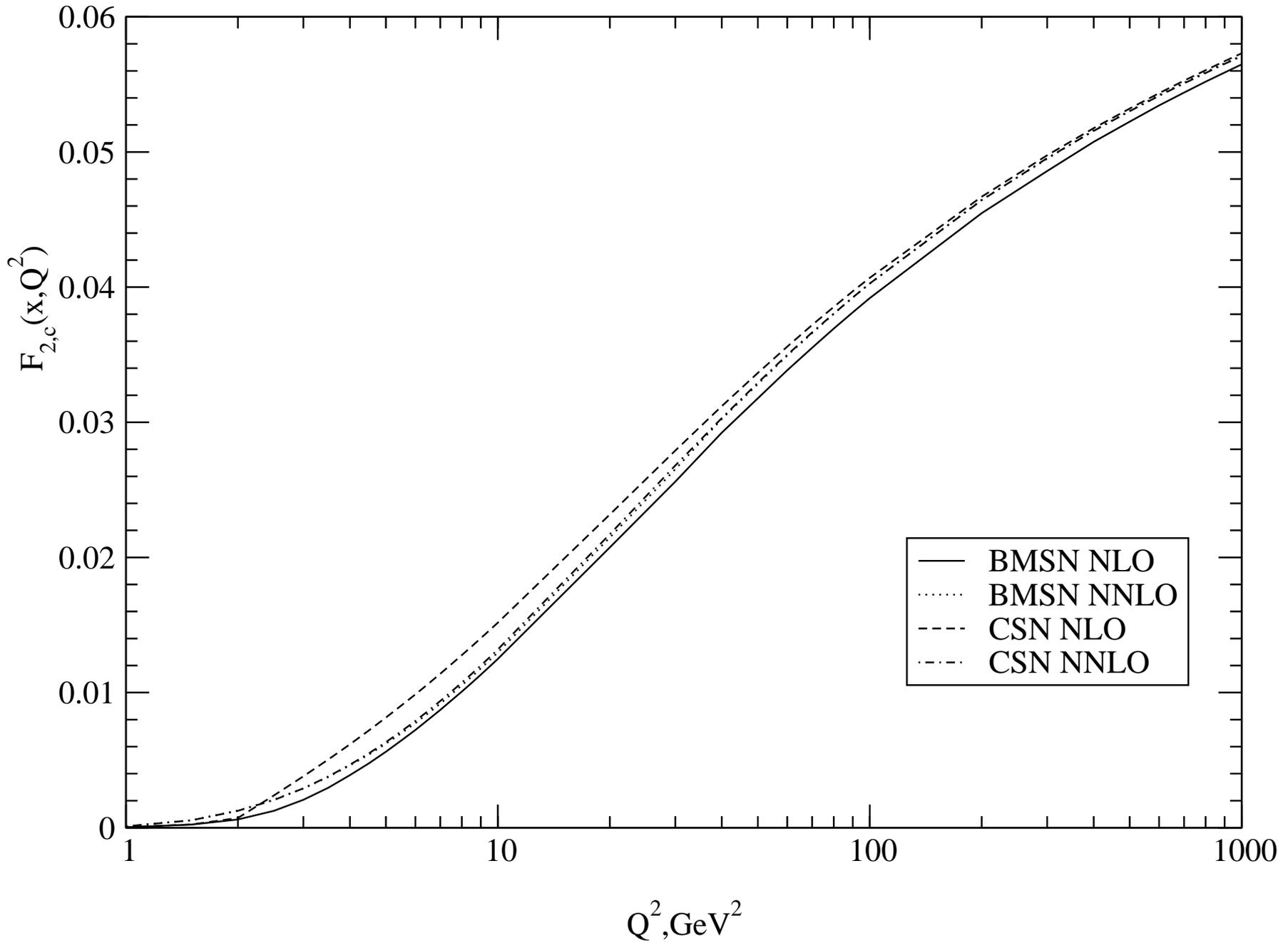
\item[Fig. 15.] 
The charm quark structure functions 
$F_{2,c}^{\rm BMSN}(n_f=4)$ in NLO (solid line), NNLO 
(dotted line) for $x=0.05$ and 
$F_{2,c}^{\rm CSN}(n_f=4)$ in NLO (dashed line), 
NNLO (dot-dashed line) for $x=0.05$
plotted as functions of $Q^2$.
%------------------------------------------------------
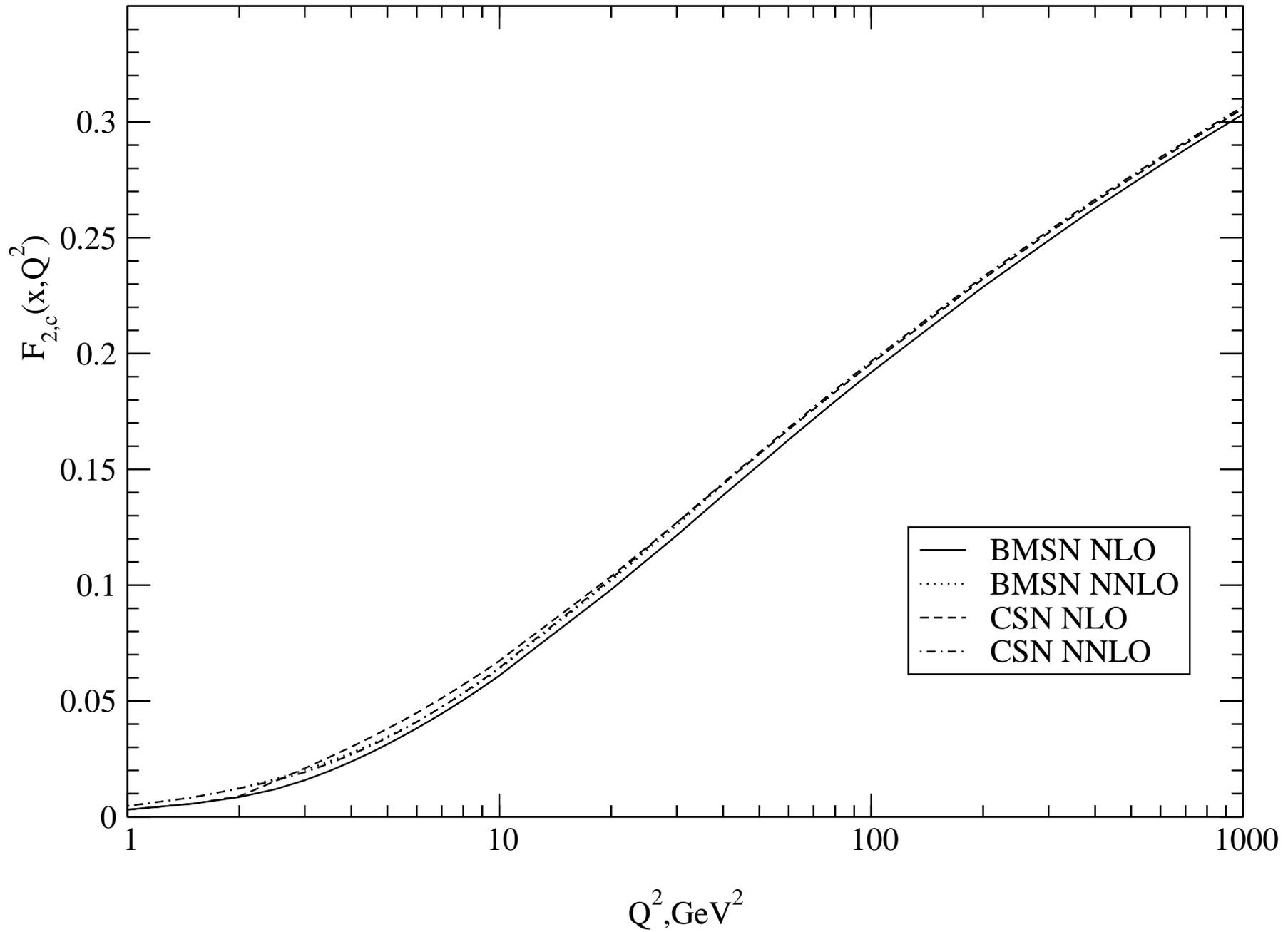
\item[Fig. 16.] Same as in Fig. 15 but now for $x=0.005$.
%-----------------------------------------------------
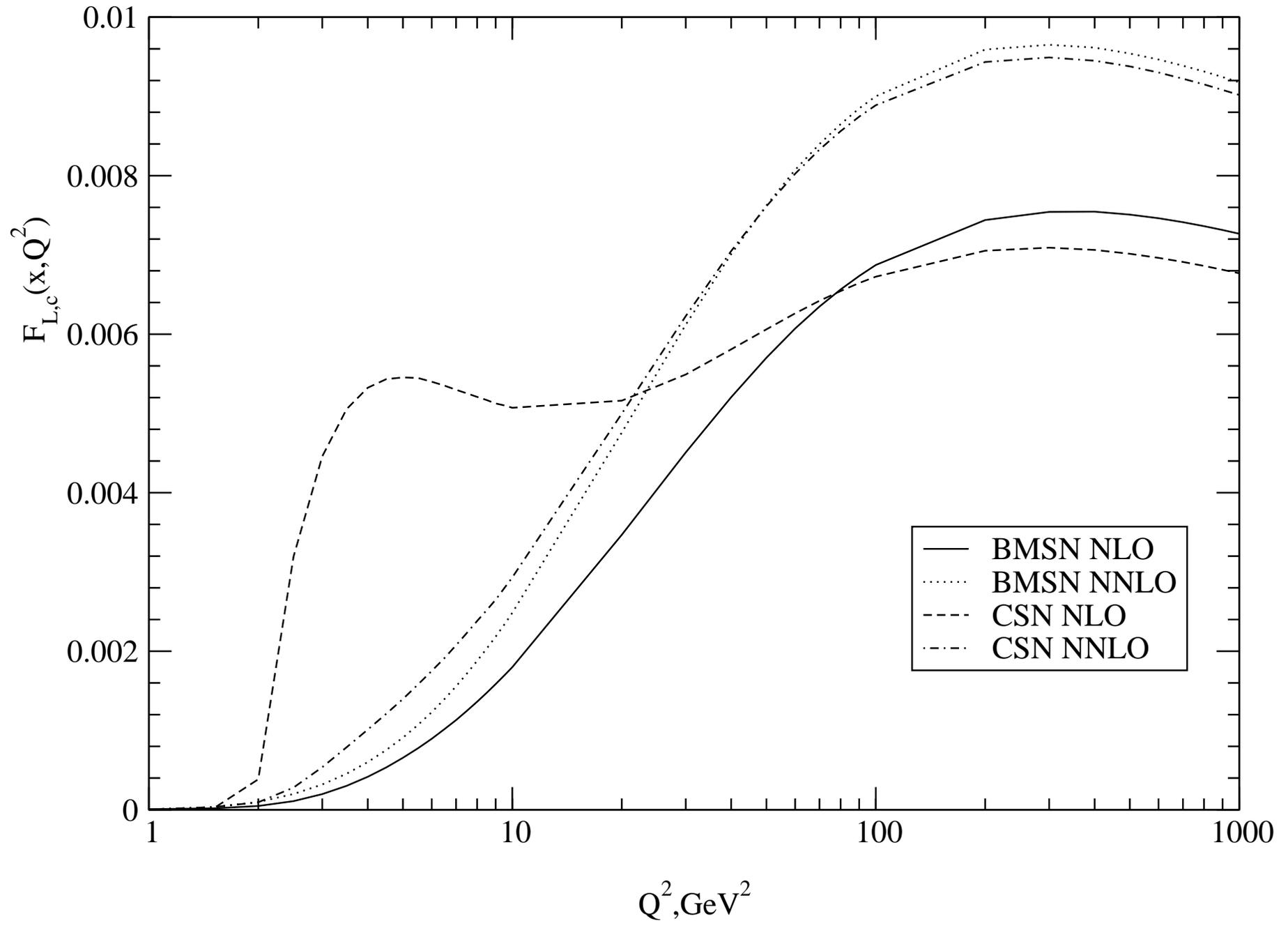
\item[Fig. 17.]
The charm quark structure functions 
$F_{L,c}^{\rm BMSN}(n_f=4)$ in NLO (solid line), NNLO 
(dotted line) for $x=0.005$ and
$F_{L,c}^{\rm CSN}(n_f=4)$ in NLO (dashed line),
NNLO (dot-dashed line) for $x=0.05$ 
plotted as functions of $Q^2$.
%----------------------------------------------------
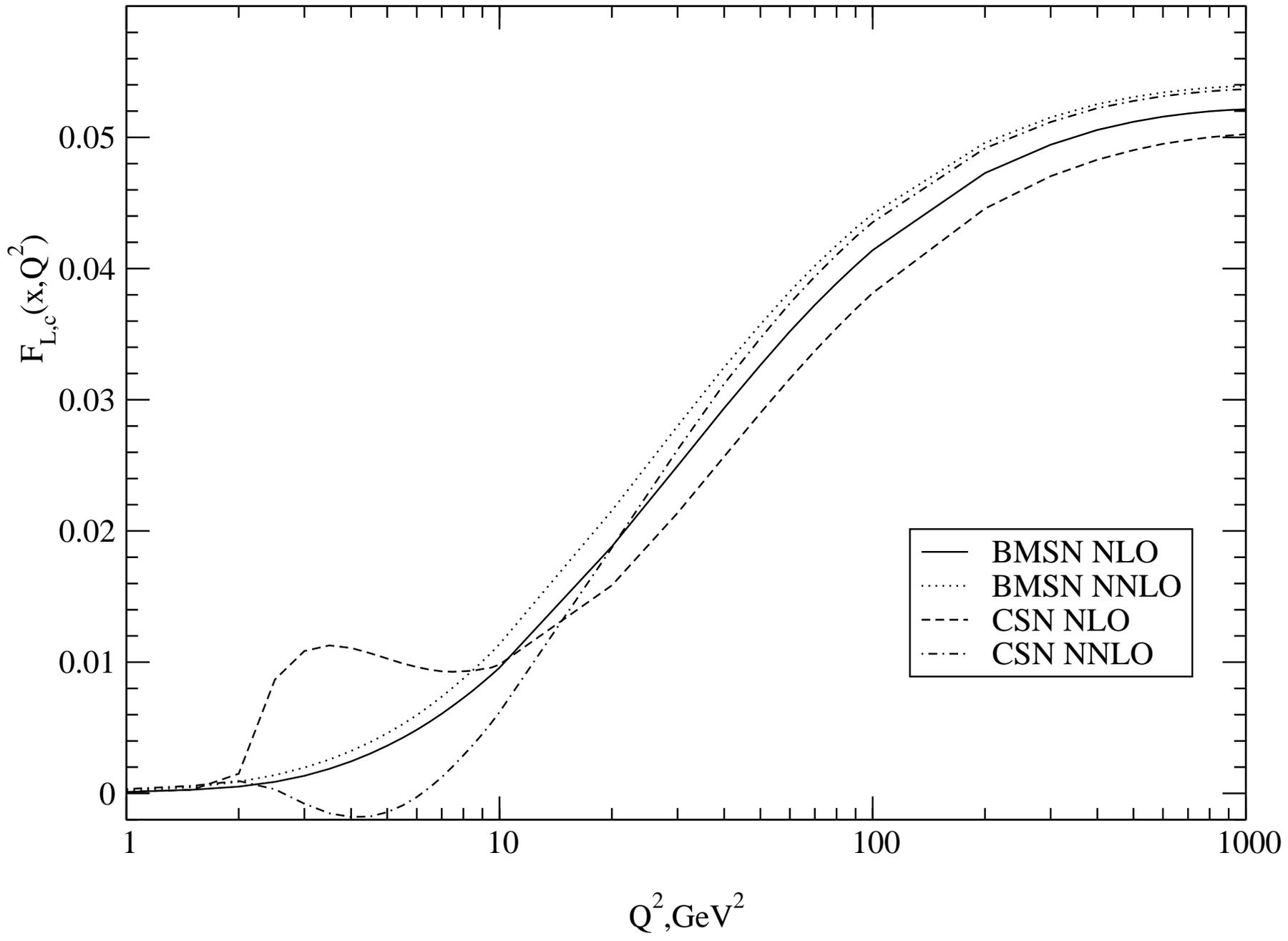
\item[Fig. 18.] Same as in Fig. 17 but now for $x=0.005$.
%---------------------------------------------------------
\end{description}
%\end{document}

%------------------------------
\end{document}